%% file: inclxx-hi-bbl.tex
\documentclass[aps,prc,twocolumn,a4paper,english,superscriptaddress,showpacs,showkeys]{revtex4-1}

\usepackage[ansinew]{inputenc}
\usepackage[T1]{fontenc}
\usepackage[english]{babel}
\usepackage[usenames,dvipsnames]{color}
\usepackage{graphicx}
\usepackage{amsmath}
\usepackage{amssymb}
\usepackage{multirow,bigdelim}
\usepackage{hyperref}
\usepackage{bm}
\usepackage{xcolor}
\usepackage{listings}
\lstdefinestyle{customc}{
  breaklines=true,
  xleftmargin=\parindent,
  language=C++,
  showstringspaces=false,
  basicstyle=\footnotesize\ttfamily,
  keywordstyle=\bfseries,
  commentstyle=\rmfamily\itshape,
  captionpos=b,
  columns=fullflexible
}
\newcommand{\code}[1]{#1}
\newcommand{\modified}[1]{#1}
\newcommand{\cxx}{\code{C++}}

\newcommand{\fortran}{\code{Fortran}}
\newcommand{\fortranss}{\code{Fortran77}}
\newcommand{\inclxx}{\code{INCL++}}
\newcommand{\inclf}{\code{INCL4.6}}
\newcommand{\inclfold}{\code{INCL4.3}}
\newcommand{\incl}{\code{INCL}}
\newcommand{\bic}{\code{BIC}}
\newcommand{\qmd}{\code{QMD}}
\newcommand{\cem}{\code{CEM}}
\newcommand{\laqgsm}{\code{LAQGSM}}

\newcommand{\abla}{\code{ABLA07}}
\newcommand{\ablabare}{\code{ABLA}}
\newcommand{\ablaold}{\code{ABLA~V3}}
\newcommand{\gem}{\code{GEM}}

\newcommand{\geh}{\code{G4ExcitationHandler}}

\newcommand{\etal}{et al.}
\newcommand{\proton}{\textit{p}}

\newcommand{\geant}{\code{Geant4}}

\newcommand{\mcnpx}{\code{MCNPX}}
\newcommand{\mcnpsix}{\code{MCNP6}}
\newcommand{\vect}[1]{\ensuremath{\boldsymbol{#1}}}

\graphicspath{{./figures/}{./figures_eps/}}

\begin{document}

\newsavebox{\spallationurl}
\sbox{\spallationurl}{\url{http://www-nds.iaea.org/spallations}}
\newsavebox{\inclurl}
\sbox{\inclurl}{\url{http://irfu.cea.fr/Sphn/Spallation/incl.html}}

\title{Extension of the Li\`ege Intranuclear-Cascade model to reactions induced
  by light nuclei}

\author{Davide Mancusi}%
\thanks{Member of the \geant\ collaboration}%
\email[corresponding author. E-mail address: ]{davide.mancusi@cea.fr}%
\affiliation{CEA, Centre de Saclay, Irfu/SPhN, F-91191
  Gif-sur-Yvette, France}
\author{Alain Boudard}%
\thanks{Former member of the \geant\ collaboration.}%
\affiliation{CEA, Centre de Saclay, Irfu/SPhN, F-91191
  Gif-sur-Yvette, France}
\author{Joseph Cugnon}%
\affiliation{Fundamental Interactions in Physics and Astrophysics, University of
  Li\`{e}ge, all\'{e}e du 6 ao\^{u}t 17, b\^{a}t.~B5, B-4000 Li\`{e}ge 1,
  Belgium}
\author{{Jean-Christophe} David}%
\thanks{Member of the \geant\ collaboration.}%
\affiliation{CEA, Centre de Saclay, Irfu/SPhN, F-91191
  Gif-sur-Yvette, France}
\author{Pekka Kaitaniemi}%
\thanks{Former member of the \geant\ collaboration.}%
\affiliation{CEA, Centre de Saclay, Irfu/SPhN, F-91191
  Gif-sur-Yvette, France}
\author{Sylvie Leray}%
\affiliation{CEA, Centre de Saclay, Irfu/SPhN, F-91191
  Gif-sur-Yvette, France}

\date{\today}

\begin{abstract}
  The purpose of this paper is twofold. First, we present the extension of the
  Li\`ege Intranuclear Cascade model to reactions induced by light ions. We
  describe here the ideas upon which we built our treatment of nucleus-nucleus
  reactions and we compare the model predictions against a vast set of
  heterogeneous experimental data. In spite of the discussed limitations of the
  intranuclear-cascade scheme, we find that our model yields valid predictions
  for a number of observables and positions itself as one of the most attractive
  alternatives available to \geant\ users for the simulation of
  light-ion-induced reactions. Second, we describe the \cxx\ version of the
  code, which it is physics-wise equivalent to the legacy version, is available
  in \geant\ and will serve as the basis for all future development of the
  model.
\end{abstract}

\pacs{24.10.Lx, 25.40.Sc, 25.70.-z, 25.70.Mn}

\keywords{spallation reactions; intranuclear cascade; light-ion-induced
  reactions; nucleus-nucleus reactions}

\maketitle

\section{Introduction and motivation}

Reactions involving light ions (defined as $A\leq18$ for the purpose of this
paper) play an important role in several applications. In hadrontherapy, for
instance, cancer patients are treated using accelerated beams of protons or
light ions \cite{kraft-radiotherapy,*kraft-radiotherapy2}. Nuclear reactions
between the beam particles and the body of the patient can be responsible for
dose deposition outside the clinical target volume, which is
undesirable. Moreover, it has been demonstrated that the production of $\beta^+$
emitters in nuclear reactions can be profitably employed to monitor dose
deposition in proton \cite{parodi-pet_protons} or carbon treatment
\cite{enghardt-pet_carbon}.

The radiation environment in space also involves energetic protons and heavy
ions \cite{durante-radiation}. The Galactic Cosmic Rays are one of the
contributing sources to radiation in the Solar System; their hadronic component
mainly consists of protons and alpha particles, but ions as heavy as iron are
known to yield sizable contributions to the equivalent dose absorbed by space
crews. Shielding against cosmic radiation relies on nuclear reactions to reduce
the health hazard.

Light-ion-induced nuclear reactions are also involved in the production of beams
of unstable nuclei. The in-flight projectile-fragmentation method
\cite{morrissey-rib_in_flight} is often realized using $^9$Be production
targets. Radioactive beams produced with the ISOL method
\cite{ravn-rib_isol_method} typically rely on light charged particles (LCPs) to
induce spallation or fission in the production target. In either case, the
luminosity of the secondary beam crucially depends on the fragment yields in
light-ion-induced reactions.

Reactions on light nuclei are also often used in fundamental research at the
limits of nuclear stability, for instance in the quest for very neutron-rich or
neutron-poor residues
\cite[e.g.][]{stolz-sn,*benlliure-south,*benlliure-xe,*alvarez_pol-u238,blank-proton_rich,*kurcewicz-u238}.
Light targets are also employed to extract information about the properties of
exotic nuclei from their response to dynamical solicitation
\cite[e.g.][]{hansen-review_direct_reactions}; see also the work described in
Ref.~\citenum{audirac-evaporation_cost}, which is particularly relevant to our
subject because it was performed with a preliminary version of the model
described in the present paper.

The applications listed above typically involve projectile energies of the order
of a few tens to several hundreds or even thousands of MeV per nucleon. Since a
great deal of reaction channels are open in this energy regime, it is unfeasible
to conduct systematic and comprehensive measurement campaigns for all the
relevant observables. Semi-empirical deterministic transport codes
\cite[e.g.][]{sihver-HIBRACnew,wilson-hzetrn} and hadrontherapy-targeted
treatment-planning systems \cite[e.g.][]{kramer-treatmentGSI_phys} can be
constructed around a restricted number of measured observables. Such codes are
usually sufficient to ensure adequate reproduction of the existing data;
however, their predictive power is essentially limited to the selected
observables in a restricted regime. Thus, there is a need for predictive,
physics-based nuclear-reaction models that can be used as all-round tools at the
bleeding edge of fundamental and applied research.

Above some 100~MeV incident energy, the nucleon-nucleus reaction dynamics can be
described as a sequence of independent nucleon-nucleon interactions taking place
in a common mean-field potential \cite{filges-handbook,bunakov-inc}. This
approximation gives rise to the intranuclear-cascade (INC) class of models,
which help shed some light on the reaction mechanism and have proven predictive
even below their nominal low-energy limit of validity. In particular, the
Li\`ege Intranuclear Cascade (\incl) \cite{boudard-incl4.6}, coupled with the
\abla\ statistical de-excitation code \cite{kelic-abla07}, has been recognized
as one of the most accurate models available on the market by the Benchmark of
Spallation Models \cite{leray-intercomparison,*intercomparison-website}, an
intercomparison of event generators for nucleon-induced reactions in the
60--3000~MeV incident-energy range, organized under the auspices of the
IAEA. The \incl\ model is a full Monte-Carlo event generator written in
\fortranss. The latest version of the \fortranss\ code is named \inclf\ and it
is described in detail in recent publications \cite{boudard-incl4.6}. As such,
it represents an ideal starting point for an extension to reactions induced by
light ions. A simple extension to light-ion-induced reactions, based on an old
\fortranss\ version of the model, was attempted a few years ago
\cite{kaitaniemi-incl_geant4}. The model yielded promising physical results, but
maintaining the code quickly grew to the proportions of a formidable task. This
was mainly due to the fact that the \fortranss\ version was monolithic, hardly
flexible and not very legible from the start.  This is one of the motivations
that has led us to redesign the \incl\ code from scratch and cast it in modern,
object-oriented \cxx.

Before describing the light-ion extension, we need to define the framework of
the model and introduce the \cxx\ redesign of the \incl\ code, named \inclxx\
(Section~\ref{sec:model-description}).
\modified{The physics of the new code is substantially equivalent to the
  reference \fortranss\ version \inclf\ for nucleon- and pion-induced reactions;
  the few minor differences will be highlighted in
  section~\ref{sec:diff-with-inclf}.}
We then introduce the extension of the \inclxx\ model to light-ion-induced
reactions (section~\ref{sec:light-ion-extension}). The differences between
\inclxx\ and the legacy \fortran\ \incl\ code are highlighted in
Section~\ref{sec:comp-with-inclf}. The predictions of the light-ion extension
are compared with a variety of experimental data in
Section~\ref{sec:comparison-exp-data}. We collect our conclusions in
section~\ref{sec:conclusions}. Some \geant-specific information about the use of
\inclxx\ within this particle-transport toolkit are given as an appendix.

\section{Model description}\label{sec:model-description}

The Li\`{e}ge Intranuclear Cascade model (\incl) \cite[][official web site:
\usebox{\inclurl}]{boudard-incl4.6} is one of the most refined existing tools
for the description of nucleon- and pion-induced reactions in the 50--3000-MeV
incident-energy range. The model is currently maintained and jointly developed
by the University of Li\`{e}ge (Belgium) and CEA-Saclay (France).

In this framework, the high-energy projectile initiates an avalanche of binary
collisions within the target nucleus. Particles (nucleons and pions) are assumed
to move in a spherical calculation volume, whose radius $R_\text{max}$ is
defined to be large enough to intersect essentially all impact parameters
leading to inelastic reactions. Binary particle-particle collisions are subject
to Pauli blocking. Emission of nucleons, pions and light clusters is possible;
light clusters, in particular, are produced via a dynamical phase-space
coalescence algorithm. The cascade stops when the remnant nucleus shows signs of
thermalization; a rather unique aspect of \incl\ is the self-consistent
determination of the cascade stopping time. The \incl\ model is not to be
considered as adjustable. It does contain parameters, but they are either taken
from known phenomenology (such as the matter density radius of the nuclei) or
have been adjusted once for all (such as the parameters of the Pauli blocking or
those that determine the coalescence module for the production of the light
charged clusters). The validity of the \incl\ model in the 50--3000-MeV
incident-energy range has been extensively demonstrated by the ``Benchmark of
Spallation Models'' \cite{leray-intercomparison,*intercomparison-website},
sponsored by IAEA.

We now turn to the description of the details that are specific to the \inclxx\
model. In what follows, and unless otherwise specified, we shall refer to
\inclxx\ \code{v5.1.14}, which is the version that was released to the public
along with \geant\ \code{v10.0} in December 2013. Subsequent patches to \geant\
\code{v10.0} introduce very few changes to the core of the model. More detailed
information about \inclxx\ in \geant\ are presented as an appendix to the
present paper.

\subsection{Differences with \inclf}\label{sec:diff-with-inclf}

We try to limit our description to those aspects of \inclxx\ that are different
from the reference \fortran\ version, \inclf\ \cite{boudard-incl4.6}. In some
cases, however, a brief presentation of the reference model needs to be
included, for clarity's sake.

We mentioned above that \inclxx\ was designed to be physically equivalent to
\inclf\ as far as nucleon- and pion-induced reactions are
concerned. Nevertheless, in some cases we deliberately chose to introduce some
minor difference for the sake of simplicity or consistency. In particular, the
treatment of pions in the two codes notably differs for the following details.

First, the radius of the pion potential in \inclf\ is taken to be
$R_\pi=(R_0+R_\text{max})/2$, where $R_0$ represents the surface half-density
radius and $R_\text{max}$ is the radius of the calculation volume. This means
that pions are assumed to quit the INC at $r=R_\pi$; however, incoming pions
still enter at $r=R_\text{max}$. This would pose some problems of consistency in
the stricter \inclxx\ code. Thus, for simplicity, pions in \inclxx\ always enter
and leave their potential (and the calculation volume) at $r=R_\text{max}$. It
is in principle possible to take into account the fact that the radius of the
pion potential is sensibly smaller than $R_\text{max}$
; however, we verified that pion spectra from nucleon-induced reactions are
insensitive to the potential radius in \inclf. Therefore, the refinement seems
unwarranted.

Second, the \inclf\ code introduced a special procedure, named ``local $E$''
\cite{boudard-incl4.6}, which tries to correct for the unrealistically large
momentum content of the nuclear surface in the nuclear model underlying
\incl. When a nucleon is involved in a collision, its kinetic energy is
preemptively reduced by an amount that depends on its position (the correction
is large at the surface of the nucleus). The nucleon momentum is rescaled
accordingly. This procedure tries to capture the fact that nucleons in the
surface are close to the turning point of their classical trajectories and,
thus, less kinetic energy is available for the collision. The local-$E$
correction is instrumental for the description of nucleon-nucleus reaction cross
sections at low incident energy \cite[Section~II.C.4.b
in][]{boudard-incl4.6}. For consistency with nucleon-induced reactions, the same
procedure is applied to the nucleon involved in the first collision of
pion-nucleus reactions in \inclxx. This has some consequences, as we shall
illustrate in Section~\ref{sec:comp-with-inclf}.

Third, the \inclf\ code introduced a dependence of the calculation-sphere radius
($R_\text{max}$) on the nucleon-nucleon ``interaction range''
\cite[Section~II.D.3 in][]{boudard-incl4.6}. In \inclxx\, the interaction range
is taken to be equal to the interaction distance used in the low-energy fusion
sector (Eq.~\eqref{eq:interaction_distance} below). This is only done for
consistency and has no physical consequence.

Finally, target preparation for $A\leq4$ is treated differently. The \inclf\
code singles out this special case and imposes that the sum of the momenta and
positions of the target nucleons should vanish, as appropriate for the
center-of-mass system; however, these conditions are not conserved during the
cascade, even in absence of collisions, due to the presence of the target
potential. Moreover, the assumed root-mean-square momenta for these targets are
inconsistent with those used when the same nuclei are considered as
projectiles. In \inclxx, instead, target preparation is consistent for all
targets. (Note however that the shape of the momentum distribution is still
taken to be different for projectiles and targets, for reasons explained in
detail in Section~\ref{sec:proj-targ-asymm}.)

Up to the four differences mentioned so far, we can state that \inclxx\ is
physically equivalent to \inclf\ as far as nucleon- and pion-induced reactions
are concerned.

Additional differences specifically concern reactions induced by light
nuclei. The value of the ``Coulomb radius'' (related to the Coulomb barrier, as
explained in Ref.~\citenum{boudard-incl4.6}) for $^3$He was found to be
inadequate and replaced with the value used for $^4$He. This will be illustrated
in Section~\ref{sec:react-induc-light}. Moreover, polarization of incident
deuterons \cite[Section~II.D.4.d]{boudard-incl4.6} is neglected in \inclxx.

The most important difference between \inclf\ and \inclxx, however, is the
ability of the latter to treat reactions induced by light ions, as detailed in
Section~\ref{sec:light-ion-extension}. Note however that the light-ion extension
led us to modify the low-energy fusion model even for incident light charged
particles ($A\leq4$), for consistency with the light-ion sector. Therefore, the
\inclxx\ predictions for LCP projectiles at low energy are \emph{not} expected
to be in agreement with those made by \inclf, as we shall illustrate in
Section~\ref{sec:comp-with-inclf} below.

The rest of this Section concerns the detailed description of the extension to
light-ion projectiles. We start by illustrating the preparation of the
projectile nucleons in the laboratory frame (Sections~\ref{sec:prep-proj} and
\ref{sec:proj-bind-lorentz}), which takes into account Coulomb deviation by the
target nucleus (Section~\ref{sec:coulomb-deviation}). Nucleons entering INC
(Section~\ref{sec:geom-part-geom-1}) are adjusted to allow for excitation energy
in the projectile pre-fragment (Section~\ref{sec:excit-energy-proj}). The INC
phase proper is rather standard and is described in
Section~\ref{sec:intr-casc-phase}. At the end of INC, a projectile-like
pre-fragment is defined if possible
(Section~\ref{sec:defin-proj-like}). Reactions at low incident energy require a
special treatment and are discussed in Section~\ref{sec:low-energy-fusion}. The
limitations of the approach we describe are discussed in
Section~\ref{sec:proj-targ-asymm}. This completes the discussion of the
light-ion extension of INC; however, we also need to discuss the relevance of
statistical de-excitation models for the complete simulation of the
nucleus-nucleus reaction (Section~\ref{sec:de-excitation-stage}).

\subsection{Extension to light-ion-induced reactions}
\label{sec:light-ion-extension}

It has been demonstrated \cite{leray-intercomparison} that the Li\`ege
Intranuclear Cascade model can successfully reproduce a vast set of observables
pertaining to nucleon-induced reactions between a few tens of MeV and a few GeV,
which suggests that the model condenses the physics that is essentially relevant
in this energy range. It is therefore natural to take it as a starting point for
the development of a new model for light-ion-induced reactions.

The treatment of nucleus-nucleus reactions in an INC framework poses several
challenges that do not apply to nucleon-nucleus reactions. First and foremost,
there is no natural way of accounting for the binding of the projectile nucleus
within the INC scheme.  The cascade takes place in a single mean-field
potential, which is typically assumed to be that of the target nucleus; this
essentially amounts to neglecting the mean-field interaction between the
projectile constituents. This approximation might be tenable for central
collisions of a light projectile on a heavy target, which rarely lead to the
emission of a projectile-like fragment; however, it is clear that no model can
describe projectile fragmentation if the binding of the projectile nucleons is
neglected. Second, INC models typically do not treat the mean-field potentials
as dynamical quantities and assume that they do not evolve during the cascade
phase. This is justifiable for nucleon-nucleus reactions, where only a
relatively small fraction of the nucleons directly participates in the reaction,
but it is clear that pre-fragments produced in nucleus-nucleus reactions can be
very different from the initial reaction partners. Therefore, any collective
rearrangement of the mean field is beyond the reach of traditional INC models.
Third, nucleons in nuclei are endowed with Fermi motion. A realistic description
of the intrinsic momentum content of both reaction partners is necessary for an
accurate description of certain observables. This is somewhat at odds with the
independent-particle Fermi-gas model that is typically used to describe the
structure of the reaction partners, especially for light nuclei. The definition
of Pauli blocking is unambiguous only if the initial momentum distribution of
the nucleons is assumed to be a hard, uniform Fermi sphere. It is well-known
however that nucleons in light nuclei exhibit smoother distributions
\cite{deWittHuberts-momentum_distributions}, which manifest themselves (among
other things) in the momenta of nucleons from the break-up of the projectile.
This point will be further developed below (see
Sections~\ref{sec:proj-targ-asymm} and \ref{sec:part-prod}).

One way to tackle the problem of binding is to separately treat projectile and
target nucleons as bound in their respective mean field. This approach is
realized e.g.\ by Isabel \cite{yariv-isabel1,yariv-isabel2}. In this model, the
reaction dynamics results from the juxtaposition of two conflicting pictures:
the nuclei are alternatively depicted as collections of nucleons or as
continuous Fermi gases. Nucleons belonging to the projectile or to the target
only feel the projectile or the target potential, respectively. Additional
assumptions are clearly necessary to determine the dynamics of cascading
particles, which do not belong to either nucleus. In this work, we follow an
alternative approach. 

We briefly repeat here that an \incl-based extension to light-ion-induced
reactions has already been attempted \cite{kaitaniemi-incl_geant4} on the basis
of an old version of the model (\inclfold). We shall not dwell on the
differences between the two approaches here, mostly because the model described
in the present work is more sophisticated in several respects and should be
considered as the reference point for any future development.

\subsubsection{Preparation of the projectile}\label{sec:prep-proj}

\begin{table}
  \caption{Assumed single-particle space and momentum densities for light
    projectile nuclei (up to $A=18$). ``MHO'' stands for ``Modified Harmonic
    Oscillator'' and $p_F=270$~MeV/$c$. For target nuclei, the same
    space densities are used; however, hard Fermi spheres are used as momentum
    distributions \cite{boudard-incl}.\label{tab:densities}}
  \begin{ruledtabular}
    \begin{tabular}{c|c|c}
      &space, $\rho_p$&momentum, $\pi_p$\\
      \hline
      deuteron & \multicolumn{2}{c}{Paris-potential wavefunction
        \cite{lacombe-deuteron_paris}}\\
      \multirow{2}{*}{$2<A\leq6$} & Gaussian, RMS & \multirow{4}{*}{Gaussian,
        $\text{RMS}=\sqrt{\frac{3}{5}}p_F$}\\
      & from \cite{de_vries-densities} & \\
      \multirow{2}{*}{$6<A\leq18$} & MHO, parameters &\\
      &  from \cite{de_vries-densities} &\\
    \end{tabular}
  \end{ruledtabular}
\end{table}

The first step in the simulation of a light-ion-induced reaction is the
preparation of the projectile and target nuclei. Since the preparation of the
target is standard, we refer the reader to Ref.~\citenum{boudard-incl} and we
limit ourselves to describing the preparation of the projectile in its
center-of-mass (CM) frame. Let $A_{p}$ and $Z_{p}$ be the mass and charge number
of the projectile. Furthermore, let $\rho_p$ and $\pi_p$ be the single-particle,
isospin-independent space and momentum densities of the projectile nucleus. The
assumed parametrizations for $\rho_p$ and $\pi_p$ are shown in
Table~\ref{tab:densities}. The table is limited to $A\leq18$, which is the
maximum mass that can be treated as a projectile in \inclxx. While this limit is
mostly dictated by the needs for applications (reactions involved in
carbon-therapy, for instance, rarely involve two nuclei heavier than $A=18$), it
is clear that INC cannot handle the collective behavior of symmetric reactions
between heavy nuclei.

In the case of deuteron, projectile preparation is trivial: the relative
distance and momentum are independently drawn at random from the Paris space and
momentum wavefunction \cite{lacombe-deuteron_paris}. The directions of the
vectors are chosen isotropically. For heavier projectiles, we first draw $A_p$
isotropically-distributed vectors $\vect{w}_i$ from the space distribution
$\rho_p$; let $\vect{W}=A_p^{-1}\sum_i \vect{w}_i$ be the mean of the
$\vect{w}_i$ vectors; then the positions of the nucleons are defined as
\[
\vect{\rho}_i = \sqrt{\frac{A_p}{A_p-1}}\left(\vect{w}_i-\vect{W}\right)\qquad i=1,\ldots,A_p\text.
\]
By construction, these positions satisfy the relation
$\sum_i\vect{\rho}_i=0$. The scaling factor $\sqrt{A_p/(A_p-1)}$ is needed to
ensure that the variance of the $\vect{\rho}_i$ vectors is equal to the variance
of the $\rho_p$ distribution. The definition of the $\vect{\rho}_i$ vectors is
such that the first and second central moments of their distribution are equal
to the corresponding moments of $\rho_p$. In general, the $\vect{\rho}_i$
vectors do not strictly follow the $\rho_p$ distribution, except if the latter
is Gaussian; deviations from the shape of $\rho_p$ are smaller if the number of
nucleons is larger.

The CM momenta of the projectile nucleons $\vect{\pi}_i$ are constructed in a
similar way. Since the momentum distributions are taken to be Gaussian for all
projectile nuclei, the generated momenta are normally distributed with the
correct width parameter and sum up to zero total momentum, as appropriate for
the CM system.

\subsubsection{Projectile binding and Lorentz boost}\label{sec:proj-bind-lorentz}

\begin{figure}
  \includegraphics[width=\linewidth]{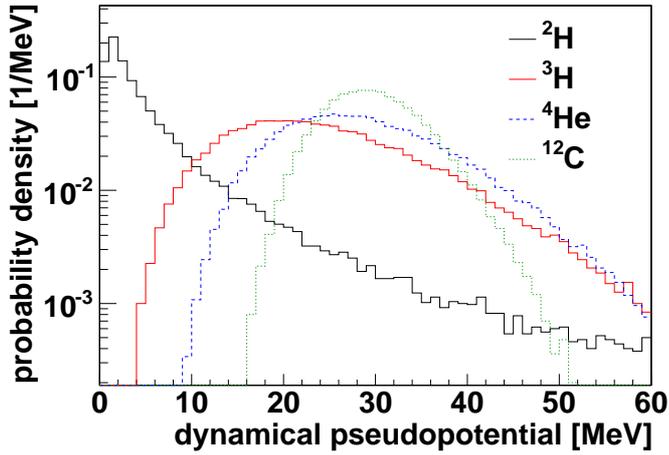}
  \caption{(Color online) Distributions of the dynamical pseudopotential used in the
    preparation of different light projectiles. Mean values and standard
    deviations are: $(8\pm18)$~MeV for deuteron, $(25\pm10)$~MeV for triton,
    $(29\pm9)$~MeV for $^4$He and $(30\pm5)$~MeV for
    $^{12}$C.\label{fig:pseudopotential}}
\end{figure}

We choose to account for the projectile binding by putting the nucleons off
their mass shell. During the INC phase, it is assumed in \incl\ that the proton
and neutron masses are equal, and they are set to the common value
$m=938.2796$~MeV\
\footnote{The exact masses of the outgoing nucleons are recovered by applying a
  correction that involves the empirical particle-emission thresholds
  \cite{boudard-incl4.6}.}.
Let $M_p$ be the mass of the projectile nucleus; we define the
\emph{dynamical pseudopotential} of the projectile as
\[
V_p=A_p^{-1}\left[\sum_{i=1}^{A_p}\sqrt{{\vect{\pi}_i}^2+m^2}-M_p\right]\text.
\]
This quantity should not be regarded as a physical potential, but rather as a
calculation device to enforce the nominal dispersion law in the laboratory
frame. The pseudopotential has the dimensions of an energy, is always positive,
and is equal to the opposite of the average potential energy that the nucleons
would feel if their total relativistic energy were to be equal to the nominal
mass of the projectile. Note that $V_p$ is a random variable because it depends
on the values of the drawn nucleon momenta. Typical distributions for the
pseudopotential are shown in Fig.~\ref{fig:pseudopotential}. The distribution
for deuteron is peculiar because its intrinsic momentum distribution is not
assumed to be Gaussian.

We define the nucleon relativistic energies in the center of mass as
\begin{equation}
  \varepsilon_i=\sqrt{{\vect{\pi}_i}^2+m^2}-V_p\text.\label{eq:off-shell_energy}
\end{equation}
The four-momenta of the projectile nucleons $(\varepsilon_i,\vect{\pi}_i)$ are
not on mass shell; however, they satisfy energy- and momentum-balance relations
that are appropriate for the center of mass of the projectile, namely:
\begin{subequations}\label{eq:center-of-mass}
  \begin{align}
    \sum \varepsilon_i&{}=M_p\text,\label{eq:center-of-mass_energy_balance}\\
    \sum \vect{\pi}_i&{}=0\text.
  \end{align}
\end{subequations}
Let $E_p$ indicate the total relativistic energy of the projectile nucleus;
assuming that the projectile moves along the positive direction of the $z$ axis,
let $\vect{P}_p=\Bigl(0,0,\sqrt{{E_p}^2-{M_p}^2}\Bigr)$ represent its momentum;
finally, let $\gamma=E_p/M_p$, $\beta=\sqrt{1-1/\gamma^2}$ and
$\vect{\beta}=(0,0,\beta)$ be the nominal Lorentz parameters of the
projectile. The four-momenta of the projectile nucleons in the laboratory frame
$(e_i,\vect{p}_i)$ are defined by a Lorentz boost on the CM four-momenta:
\begin{subequations}\label{eq:boost}
  \begin{align}
    e_i&{}=\gamma(\varepsilon_i+\vect{\beta}\cdot\vect{\pi}_i)\label{eq:nucleon_energy}\\
    \vect{p}_i&{}=\gamma(\vect{\pi}_i+\vect{\beta} \varepsilon_i)\text.\label{eq:nucleon_momentum}
  \end{align}
\end{subequations}
Eqs.~\eqref{eq:center-of-mass} guarantee that the energy and momentum balance
are correct:
\begin{subequations}\label{eq:balance}
  \begin{align}
    \sum e_i&{}=E_p\label{eq:lab_energy_balance}\\
    \sum \vect{p}_i&{}=\vect{P}_p\text.\label{eq:lab_momentum_balance}
  \end{align}
\end{subequations}
The positions of the nucleons in the laboratory frame take into account Lorentz
contraction along the $z$ axis.

\begin{figure}
  \includegraphics[width=\linewidth]{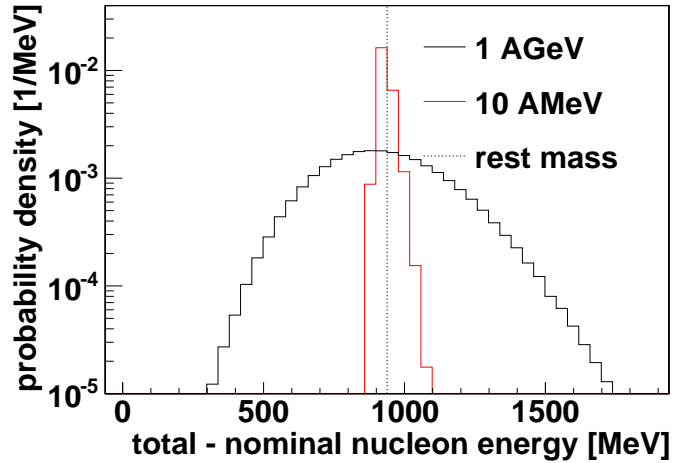}
  \caption{(Color online) Distribution of the total relativistic nucleon energy in the
    laboratory frame $e_i$ minus the nominal kinetic energy per nucleon of the
    projectile, for a $^{12}$C nucleus at 1~$A$GeV (black line) or 10~$A$MeV
    (red line).\label{fig:nucleon_energies}}
\end{figure}

We illustrate the procedure for the preparation of the nucleons in
Fig.~\ref{fig:nucleon_energies} with the distributions of nucleon energies for a
$^{12}$C nucleus at 10~$A$MeV and 1~$A$GeV. The nominal kinetic energy per
nucleon of the $^{12}$C nucleus was subtracted from the total nucleon energy. In
absence of Fermi motion, the distributions would collapse to a Dirac delta
function centered around the nucleon rest mass.

Note that Fermi motion induces a \emph{larger} spread at 1~$A$GeV than at
10~$A$MeV. This is a direct consequence of Eq.~\eqref{eq:nucleon_energy} which
is easy to visualize for non-relativistic velocities. Indeed, in this limit
Eq.~\eqref{eq:nucleon_energy} reduces to
\[
e_i=m+\frac{\vect\pi_i^2}{2m}+\vect{\pi}_i\cdot\vect\beta+\frac{m\vect{\beta}^2}{2}\text.
\]
For fixed absolute values of the boost speed $|\vect\beta|$ and of the nucleon
momentum $|\vect\pi_i|$, the fluctuations in $e_i$ are generated by the only
non-constant term on the r.h.s., namely the scalar product
$\vect\pi_i\cdot\vect\beta$, and are therefore proportional to $|\vect\beta|$,
i.e. they are more important at high energy than at low energy.

Summarizing, the procedure described above defines positions and four-momenta
for the $A_p$ projectile nucleons in the laboratory frame. The sum of the
nucleon four-momenta is equal to the nominal four-momentum of the projectile
nucleus. However, the nucleon four-momenta are off mass shell.

Finally, we have also verified that the projectile preparation algorithm is
relatively robust with respect to the choice of the reference frame: the nucleon
energies are essentially unchanged if we choose to introduce the dynamical
pseudopotential in the laboratory frame.

\subsubsection{Coulomb deviation}\label{sec:coulomb-deviation}

The projectile preparation step results in the definition of the (off-shell)
four-momenta of $A_p$ nucleons in the laboratory frame. The relative positions
of the nucleons in the laboratory frame are also defined. The initial positions
of the nucleons with respect to the target nucleus are defined by the impact
parameter and by an algorithm that takes into account the Coulomb deviation of
the projectile trajectory. The procedure used in \inclxx\ closely resembles the
one used in \inclf\ \cite{boudard-incl4.6}, to which the reader is referred. The
result of the algorithm is to define entrance positions and times for all
projectile nucleons into the calculation volume.

The main ingredient is the Coulomb radius $R_\text{Coul}$, a function of the
projectile and target species, which essentially defines the height of the
barrier. Compared to the \inclf\ algorithm, we have a different parametrisation
of $R_\text{Coul}$ for $^3$He projectiles --- we use the same formula for $^3$He
and $^4$He. For projectiles with $Z>2$, which lie outside the scope of the
\inclf\ model, a new prescription has to be given. We choose the following:
\[
R_\text{Coul}=\frac{e^2Z_pZ_t}{B_\text{Shen}}\text,\qquad(Z_p>2)
\]
where $Z_t$ is the target charge number and $B_\text{Shen}$ is the Coulomb
barrier calculated using Shen's parametrisation \cite{shen-xsec}:
\[
B_\text{Shen}=\frac{e^2Z_pZ_t}{R_p+R_t+3.2\text{~fm}}-a\frac{R_pR_t}{R_p+R_t}\text,
\]
with $R_i=(1.12A_i^{1/3}-0.94A_i^{-1/3})$~fm and $a=1$~MeV/fm.

\subsubsection{Geometrical participants, geometrical spectators and dynamical
  spectators}\label{sec:geom-part-geom-1}

An important ingredient of the nucleus-nucleus extension is the assumption that
projectile nucleons propagate with the (Coulomb-distorted) collective velocity
of the projectile beam until they undergo a collision. This has two
consequences. First, projectile nucleons can immediately be divided in two
classes: those whose trajectory intersects the \incl\ calculation volume are
labeled as \emph{geometrical participants}; the others are called
\emph{geometrical spectators}. If there are no geometrical participants, the
event is considered as transparent (no reaction). Second, the entrance times of
the geometrical participants in the calculation volume can be analytically
predicted. The entrance time of the first nucleon is taken as the start of the
intranuclear cascade.

It should be stressed that the distinction between geometrical participants and
spectators is not physical, because it is a consequence of the finite radius of
the \incl\ calculation volume, $R_\text{max}$, which is \emph{not} a physical
parameter.  Ideally, the model predictions (e.g.\ cross sections) should be
completely independent of $R_\text{max}$ (for sufficiently large values of
$R_\text{max}$). However, geometrical spectators never enter the calculation
volume, and thus cannot undergo any interaction. For continuity, the radius
$R_\text{max}$ must be taken sufficiently large so that the probability that a
geometrical participant entering close to $R_\text{max}$ undergoes a collision
is negligibly small. Still, this condition is not sufficient to ensure that the
model predictions are independent of $R_\text{max}$. Indeed, geometrical
participants can traverse the calculation volume without undergoing any
collision. Such particles, which we call \emph{dynamical spectators}, must be
treated on the same footing as the geometrical spectators. We shall discuss in
section~\ref{sec:defin-proj-like} to what extent this goal has been attained in
the current incarnation of \inclxx.

\subsubsection{Excitation and kinetic energy of the projectile-like pre-fragment}\label{sec:excit-energy-proj}

The intranuclear-cascade phase starts with one of the projectile nucleons
entering the calculation volume. This event can actually be seen as the transfer
of a nucleon from the projectile to the target nucleus. If we seek to conserve
energy during the whole intranuclear-cascade phase, the $Q$-value for nucleon
transfer must somehow be taken into account in the treatment of the incoming
nucleon. In the context of nucleon-induced reactions, this observation has led
us to introduce empirical thresholds for particle emission and absorption
\cite{boudard-incl4.6}: the energy of a particle entering and leaving the
nucleus is corrected according to differences of masses taken from tables
\cite{tuli-nuclear_wallet_cards}. In nucleus-nucleus reactions, the situation is
complicated by the possibility that nucleon transfer from the projectile to the
target may leave the projectile in an excited state. The intranuclear-cascade
model does not offer any natural prescription to fix the excitation energy of
the projectile-like pre-fragment. The reader should contrast this with the
excitation energy of the target nucleus, which can be naturally defined as a sum
over particle-hole excitations.  Therefore, we need to resort to a model to
define the excitation energy of the projectile-like pre-fragment.

We postulate that nucleon removal leads to a particle-hole-like excitation
energy in the projectile, too. More precisely, assume that only the $A$ nucleons
labeled by $i=1,\ldots,A$ are left in the projectile; then we define the
excitation energy as
\begin{subequations}\label{eq:proj-prefrag}
  \begin{equation}
    E^*_{A}=\sum_{j=1}^A\varepsilon_j - \sum_{j=1}^A\varepsilon_{i_j}\text.\label{eq:estar-proj}
  \end{equation}
  Here the second summation is intended to run over the $A$ smallest values of
  the CM energies $\varepsilon_i$ (Eq.~\eqref{eq:off-shell_energy}), which are
  collectively meant to represent a reference state for the $A$-nucleon
  pre-fragment. The excitation energy is computed as the difference between the
  total energy left in the pre-fragment CM and the energy of the reference
  state. It has the desirable properties of always being non-negative and of
  vanishing for $A=A_p$.

  The state of motion of the projectile pre-fragment is also perturbed by
  nucleon removal. Let $(E_{A},\vect{P}_{A})$ and $(E_{A-1},\vect{P}_{A-1})$ be
  the four-momenta of the pre-fragment before and after nucleon removal, $A$
  being the running mass of the projectile pre-fragment. At the beginning of
  INC, we have $A=A_p$, $E_{A}=E_{A_p}=E_p$ and
  $\vect{P}_A=\vect{P}_{A_p}=\vect{P}_p$.  Without lack of generality, we assume
  that the nucleons are removed from the projectile in decreasing index order
  (the $A_p$-th nucleon first, then the $(A_p-1)$-th, \ldots). When removing one
  nucleon, i.e. going from mass $A$ to mass $A-1$, the change in total momentum
  is taken equal to minus the momentum (Eq.~\eqref{eq:nucleon_momentum}) of the
  removed nucleon:
  \begin{equation}
    \vect{P}_{A-1}=\vect{P}_{A}-\vect{p}_{A}\text.\label{eq:momentum-proj}
  \end{equation}
  The total energy $E_{A-1}$ is defined by the dispersion relation:
  \begin{equation}
    E_{A-1}=\sqrt{{(M_{A-1}+E^*_{A-1})}^2+{\vect{P}_{A-1}}^2}\text,\label{eq:energy-proj}
  \end{equation}
\end{subequations}
where $M_{A}$ is the tabulated mass of the pre-fragment and the excitation
energy $E^*_A$ is given by Eq.~\eqref{eq:estar-proj} above.  Finally, if there
is more than one geometrical participant, the procedure is applied to each
nucleon transfer.


\subsubsection{Intranuclear-cascade phase}\label{sec:intr-casc-phase}

The excitation energy of the projectile-like pre-fragment,
Eq.~\eqref{eq:estar-proj}, was introduced ``by hand''. If we wish to enforce
energy conservation at all steps of the INC, we must compensate for it by
correcting the energy of the transferred nucleon.  This is necessary even when
the excitation energy of the projectile-like pre-fragment does not change,
because the nucleon transfer is in general associated with a non-vanishing
$Q$-value.

It is assumed that the mean field of the target nucleons acts on the projectile
nucleon as soon as it enters the calculation volume. Given the total
relativistic energy of the projectile nucleon $e_i$
(Eq.~\eqref{eq:nucleon_energy}), we now seek the total relativistic energy $E_i$
\emph{inside} the target potential. The task is complicated by the fact that the
potentials adopted in \incl\ are not constant but depend on the energy of the
nucleon itself, in the spirit of the phenomenology of the optical-potential
model \cite{aoust-potential}. Therefore, the energy $E_i$ must be sought as a
numerical solution to the equation
\begin{equation}
  E_i=e_i+V(E_i)+\Delta Q+\Delta E^*_p\text,\label{eq:particle_enters}
\end{equation}
where $\Delta Q$ is a correction due to the difference between the real
$Q$-value for nucleon transfer and \incl's internal value, and $\Delta E^*_p$ is
a correction that allows for a change in projectile excitation energy. If the
excitation energy of the projectile pre-fragment is unchanged by the nucleon
transfer, then $\Delta E^*_p=0$.

As customary, it is assumed in the INC framework that cascading nucleons are
on mass shell. Therefore, once the energy $E_i$ is determined as the solution of
Eq.~\eqref{eq:particle_enters}, the magnitude of the nucleon momentum inside the
target potential is defined by the on-shell dispersion relation
\[
\vect{P}_i^2=E_i^2-m^2\text,
\]
$m$ being the \incl\ nucleon mass. The direction of the $\vect{P}_i$ vector is
taken to be parallel to $\vect{p}_i$ (Eq.~\eqref{eq:nucleon_momentum}), the
nucleon momentum outside the target potential (i.e.\ no refraction takes place
at the surface).

We draw the attention of the reader to an important detail. As long as the
nucleon has not undergone any collision, it is taken to propagate inside the
target potential with the \emph{collective} velocity of the projectile
nucleus. The intrinsic Fermi motion of the projectile is frozen during
propagation. The nucleon four-momentum $(E_i,\vect{P}_i)$ is however correctly
used in the computation of the elementary cross sections and in the kinematics
of the binary collisions. Once the nucleon has experienced a (non-Pauli-blocked)
binary collision, it resumes its normal propagation. Note also that this
prescription effectively forbids collisions between projectile nucleons (because
their relative distance does not change) until they undergo a collision with a
target nucleon.

The intranuclear cascade unfolds normally until another projectile nucleon
reaches the surface of the calculation volume. The procedure is then applied to
the new nucleon and normal cascade is resumed. Once all the nucleons have
entered the calculation volume, the usual conditions for cascade stopping apply
\cite{boudard-incl}.

\subsubsection{Definition of the projectile-like
  pre-fragment}\label{sec:defin-proj-like}

At the end of the intranuclear cascade, a projectile pre-fragment may be defined
if some nucleons missed the calculation volume (geometrical spectators) or
traversed the calculation volume without undergoing any collision (dynamical
spectators). If no dynamical spectators are present, the mass, charge,
excitation energy and state of motion of the projectile pre-fragment are already
defined (Eqs.~\eqref{eq:proj-prefrag} above) and are guaranteed to satisfy
four-momentum conservation.

However, if dynamical spectators are present and are to be merged back into the
projectile-like pre-fragment, some adjustment is necessary to make sure that the
resulting pre-fragment is well-defined. Indeed, the non-negativity condition on
the excitation energy (Eq.~\eqref{eq:estar-proj}) is not sufficient because a
net energy transfer between the dynamical spectators and the target is always
possible because of the application of empirical thresholds for particle
absorption/emission.

We then tentatively define the pre-fragment four-momentum as the sum of the
four-momenta of the dynamical and geometrical spectators. If the resulting
four-momentum leads to a negative excitation energy, we apply an iterative
procedure to determine the maximal number of dynamical spectators that can be
incorporated in the pre-fragment without leading to negative excitation
energy.

From our discussion it clearly emerges that, despite our efforts, dynamical and
geometrical spectators are not (and cannot) be treated on exactly the same
footing. The crucial reason for this is that the four-momenta of dynamical
spectators are perturbed when they enter the target nucleus. Indeed, their
energy is corrected to keep the energy balance satisfied and to possibly make
room for some excitation energy of the projectile-like pre-fragment
(Eq.~\eqref{eq:estar-proj}).


\subsection{Low-energy fusion model}\label{sec:low-energy-fusion}

So far we have implicitly assumed that the transfer of one nucleon from the
projectile to the target is always possible. However, serious conceptual and
technical complications arise if the kinetic energy of one of the entering
nucleons is lower than the Fermi energy of the target. One would expect such a
process to be forbidden by the Pauli exclusion principle, especially for the
first projectile nucleon entering the unperturbed target Fermi sea. This
difficulty has already been encountered in the extension of the Fortran version
of \incl\ to light incident clusters \cite{boudard-incl}. In that case it was
observed that the problematic circumstance is most likely to occur when the
projectile kinetic energy per nucleon is comparable to or smaller than the
dynamical projectile pseudopotential. Under these conditions, it seems
reasonable to assume that, independently of the details of the dynamics, most of
the incoming nucleons will be trapped by the target potential well, resulting in
(possibly incomplete) fusion of the projectile and the target. This argument is
especially cogent for reactions between a light composite particle ($A\leq4$)
and a large nucleus. Therefore, for problematic events, \inclf\ abandons normal
INC in favor of a simple geometrical fusion model.

The application of \inclf\ to low-energy (in the sense outlined above)
composite-particle-induced reactions has been proven to produce surprisingly
good results \cite{boudard-incl4.6}. Yet, \inclf's fusion model is unsatisfactory
inasmuch as only the geometrical participants of the projectile (see
section~\ref{sec:geom-part-geom-1}) are taken to fuse with the target
nucleus. The distinction between geometrical participants and spectators has no
physical meaning because it is determined by the radius of the calculation
volume, $R_\text{max}$. In \inclf, this parameter must be considered as an
additional physical ingredient of the model, at least as far as low-energy
fusion is concerned.

We were therefore led to revise the low-energy fusion sector in our extension of
\inclxx\ to light incident ions. Admittedly, the fundamental assumption that the
low-energy dynamics is dominated by fusion is more difficult to defend for
reactions between light ions. This limitation is partly mitigated by the fact
that our fusion model naturally yields some ``incomplete fusion'', as we shall
now explain.

The fusion algorithm is triggered if, at any moment during the intranuclear
cascade, the particle-entry procedure (section~\ref{sec:intr-casc-phase}) endows
the entering projectile nucleon with a kinetic energy lower than the target
Fermi energy. Normal intranuclear cascade is then abandoned, but the information
about the initial position and momenta of the projectile nucleons is retained.

In the spirit of critical-distance fusion models
\cite{galin-fusion_critical_distance,bass-fusion}, we define an
\emph{interaction radius} $R_\text{int}$ and we prescribe that only nucleons
whose collective trajectory intersects the sphere of radius $R_\text{int}$ shall
fuse with the target nucleus. The interaction radius is defined as
\[
R_\text{int}{}=R_0+d_\text{int}
\]
in terms of the \emph{interaction distance} $d_\text{int}$,
\begin{equation}
  d_\text{int}{}=\sqrt{\max\left(\sigma_{pp}, \sigma_{nn},
      \sigma_{pn}\right)/\pi}\text,
  \label{eq:interaction_distance}
\end{equation}
where the elementary nucleon-nucleon cross sections $\sigma_i$ are calculated at
the nominal kinetic energy per nucleon of the light-ion projectile.

Nucleons that miss the interaction sphere are assumed not to fuse with the
target and are collectively considered as a projectile-like pre-fragment,
defined by Eqs.~\eqref{eq:proj-prefrag}. This defines another (possibly excited)
source and is expected to mimic incomplete fusion. The four-momentum of the
compound nucleus (the source composed of the target and the fusing nucleons) is
defined as the difference between the initial total four-momentum and the
four-momentum of the projectile-like pre-fragment. If the compound-nucleus
four-momentum corresponds to negative excitation energy, the event is discarded
and treated as a non-reaction. As a consequence, and in accordance with known
phenomenology, incomplete fusion at low projectile kinetic energy is
automatically suppressed because energetically forbidden.

The result of the new fusion algorithm is entirely independent of the size of
the calculation volume, $R_\text{max}$; in this respect, it is more satisfactory
than the algorithm used in \inclf. However, the condition that triggers the
fusion algorithm (energy of the entering nucleon below the Fermi level) is only
checked for geometrical participants, and thus still depends on $R_\text{max}$,
although in a much weaker way. One way to avoid this would be to define the
shape of the calculation volume in order to suppress the existence of
geometrical spectators; this solution would however require a deep revision of
the model logic and will not be pursued here.

The differences between \inclf's and \inclxx's fusion sectors will be
illustrated below (Section~\ref{sec:comp-with-inclf}).

\subsection{Projectile-target asymmetry}\label{sec:proj-targ-asymm}

\modified{The model description above shows that the new nucleus-nucleus
  capabilities add several new parameters/ingredients to the core of the
  model. While \inclxx's treatment of nucleon- and pion-induced reactions can be
  considered to be essentially parameter-free, the same cannot be said for the
  nucleus-nucleus sector. The nucleus-nucleus extension is admittedly more
  phenomenological.}

We turn now to a detailed discussion of the limitations of the extended \inclxx\
model. First and foremost, already at the level of the preparation of the
reaction partners, we have to stress that the momentum content of the projectile
and the target is different. The momentum distribution of target nuclei is
assumed to be a hard Fermi sphere of radius $p_F=270$~MeV/$c$
\cite{boudard-incl}, whereas projectile nuclei are assigned a Gaussian
distribution with the same RMS momentum ($\sqrt{3/5}\,p_F$). There are two
reasons for this difference.  First, only the hard-sphere distribution allows a
straightforward definition of Pauli blocking. Even in nucleon-induced reactions,
we need to assign a hard-sphere momentum distribution to target nuclei for Pauli
blocking to be unambiguously defined. However, hard Fermi spheres are
undoubtedly inadequate to describe momentum distributions in light nuclei
\cite{deWittHuberts-momentum_distributions}. Experimental handles on the
intrinsic momentum distribution are provided by the momentum distribution of
spectator nucleons emitted in peripheral reactions. These observables are better
described if realistic momentum distributions are assumed for the projectile
nucleus \cite[see e.g.\ Fig.~22 in Ref.][]{boudard-incl}. Thus, our asymmetric
choice strikes a compromise between the limitations of the unavoidable Fermi-gas
nuclear model and an attempt to improve the quality of the model predictions by
the inclusion of known phenomenology.

One of the weaknesses of the light-ion extension here described is that it
clearly introduces a projectile-target asymmetry. We can identify a few crucial
differences between the treatment of the projectile and of the target:
\begin{itemize}
\item the Fermi-momentum distribution is taken to be different for projectiles
  and targets, as we just discussed;
\item the projectile nucleus is essentially treated as a collection of free
  off-mass-shell nucleons, while the target nucleus is endowed with a mean-field
  potential;
\item Fermi motion in the projectile is frozen, in the sense outlined in
  Section~\ref{sec:intr-casc-phase};
\item projectile nucleons can miss the calculation volume (geometrical
  spectators, Section~\ref{sec:geom-part-geom-1}), while target nucleons cannot;
\item we neglect Pauli blocking of the first collision in the projectile Fermi
  sea;
\item participant nucleons can escape or finish the reaction in the target-like
  pre-fragment (they can be trapped by the mean-field potential), but they can
  never finish in the projectile-like pre-fragment. In the language of the
  abrasion-ablation picture, the projectile spectator does not receive any
  energy from the participant zone (final-state interactions);
\item the excitation energy assigned to the projectile-like pre-fragment is
  based on a simple particle-hole model, while that assigned to the target-like
  pre-fragment results from and carries information about the full INC dynamics;
\item the calculation is performed in the target rest frame \emph{and} the
  dynamics is not Lorentz-covariant because it singles out a global time
  variable. It has however been shown that the violations introduced by suitable
  non-covariant dynamics are not necessarily severe, even around 1~$A$GeV
  \cite{mancusi-rjqmd}.
  \modified{Note also that there exist no covariant INC models, not even for
    nucleon- and pion-induced reactions;}
\item in the low-energy fusion sector, projectile nucleons can elude fusion if
  their impact parameter is large enough, while target nucleons cannot.
\end{itemize}

One practical consequence of the projectile-target asymmetry is that the cross
sections for producing a given nuclide as a projectile-like fragment or as a
target-like fragment will in general not be equal, even for a symmetric reaction
(e.g.\ $^{12}$C+$^{12}$C). Consider however that the predictions for target-like
fragment production should be closer to the experimental data, given the
superior physical modeling of the target nucleus. This is unfortunate if
projectile-like fragmentation is more important than target-like fragmentation
for a specific application. However, if both reaction partners are light, one
can consider swapping the roles of projectile and target in the simulation: in
other words, the reaction can be simulated in inverse kinematics (i.e.\ as
target on projectile), with the reaction products being boosted back to the
laboratory frame at the end of the simulation. We refer to this calculation
method as \emph{accurate-projectile mode}, while we use the expression
\emph{accurate-target mode} to refer to the normal \inclxx\ calculation
mode. The naming convention reflects our expectations about the accuracy of the
predictions for projectile- and target-like fragments, which are kinematically
well separated. However, the statement about the calculation accuracy should be
tempered for lighter particles, and nucleons in particular, whose origin cannot
be clearly discriminated on a kinematical basis. We shall illustrate the
differences between the two calculation modes in Section~\ref{sec:comparison-exp-data}.

We should stress that the choice between accurate-target and accurate-projectile
mode is application-dependent. If the user is interested e.g.\ in
projectile-like fragments for radiation-protection and hadrontherapy
simulations, they should use accurate-projectile mode. A universal choice is not
possible; however, we believe that accurate-projectile mode provides a better
description of particle transport for several applications where \inclxx\ is
likely to give accurate results. Therefore, \geant\ uses \inclxx\ in
accurate-projectile mode by default. The \geant\ user can switch to
accurate-target mode using a macro.

\begin{table}
  \caption{Choices for the internal reaction kinematics in \inclxx\ for nucleon-
    and nucleus-nucleus reactions, when the model is used within \geant; $A_p$
    and $A_t$ represent the projectile and target mass numbers within particle
    transport. The table entry indicates which nucleus is internally treated as
    the projectile in \inclxx. Note that reactions with $A_p>18$ and $A_t>18$
    are delegated to \bic.\label{tab:g4-kinematics}}
  \begin{ruledtabular}
    \begin{tabular}{c|cc}
      & accurate- & accurate- \\
      & projectile mode & target mode\\
      \hline
      $A_p<A_t\leq4$ & projectile & projectile \\
      \hline
      $A_t\leq A_p\leq4$ & target & target \\
      \hline
      $A_p\leq4<A_t$ & projectile & projectile \\
      \hline
      $A_t\leq4<A_p$ & target & target \\
      \hline
      $4\leq A_p\leq18$& \multirow{2}{*}{target}  & \multirow{2}{*}{projectile} \\
      and $4\leq A_t\leq18$ &  &\\
      \hline
      $A_p\leq 18<A_t$ & projectile & projectile\\
      \hline
      $A_t\leq18<A_p$ & target & target \\
    \end{tabular}
  \end{ruledtabular}
\end{table}

The accurate-projectile/target option should be contrasted with the approach
used by many INC models (including \geant's \bic\ model) when treating composite
projectiles, in which one identifies the lighter nucleus with the projectile and
the heavier nucleus with the target of the INC. The rationale behind the
``light-on-heavy'' criterion is that the largest nucleus is expected to dominate
the mean-field potential. However, this paradigm does not provide clear
guidelines for symmetric reactions; furthermore, even in quasi-symmetric
reactions (e.g.\ $^{12}$C+$^{16}$O), one can hardly expect the mean field to be
dominated by the heavier partner. Therefore, it seems unwarranted to
systematically select the light-on-heavy option, especially if reactions between
light nuclei (such as those encountered in hadrontherapy) are
involved. Nevertheless, it is probably reasonable to always treat the lightest
nuclei as projectiles. Therefore, \inclxx\ in \geant\ runs calculations as
light-on-heavy if either reaction partner has $A\leq4$, regardless of the
calculation mode chosen by the \geant\ user. The light-on-heavy mode is also
selected if either reaction partner is a heavy nucleus ($A>18$). The choices for
the reaction kinematics are summarized in Table~\ref{tab:g4-kinematics}. In
summary, the user-specified accurate-projectile/target option is honored only if
both projectile and target masses satisfy $4\leq A\leq18$.

As an alternative to the accurate-projectile/target dichotomy, it would be
possible to palliate the model asymmetry by randomly choosing to simulate the
reaction in the rest frame of the projectile or of the target. It is fair to
assume that symmetric reactions should result in a straightforward 50-50 split
between the two kinematical choices; it is however unclear what should be done
of asymmetric reactions such as $^{12}$C+$^{16}$O, especially because reactions
induced by $A\leq4$ projectiles should always be described as light-on-heavy and
the model should behave continuously as a function of the projectile
mass. Therefore, additional prescriptions would be necessary in this
case. Nevertheless, we shall illustrate random symmetrization in
Section~\ref{sec:comparison-exp-data} with a few selected examples for symmetric
reactions.

\subsection{De-excitation stage}\label{sec:de-excitation-stage}

Before turning to the comparison of the new \inclxx\ model results with
nucleus-nucleus experimental data, we need to spend a few words about the
coupling with de-excitation models. Historically, the \incl\ model has been
coupled to statistical evaporation/fission models such as \ablaold\
\cite{gaimard-abla,*junghans-abla,*schmitt-cxx_translation} or \abla\
\cite{kelic-abla07}. This was motivated by the typical application of \incl\ to
spallation reactions, and in particular to reactions induced by nucleons on
relatively large nuclei ($A\gtrsim50$); for these systems, the excitation energy
is relatively low and evaporation/fission models are indeed capable of providing
a very good description of most observables \cite{leray-intercomparison}. It is
legitimate to ask whether these models would perform equally well on reactions
between light nuclei.

One peculiarity of such reactions is that the binding energy and the excitation
energy of the remnants produced by the INC stage may be of the same order of
magnitude. Under these conditions, de-excitation becomes a relatively fast
process and it is questionable to make use of the statistical hypothesis, or at
the very least it seems inappropriate to describe the de-excitation step as a
well-defined sequence of binary, evaporation-like splits. This issue is even
more pressing as the sensitivity of the model predictions to the details of
de-excitation in general increases with the excitation energy.

An alternative picture is provided by Fermi break-up (FBU), a model that was
initially developed to describe the production of pions in high-energy
nucleon-nucleon collisions \cite{fermi-break_up} and that was subsequently
adapted to the description of fragmentation of excited light nuclei
\cite{zhdanov-fermi_breakup}. The model does not provide a time-like description
of the de-excitation chain, but limits itself to providing probabilities for the
final configurations, which are specified by the masses, charges and momenta of
the observed cold fragments. The crucial assumption of the model is that the
probability to observe a given fragment configuration is simply proportional to
the density of phase-space states around it. This amounts to assuming that the
transitions from the excited pre-fragment to all the final configurations are
described by the same matrix elements; in this sense, the Fermi model represents
the simplest possible description of simultaneous nuclear break-up. More
sophisticated approaches are provided by the family of statistical
multifragmentation models, which will not be discussed here; we refer the reader
to Ref.~\citenum{carlson-fermi_bu_smm} for an account of the relations that the
two model classes bear to each other.

The default \geant\ de-excitation model
\cite[\geh,][]{quesada-g4excitationhandler} implements FBU as one of the
possible channels. However, standard FBU does not provide absolute decay widths,
but only yields probabilities for each break-up configuration; for this reason,
it is non-trivial to introduce direct competition between FBU and other
de-excitation mechanisms, such as particle evaporation.  The developers of \geh\
made the choice of applying FBU for any de-exciting nucleus with
\modified{$A\leq16$ and $Z\leq8$; note that the choice for the threshold values
  can affect the calculated cross sections by as much as a factor of two
  \cite{mashnik-fragmentation_light_nuclei}}. The FBU mechanism can also be
triggered during the de-excitation chain, if particle evaporation or other
mechanisms bring the excited nucleus in the $A$-$Z$ region indicated above.

In what follows (Sections~\ref{sec:comp-with-inclf} and
\ref{sec:comparison-exp-data}) we shall compare the results of calculations
performed with \inclxx\ coupled with \abla, \ablaold\ and \geh. Although the
\abla\ model does not include a FBU module, it does include a semi-empirical
treatment of multifragmentation \cite{kelic-abla07}. We will discuss below to
what extent this makes it applicable to the highest excitation energies.  Also,
we draw the reader's attention to the fact that \abla\ is not available for
transport calculations in the official \geant\ code;
\modified{a \cxx/\fortran\ interface to \abla\ can however be privately provided
  on request}.
The \inclf/\abla\ code has been included in a private version of \mcnpx\ and is
expected to be distributed with a future release of the \mcnpsix\ code
\cite{goorley-mcnp6}; that version is however incapable of handling
light-ion-induced reactions.

\section{Comparison with \inclf}\label{sec:comp-with-inclf}

\begin{figure}
  \includegraphics[width=\linewidth]{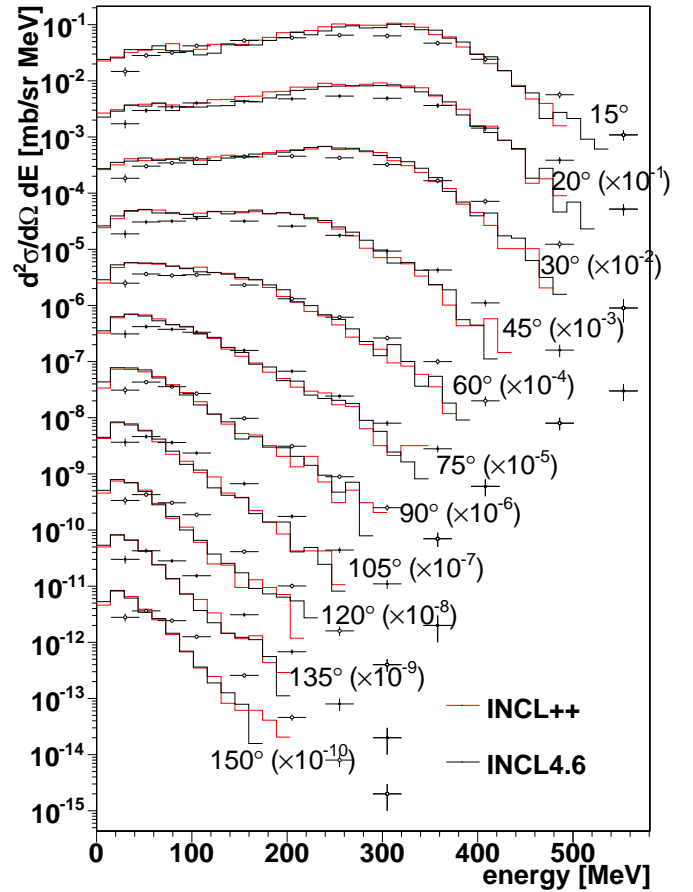}
  \caption{(Color online) Double-differential cross section for $\pi^+$ production from the
    730-MeV \textit{p}+Cu reaction. Red (black) lines represent the \inclxx\
    (\inclf) result. Data taken from
    Ref.~\citenum{cochran-pions}.\label{fig:ddxs_pi+}}
\end{figure}

\begin{figure}
  \includegraphics[width=\linewidth]{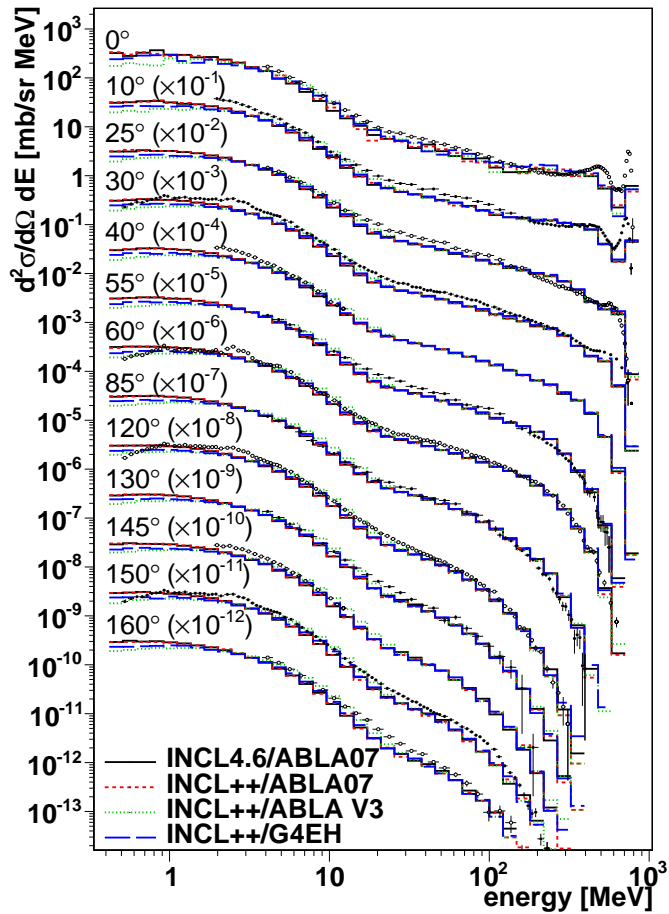}
  \caption{(Color online) Double-differential cross section for neutron production from the
    800-MeV \textit{p}+Pb reaction. The different model calculations are
    described in the text (\emph{G4EH} in the plot legend stands for \geh). Data
    taken from Refs.~\citenum{leray-neutrons,amian-neutrons}.\label{fig:ddxs_n}}
\end{figure}

\begin{figure}
  \includegraphics[width=\linewidth]{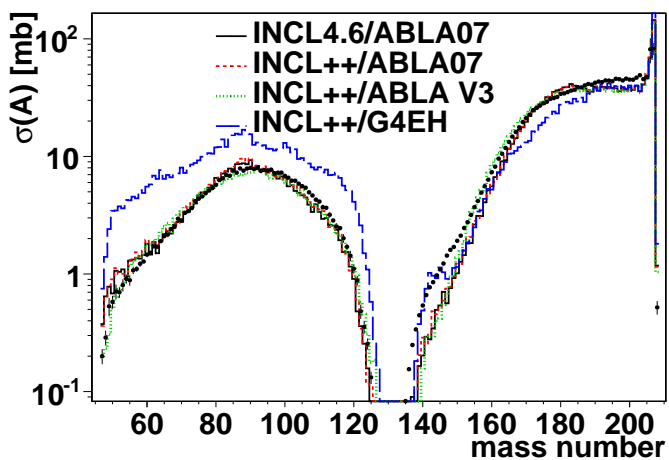}
  \caption{(Color online) Fragmentation cross sections for the 1-$A$GeV $^{208}$Pb+$^1$H
    reaction, as a function of the fragment mass number. The different model
    calculations are described in the text (\emph{G4EH} in the plot legend
    stands for \geh). Data taken from Refs.~\citenum{enqvist-lead} and
    \citenum{kelic-bi}.\label{fig:isot_Pb}}
\end{figure}

We shall now document the physical equivalence of the \inclf\ and \inclxx\
codes. Figure~\ref{fig:ddxs_pi+} shows double-differential cross sections for
the production of positive pions from a 730-MeV proton colliding with a copper
target; this observable is entirely due to the intranuclear-cascade stage of the
reaction. Figure~\ref{fig:ddxs_n} shows double-differential cross sections for
the production of neutrons from a 800-MeV proton colliding with a lead
target. Finally, Fig.~\ref{fig:isot_Pb} shows the mass distribution of the
fragments produced in a 1-GeV $^{208}$Pb+$^1$H reaction. The observables
depicted in Figs.~\ref{fig:ddxs_n} and \ref{fig:isot_Pb} are also sensitive to
the de-excitation stage of the nuclear reaction. Specifically, de-excitation
dominates the low-energy part of the double-differential neutron spectrum (say
up to 20~MeV) and is entirely responsible for the mass distribution of
Fig.~\ref{fig:isot_Pb}, albeit the production of residues with the largest
masses is dominated by INC. For the purpose of these comparisons, we coupled our
cascade models with the \abla\ de-excitation model \cite{kelic-abla07}. All
plots show perfect agreement between \inclf\ and \inclxx.

We also show calculations performed by coupling \inclxx\ with the \geh\ and
\ablaold\ de-excitation models, available in \geant. De-excitation is the
dominant mechanism for the production of the low-energy neutrons in
Fig.~\ref{fig:ddxs_n}; one indeed remarks that the \geh\ yields around 1~MeV are
intermediate between those predicted by \ablaold\ (lowest) and \abla\
(highest). There is a difference of about a factor of 2 between \ablaold\ and
\abla, with the latter being closer to the experimental data. Note however that
\ablaold\ also results in larger yields around 10~MeV, which seems to improve
the agreement with the experimental data in that region. This difference
probably indicates the average kinetic energy of the emitted neutron is higher
in \ablaold\ than in \abla.

Figure~\ref{fig:isot_Pb} provides a somewhat complementary picture for a similar
system. Although \ablaold\ and \abla\ predict rather different neutron yields,
this seems to have little impact on the fission cross section. However, \abla's
fission sector is substantially different from \ablaold's model and was probably
readjusted to fit the data shown here. \inclxx/\geh\ largely overpredicts the
fission cross section and underestimates the yields for heavy spallation
residues ($A\simeq175$). Note however that the parameters of \geh\ were tuned to
yield a correct reproduction of the data in Fig.~\ref{fig:isot_Pb} when coupled
with \bic\ \cite{quesada-g4excitationhandler}.

\subsection{Reactions induced by pions}
\label{sec:react-induc-pions}

As mentioned in Section~\ref{sec:diff-with-inclf}, \inclf\ and \inclxx\ mainly
diverge in the treatment of reactions induced by pions and composite
particles. The most prominent difference in pion-induced reactions is the
application of the local-$E$ correction on the first pion-nucleon collision. The
net effect of the correction is to reduce the center-of-mass energy at which the
pion-nucleon collision takes place. We can reasonably expect this to have an
effect on the pion-nucleus reaction cross section, inasmuch as the latter tracks
the energy dependence of the elementary pion-nucleon cross section. A similar
argument explains the effect of the local-$E$ correction on nucleon-nucleus
reaction cross sections \cite[Section~III.B in][]{boudard-incl4.6}.

\begin{figure}
  \includegraphics[width=\linewidth]{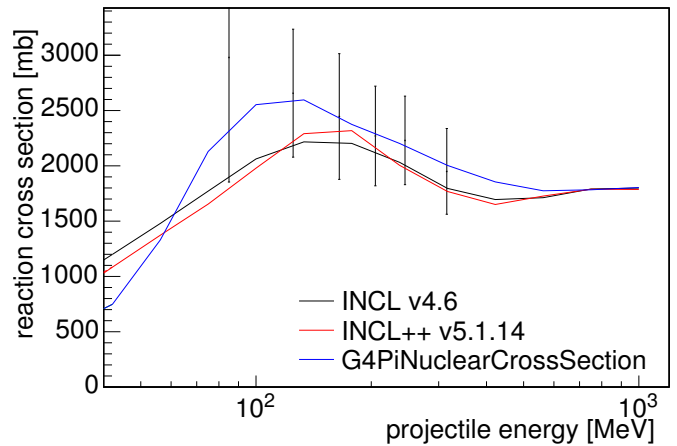}
  \caption{(Color online) Excitation function for the $\pi^+$+$^{209}$Bi reaction cross
    section, calculated with \inclxx\ (red line), \inclf\ (black line) and
    \geant's semi-empirical reaction-cross-section model (blue
    line). Experimental data taken from
    Ref.~\citenum{ashery-pions}.\label{fig:reaction_pi+}}
\end{figure}

\begin{figure}
  \includegraphics[width=\linewidth]{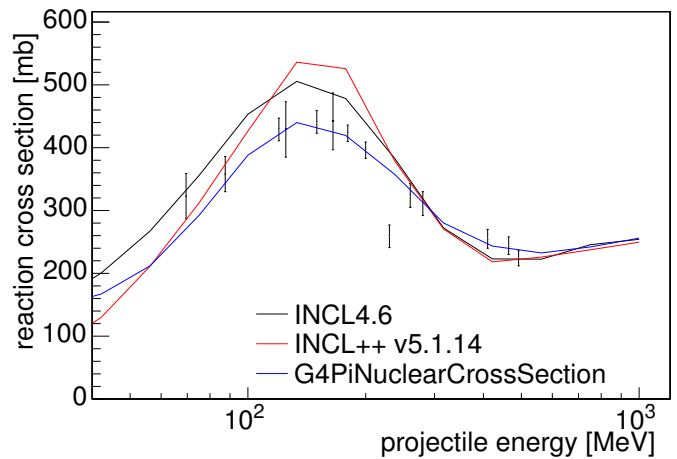}
  \caption{(Color online) Same as Fig.~\ref{fig:reaction_pi+}, for $\pi^-$ +
    $^{12}$C. Experimental data taken from
    Refs.~\citenum{gelderloos-pions,ashery-pions,binon-pions,crozon-pions}.\label{fig:reaction_pi-}}
\end{figure}

The effect of the local-$E$ correction on pion-nucleus reaction cross sections
is illustrated in Figs.~\ref{fig:reaction_pi+} and \ref{fig:reaction_pi-}. The
reaction cross section used by the \geant\ particle transport is also shown for
comparison. The difference is mostly visible at low energy, which is the region
where the elementary pion-nucleon reactions varies most quickly due to the
presence of the $\Delta(1232)$ resonance, but it stays very small in all
cases. Note that the calculations in Fig.~\ref{fig:reaction_pi-} were performed
with \inclxx\ \code{v5.1.14}, corrected for a small bug in the Coulomb deviation
of incoming negative particles. The bug is fixed in \geant\ \code{v10.0.p03} and
\code{v10.1$\beta$}.

\begin{figure}
  \includegraphics[width=\linewidth]{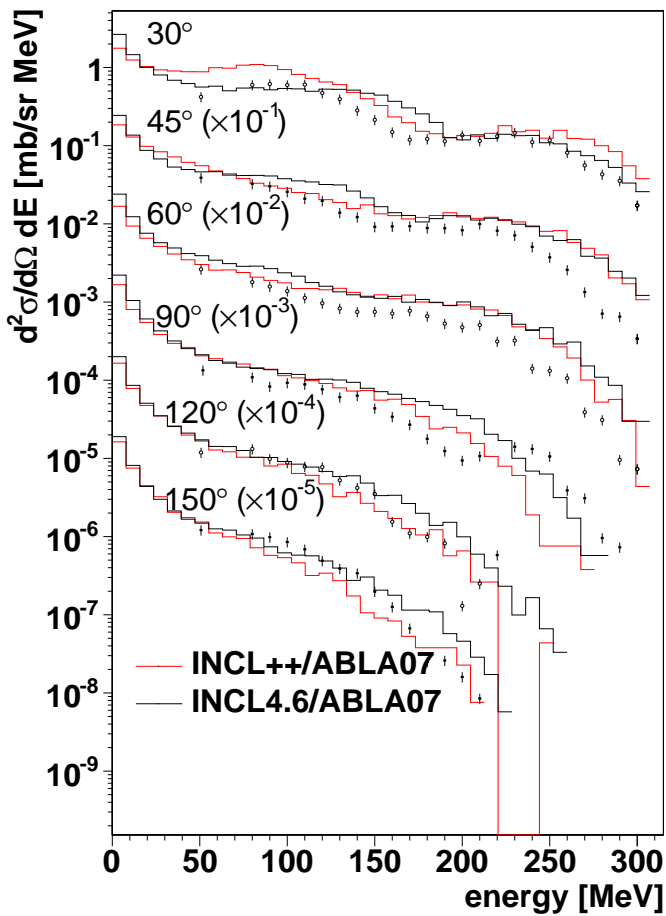}
  \caption{(Color online) Double-differential cross section for proton emission in 220-MeV
    $\pi^+$+$^{12}$C, calculated with \inclxx\ (red line) and \inclf\ (black
    line). Experimental data taken from
    Refs.~\citenum{mckeown-pions}.\label{fig:ddxs_pi+_C_p}}
\end{figure}

\begin{figure}
  \includegraphics[width=\linewidth]{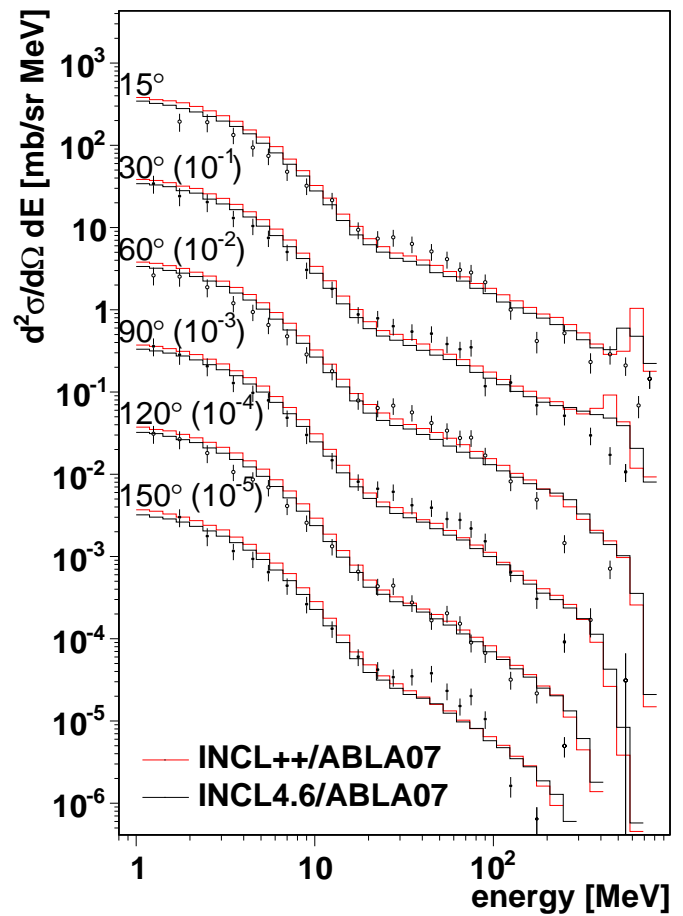}
  \caption{(Color online) Double-differential cross section for neutron emission in 870-MeV
    $\pi^-$+$^{208}$Pb, calculated with \inclxx\ (red line) and \inclf\ (black
    line). Experimental data taken from
    Refs.~\citenum{iwamoto-pions}.\label{fig:ddxs_pi-_Pb_n}}

\end{figure}

Note that the reaction cross sections calculated by \inclxx\ are \emph{not} used
for particle transport in \geant. The transport algorithm relies on independent,
semi-empirical cross-section parametrisations, which are generally more accurate
than the cross sections predicted by the nuclear-reaction models. (This remark
also applies to the nucleus-nucleus reaction cross sections depicted below, in
Figs.~\ref{fig:reaction_d}, \ref{fig:reaction_He4} and \ref{fig:reaction_C12}.)
An unreasonable prediction for the reaction cross section is however a sign that
some physics is not suitably accounted for.

The local-$E$ correction for pion-induced reactions also manifests itself in
other observables, such as double-differential cross sections for proton
(Figs.~\ref{fig:ddxs_pi+_C_p}) and neutron emission
(Fig.~\ref{fig:ddxs_pi-_Pb_n}). The 220-MeV experimental data show some peculiar
structure at forward angles that is typical of pion absorption. At 30$^\circ$
one can distinguish two humps centered around 110 and 230~MeV. The 110-MeV peak
corresponds to the emission of a proton by intermediate excitation and decay of
a $\Delta$ resonance\
\footnote{We remind the reader that the production of the $\Delta$ resonance is
  the only open channel for pions in \inclf\ and \inclxx\ \code{v5.1.14}.}:
\[
\pi^++N\to \Delta\to p\text{ (escapes)}+\pi\text.
\]
The 230-MeV peak corresponds instead to the absorption of the intermediate resonance:
\[
\pi^++N\to \Delta,\qquad\Delta+N\to p\text{ (escapes)}+N\text.
\]
It is clear that the second mechanism leads to higher proton kinetic energies
(on average) because of the absorption of the pion mass.

The energy distributions for these components are smeared out by the Fermi
motion of the nucleons in the target. Since the local-$E$ correction suppresses
the importance of Fermi motion in the nuclear surface, we observe that the peaks
are somewhat sharper in the \inclxx\ calculations. A similar consideration can
be made concerning Fig.~\ref{fig:ddxs_pi-_Pb_n}, where one observes that
\inclxx\ leads to a sharper peak around
600~MeV\modified{, which is less satisfactory}.

\subsection{Reactions induced by light composite particles}
\label{sec:react-induc-light}

\begin{figure}
  \includegraphics[width=\linewidth]{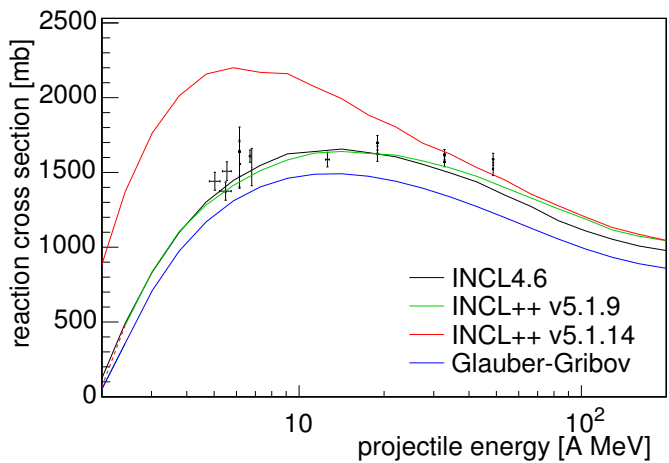}
  \caption{(Color online) Excitation function for the $d$+$^{56}$Fe reaction
    cross section, calculated with two versions of \inclxx\ (red and green
    lines), \inclf\ (black line) and \geant's Glauber-Gribov semi-empirical
    reaction-cross-section model (blue line). Experimental data refer to
    $^{56\text{--}58}$Fe and $^{58\text{--}60}$Ni targets and are taken from
    Refs.~\citenum{auce-deuterons,dubar-deuterons,brown-deuterons,bearpark-deuterons}.\label{fig:reaction_d}}
\end{figure}

\begin{figure}
  \includegraphics[width=\linewidth]{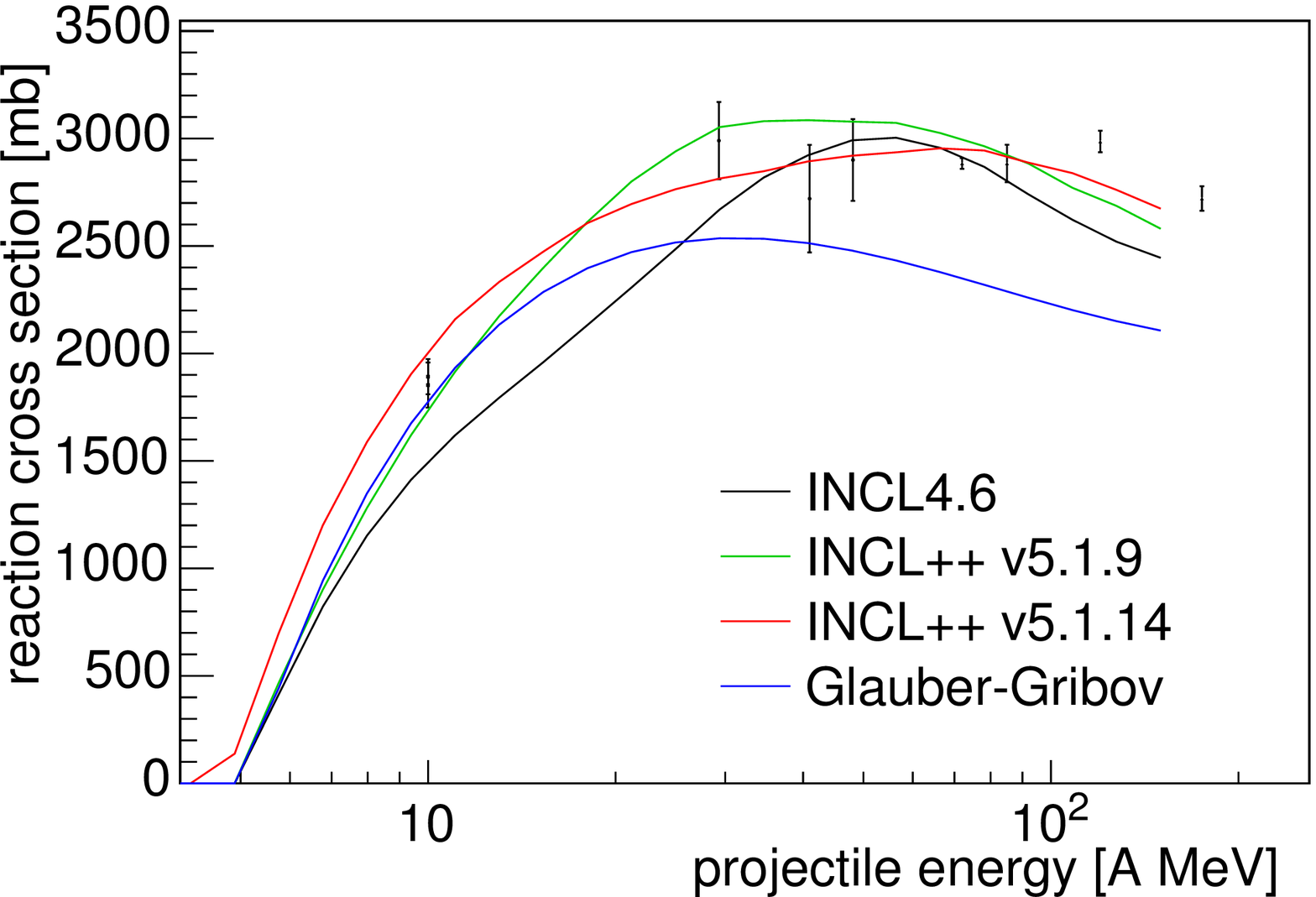}
  \caption{(Color online) Excitation function for the $^4$He+$^{208}$Pb/$^{209}$Bi reaction
    cross section, calculated with two versions of \inclxx\ (red and green
    lines), \inclf\ (black line) and \geant's Glauber-Gribov semi-empirical
    reaction-cross-section model (blue line). Experimental data refer to
    $^{208}$Pb and $^{209}$Bi targets and are taken from
    Refs.~\citenum{bonin-reaction,auce-reaction,ingemarsson-reaction}.\label{fig:reaction_He4}}
\end{figure}

\begin{figure}
  \includegraphics[width=\linewidth]{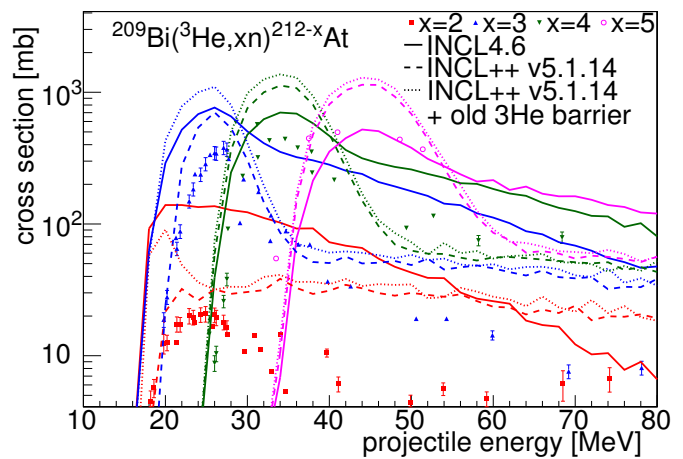}
  \caption{(Color online) Excitation functions for $^{209}\text{Bi}(^3\text{He},x\text{n})$
    cross sections. Different colors refer to different values of $x$, while the
    line styles denote calculations performed with \inclf\ (solid), \inclxx\
    \code{v5.1.14} (dashed) and \inclxx\ \code{v5.1.14} with the $^3$He Coulomb
    radius as in \inclf\ (dotted). Data taken from
    Refs.~\citenum{nagame-he3_bi,szucs-he3_bi,vysotsky-he3_bi209}.\label{fig:He3_Bi209_xn}}
\end{figure}

We mentioned in Section~\ref{sec:diff-with-inclf} that \inclxx\ and \inclf\
differ in how they handle composite projectiles, especially at low energy. This
is illustrated by Figs.~\ref{fig:reaction_d} and \ref{fig:reaction_He4}, which
show a comparison of the predicted reaction cross section for the $d$ +
$^{56\text{--}58}$Fe/$^{58\text{--}60}$Ni and $^4$He+$^{208}$Pb/$^{209}$Bi
system. \inclxx\ performs sensibly worse than \inclf\ for the deuteron-induced
reaction, while predictions for $^4$He are similar. The degradation is
essentially due to the fact that \inclxx\ uses a unique parameter set to
describe reactions induced by composite particles up to $A=18$, while \inclf\
was limited to $A\leq4$.

Figures~\ref{fig:reaction_d} and \ref{fig:reaction_He4} also show the
predictions of \inclxx\ \code{v5.1.9} (green lines), which is the version that
was distributed with \geant\ \code{v9.6.p02}. The similarity to the \inclf\
results is due to the fact that the two models have very similar low-energy
fusion sectors.  We will show in Section~\ref{sec:comparison-exp-data} that \inclxx\
\code{v5.1.9} is unsuitable for light-ion-induced reactions.

Another difference between \inclf\ and \inclxx\ is illustrated by
Fig.~\ref{fig:He3_Bi209_xn}, which shows excitation functions for the
$^{209}\text{Bi}(^3\text{He},x\text{n})$ reactions. As for
Figs.~\ref{fig:reaction_d} and \ref{fig:reaction_He4}, the projectile energies
are rather low and we mostly probe the fusion sector of the \incl\ model; this
is why the \inclf\ and \inclxx\ calculations are in disagreement. However,
Fig.~\ref{fig:He3_Bi209_xn} also illustrates the effect of the modification of
the Coulomb barrier for incoming $^3$He nuclei. The calculations for \inclxx\
\code{v5.1.14} are in better agreement with the experimental data than the
modified calculations with the old Coulomb barrier, and even than the
calculations performed with the legacy \inclf\ model.


\section{Comparison with nucleus-nucleus experimental
  data}\label{sec:comparison-exp-data}

We now turn to the verification of the most prominent new feature of \inclxx,
namely the capability to handle light-ion-induced reactions. The observables
selected for verification reproduce the choices made for nucleon-nucleus
reactions \cite{boudard-incl,boudard-incl4.6}, a strategy that proved successful
\cite{leray-intercomparison,*intercomparison-website}.  We start by considering
reaction cross sections, which capture global aspects of the model
(Section~\ref{sec:react-cross-sect}). We will then proceed to investigate
double-differential cross sections for the production of nucleons and LCPs
(Section~\ref{sec:part-prod}). The rationale for this choice lies in the fact
that particle emission during INC proceeds more or less directly from hard
nucleon-nucleon scattering events, which constitute the core of the cascade
mechanism. In certain kinematical regions, de-excitation of the pre-fragments
contributes to (or dominates) particle production; therefore,
double-differential cross sections indirectly verify some global characteristics
of the cascade pre-fragments, too.

Finer details about the distribution of cascade pre-fragments are emphasized by
fragmentation cross sections (Section~\ref{sec:fragm-cross-sect}), especially if
per-isotope information is available. Although it may be non-trivial to
disentangle the contributions of cascade and de-excitation, the study of
isotopic fragmentation cross sections for different systems and energies has
proven extremely valuable in the development of the proton-nucleus model
\cite{boudard-incl,boudard-incl4.6}.

We remark in passing that most of the experimental data were analyzed with other
models; see for example the vast validation effort of the \mcnpx/\mcnpsix\ event
generators \cem\ and \laqgsm\
\modified{\cite{mashnik-cem,mashnik-pion,iwase-spectra,mashnik-fragments,mashnik-gsi,mashnik-laqgsm_upgrade,mokhov-hadron}}.
We will however not enter into a detailed comparison because these calculations
have no direct bearing upon \geant.

\subsection{Reaction cross sections}\label{sec:react-cross-sect}

\begin{figure}
  \includegraphics[width=\linewidth]{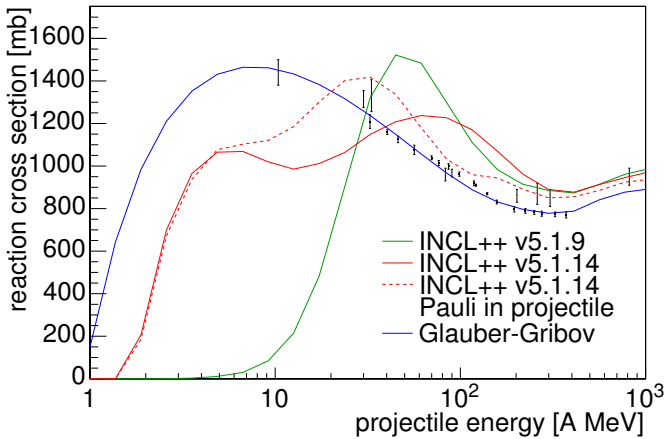}
  \caption{(Color online) Excitation function for the $^{12}$C+$^{12}$C reaction cross section,
    calculated with \inclxx\ \code{v5.1.9} (green line), \inclxx\ \code{v5.1.14}
    (solid red line), \inclxx\ \code{v5.1.14} with Pauli blocking and a hard
    Fermi sphere in the projectile (dashed red line) and \geant's Glauber-Gribov
    semi-empirical reaction-cross-section model (blue line).  Experimental data
    taken from
    Refs.~\citenum{takechi-reaction,fukuda-reaction,takechi-phd,takechi-reaction2,kox-reaction,kox-reaction2}.\label{fig:reaction_C12}}
\end{figure}

Figure~\ref{fig:reaction_C12} shows an excitation function for the
$^{12}$C+$^{12}$C reaction cross section. The agreement with the experimental
data is far from perfect. More precisely, we can observe that the double-humped
\inclxx\ excitation function clearly exhibits two distinct regimes. The
low-energy peak (around 5~$A$MeV) is due to the fusion model. In fact, pure INC
plays essentially no role as long as at least one projectile nucleon enters the
calculation volume below the Fermi energy. The importance of the fusion
mechanism starts to decrease above 5~$A$MeV and gradually leaves room for the
pure INC mechanism, which is responsible for the second peak (around 70~$A$MeV).

Particle transport in \geant\ is not seriously affected by this deficiency,
because the reaction cross section is imposed during the transport step;
however, the disagreement clearly indicates a failure to correctly describe the
physics of this reaction, especially at low energy. It might be argued that the
$^{12}$C+$^{12}$C reaction does not represent a fair benchmark for
intranuclear-cascade models, which assume that the larger reaction partner is
left relatively unperturbed by the cascade; however, the reaction cross section
is determined by the \emph{first} non-Pauli-blocked nucleon-nucleon collision,
which typically involves surface nucleons at an early, relatively unperturbed
stage of the reaction.

In spite of the disappointing result of Fig.~\ref{fig:reaction_C12}, the
comparison with the double-differential and residue-production data
(Sections~\ref{sec:part-prod} and \ref{sec:fragm-cross-sect} below) shows that
\inclxx\ in general captures the essential aspects of the fragmentation in the
$^{12}$C+$^{12}$C reaction.

Note that the INC approximation is expected to be valid above some 150~$A$MeV.
In this energy range, the contribution from the (admittedly empirical) fusion
sector is negligible, thereby simplifying the interpretation of the resulting
cross section. We see that the model overestimates the experimental data by
about 25\%; part of the overestimation is due to the fact that we neglect strict
Pauli blocking of the first collision in the Fermi sea of the projectile. This
analysis is corroborated by the observation that the nucleon-$^{12}$C reaction
cross sections are correctly predicted in the same energy-per-nucleon range
\cite{boudard-incl4.6}. As mentioned in Section~\ref{sec:proj-targ-asymm}, the
use of realistic (Gaussian) momentum distributions for the projectile is
somewhat irreconcilable with the definition of Pauli blocking. We therefore
performed a test calculation with a hard Fermi sphere for the projectile
momentum distribution; strict Pauli blocking in the projectile Fermi sea was
applied on the first collision. The resulting excitation function is displayed
in Fig.~\ref{fig:reaction_C12} and is in much better agreement with the
experimental data. Note also that the refined calculations yields a
\emph{larger} reaction cross section between $\sim5$ and $\sim60$~$A$MeV; this
is an effect of the hard Fermi sphere and is of course not due to the
introduction of Pauli blocking.

Finally, Fig.~\ref{fig:reaction_C12} also shows the prediction of an older
version of \inclxx\ (\code{v5.1.9}). As mentioned above, the interest of this
comparison mainly lies in the fact that the low-energy fusion sector of
\code{v5.1.9} is a straightforward extension to $A\leq18$ of the \inclf\
approach. The reaction cross section at low energy (say below 10~$A$MeV) is
largely suppressed because almost all impact parameters result in incomplete
fusion, which is energetically forbidden by the tight binding of $^{12}$C
nuclei. It is apparent that \inclxx\ \code{v5.1.9} is inadequate, which
justifies the revision of \inclxx's low-energy sector that was described in
Section~\ref{sec:low-energy-fusion}.


\subsection{Caveat about cross-section
  normalization}\label{sec:caveat-about-cross}

Before turning to double-differential cross sections for particle production, a
word of caution should be said about the comparisons shown in the following
sections between \inclxx\ and the other models available in \geant. As already
mentioned above (Section~\ref{sec:react-induc-pions}), most nuclear-reaction
models are able to predict absolute reaction cross sections; however, these
quantities are not directly used in particle transport, because more accurate
semi-empirical parametrizations are usually available. Nevertheless, a
misprediction of the reaction cross section might indicate that the model fails
to describe some particular channel. We try to make our point clearer by
referring to Fig.~\ref{fig:reaction_C12} above. We showed that the
overprediction of the $^{12}$C+$^{12}$C reaction cross section at high energy is
largely due to the lack of Pauli blocking on the first collision in the
projectile Fermi sea. This defect should mostly lead to an overestimation of the
cross sections associated with peripheral collisions. Therefore, even though the
gross overestimation is only 25\% of the reaction cross section, the relative
overprediction may be much more conspicuous in channels associated with
peripheral collisions.

The \geant\ nuclear-reaction models discussed below (\qmd, \bic,
Bertini+PreCompound) are only accessible through their \geant\ interface
classes. Because of the way nuclear-reaction models are used in particle
transport, the interface iterates calls to the model engine until an inelastic
event is generated. Therefore, the absolute reaction cross sections predicted by
the \geant\ models are \emph{not} available to us. We chose to normalize the raw
model predictions (counts) using the Shen nucleus-nucleus cross section
\cite{shen-xsec}, which is available in \geant\ through class
\code{G4IonsShenCrossSection}.

\subsection{Particle-production cross sections}\label{sec:part-prod}

\begin{figure*}
  \includegraphics[width=\linewidth]{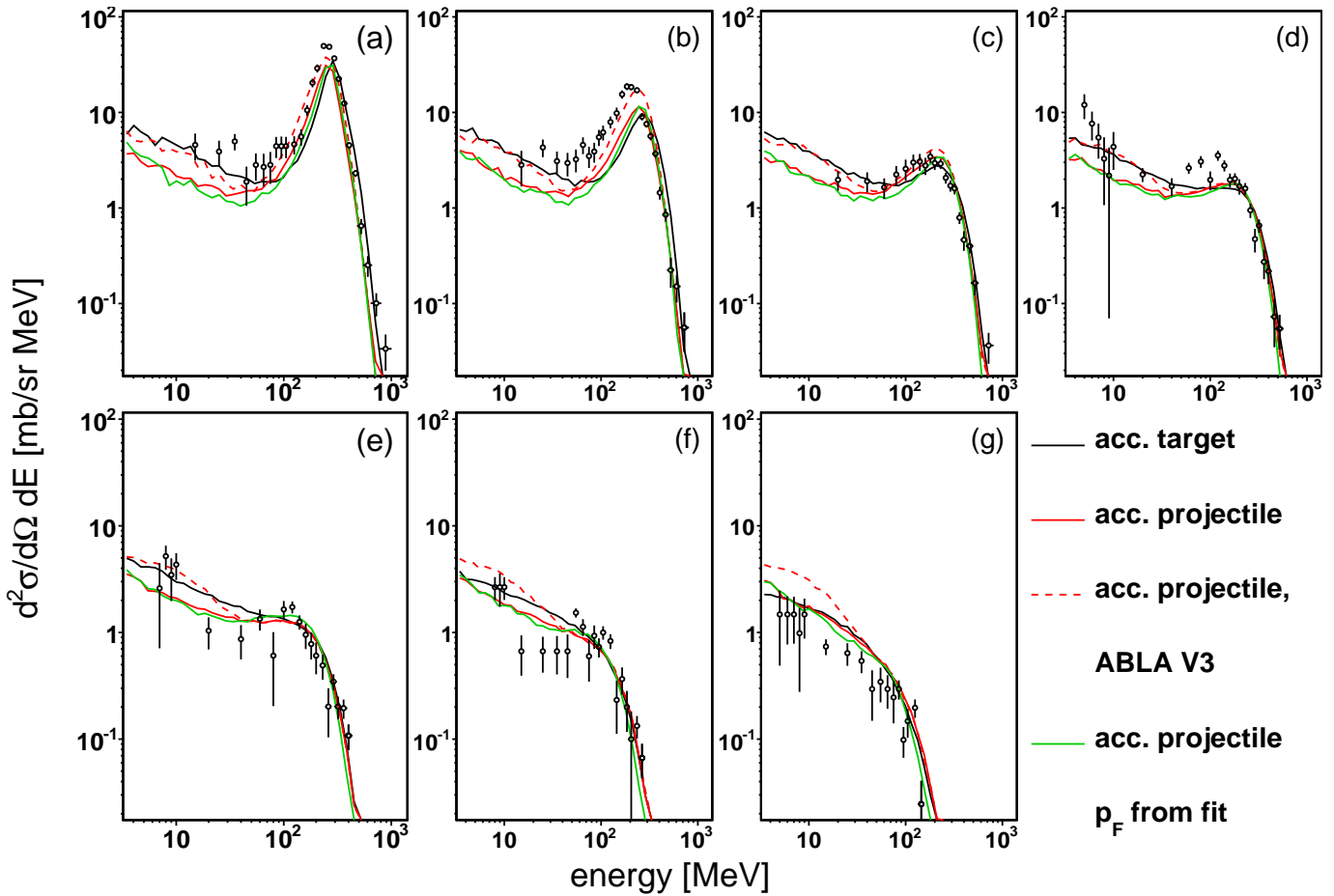}
  \caption{(Color online) Double-differential cross sections for neutron
    production at (a) $5^\circ$, (b) $10^\circ$, (c) $20^\circ$, (d) $30^\circ$,
    (e) $40^\circ$, (f) $60^\circ$ and (g) $80^\circ$, from a 290-$A$MeV
    $^{12}$C+C reaction. The \inclxx\ calculations are presented in
    accurate-target (black lines) and accurate-projectile (solid red lines)
    mode, coupled with the \geh\ de-excitation model. We also show an
    accurate-projectile calculation with \ablaold\ (dashed red lines) and a
    calculation with a modified value of the Fermi momentum (green lines, see
    text). Experimental data are taken from
    Ref.~\citenum{iwata-ddxs_neutron}.\label{fig:ddxs_C_C_n_inclxx}}
\end{figure*}

Figure~\ref{fig:ddxs_C_C_n_inclxx} demonstrates the difference between
accurate-projectile and accurate-target mode (see
Section~\ref{sec:proj-targ-asymm}) using double-differential cross sections for
neutron production from the symmetric 290-$A$MeV $^{12}$C+C
\cite{iwata-ddxs_neutron} reaction. Note that the incident energy is large
enough so that the low-energy fusion sector can be neglected.  Both calculations
were coupled to the native \geant\ de-excitation model
\cite{quesada-g4excitationhandler}. Differences are mostly visible at forward
angles and low energy; the predictions for the largest angles are very close to
each other. In general, the shapes of the experimental spectra are quite well
reproduced by both \inclxx\ calculations. Therefore, we conclude that neutron
emission is nevertheless projectile-target symmetric to a good degree.

Note that the experimental data show a peak at forward angles roughly centered
around the nominal energy per nucleon of the projectile and corresponding to
neutrons with a rather small energy in the projectile rest frame. In \inclxx,
they mainly originate from the break-up of the projectile nucleus. The shape and
the height of the peak depend on the selected de-excitation model; this is
illustrated again by Fig.~\ref{fig:ddxs_C_C_n_inclxx}, where the
accurate-projectile calculation coupled with \geh\ (which for this system
reduces to Fermi break-up, see Section~\ref{sec:de-excitation-stage}) is
contrasted to an \inclxx/\ablaold\ calculation (solid and dashed red lines,
respectively). The \ablaold\ model yields a larger peak, in better agreement
with the experimental data at forward angles, but also affects the low-energy
neutron yields.

The shape of the projectile-fragmentation peak is also sensitive to the assumed
Fermi momentum of the projectile nucleus. This is illustrated by an
accurate-projectile \inclxx\ calculation using a mass-dependent Fermi momentum
given by
\begin{gather*}
  p_F(A)=\alpha-\beta \exp{\left(-\gamma A\right)}\\
  \alpha=259.416\text{~MeV}/c\\\beta=152.824\text{~MeV}/c\\\gamma=9.5157\cdot10^{-2}\text.
\end{gather*}
This formula is a fit to Moniz \etal's direct measurements by quasi-elastic
electron scattering \cite{moniz-fermi_momenta}. For $^{12}$C, the formula yields
$p_F(^{12}\text{C})\simeq 210\text{~MeV}/c$ (Moniz \etal's measurement is
actually $(221\pm5)\text{~MeV}/c$), which is not very different from the default
\inclxx\ value of 270~MeV/$c$ (see Table~\ref{tab:densities}). Nevertheless,
Fig.~\ref{fig:ddxs_C_C_n_inclxx} shows that the neutron spectra are roughly
equally sensitive to the de-excitation model and to the Fermi momentum.

\begin{figure*}
  \includegraphics[width=\linewidth]{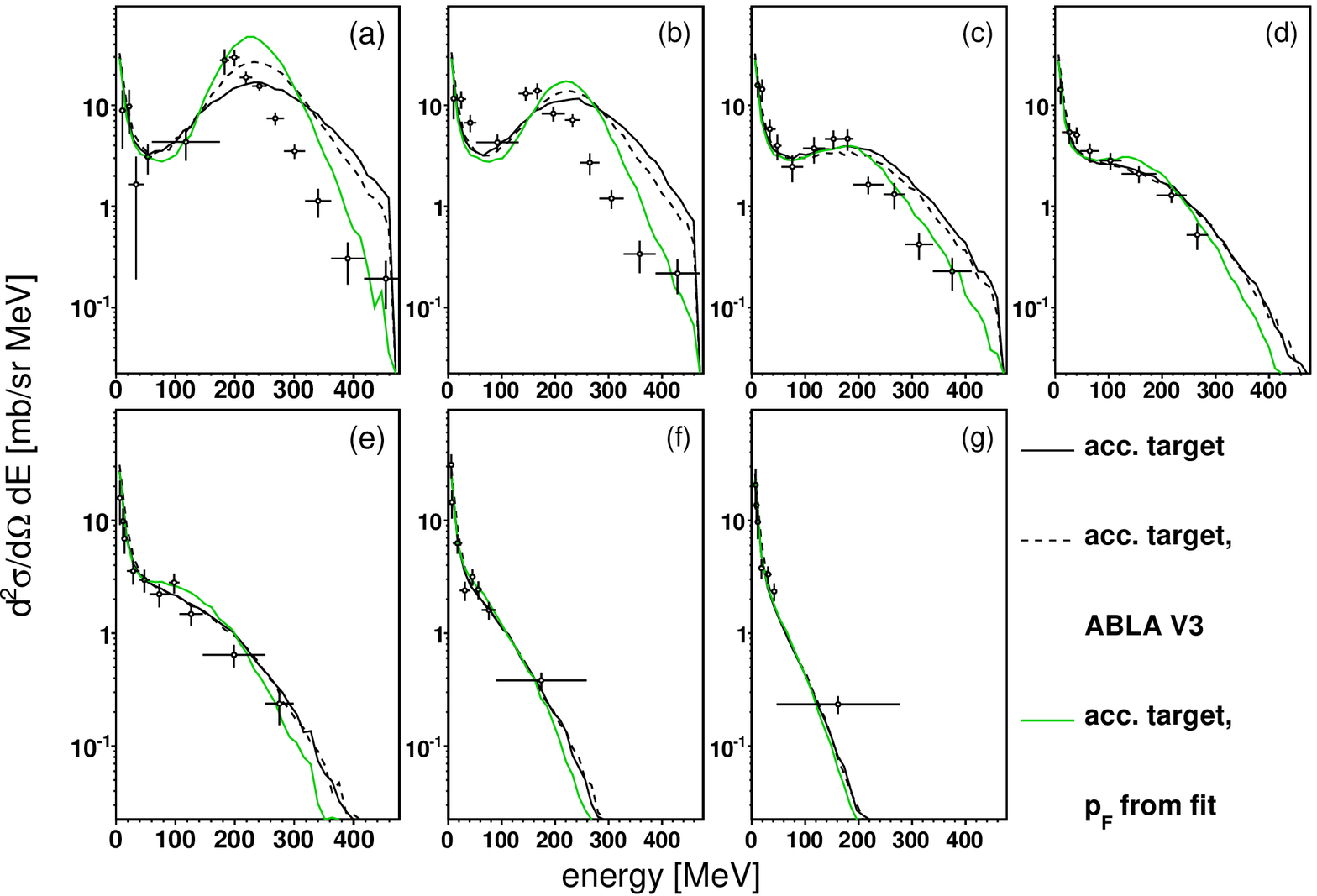}
  \caption{(Color online) Double-differential cross sections for neutron
    production at (a) $4^\circ$, (b) $9^\circ$, (c) $20^\circ$, (d) $30^\circ$,
    (e) $40^\circ$, (f) $60^\circ$ and (g) $80^\circ$, from a 230-$A$MeV
    $^4$He+Cu reaction. The \inclxx\ calculations are presented in
    accurate-target mode (solid black lines), coupled with the \geh\
    de-excitation model. We also show a calculation with \ablaold\ (dashed black
    lines) and a calculation with a modified value of the Fermi momentum (green
    lines, see text). Experimental data are taken from
    Ref.~\citenum{heilbronn-alpha_neutrons}.\label{fig:ddxs_a_Cu_n_inclxx}}
\end{figure*}

The sensitivity to $p_F$ can be enhanced by looking at lighter projectiles, such
as $^4$He in the 230-$A$MeV $^4$He+Cu reaction depicted in
Fig.~\ref{fig:ddxs_a_Cu_n_inclxx}. Here $p_F(^4\text{He})=155\text{~MeV}/c$,
almost a factor of two smaller than the nominal \inclxx\ value. For this system,
standard \inclxx\ fails to describe the part of the spectrum above
200~MeV. However, the projectile-fragmentation peak at forward angles is much
better reproduced using the empirical Fermi momentum. Nevertheless, since we
have not extensively tested the implications of empirical Fermi momenta in
\inclxx, we keep $p_F=270\text{~MeV}/c$ as the default value. We reserve a
detailed study to a future publication.

\begin{figure*}
  \includegraphics[width=\linewidth]{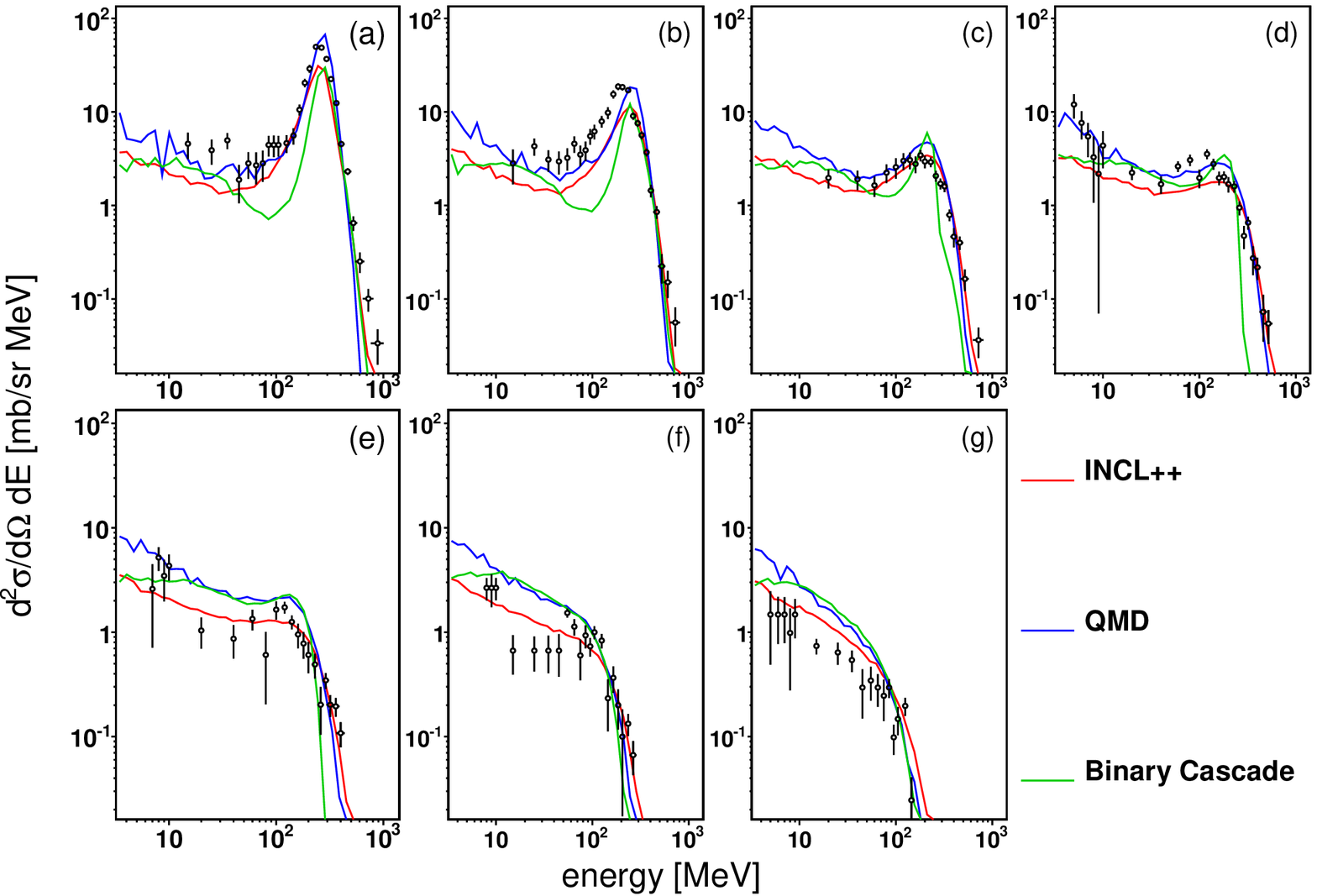}
  \caption{(Color online) Same as Fig.~\ref{fig:ddxs_C_C_n_inclxx} for \inclxx\
    (accurate-projectile mode, red lines), \geant's \qmd\ model (blue lines) and
    \bic\ model (green lines). All models are coupled to
    \geh.\label{fig:ddxs_C_C_n_models}}
\end{figure*}

\begin{figure*}
  \includegraphics[width=\linewidth]{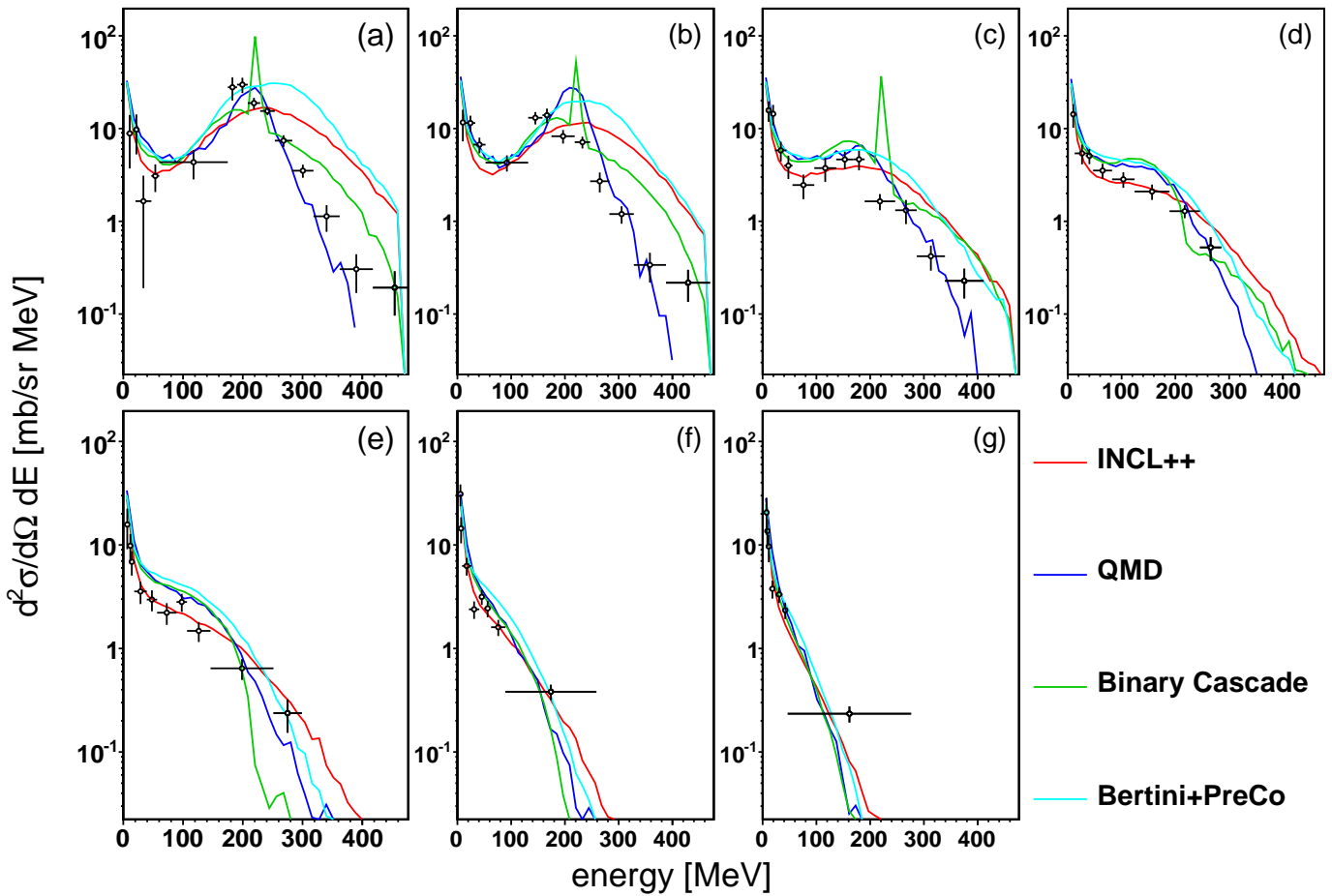}
  \caption{(Color online) Same as Fig.~\ref{fig:ddxs_a_Cu_n_inclxx} for \inclxx\
    (accurate-projectile mode, red lines), \geant's \qmd\ model (blue lines),
    \bic\ (green lines) and Bertini+PreCompound (cyan lines) models. All models
    except Bertini+PreCompound are coupled to
    \geh.\label{fig:ddxs_a_Cu_n_models}}
\end{figure*}

\begin{figure*}
  \includegraphics[width=\linewidth]{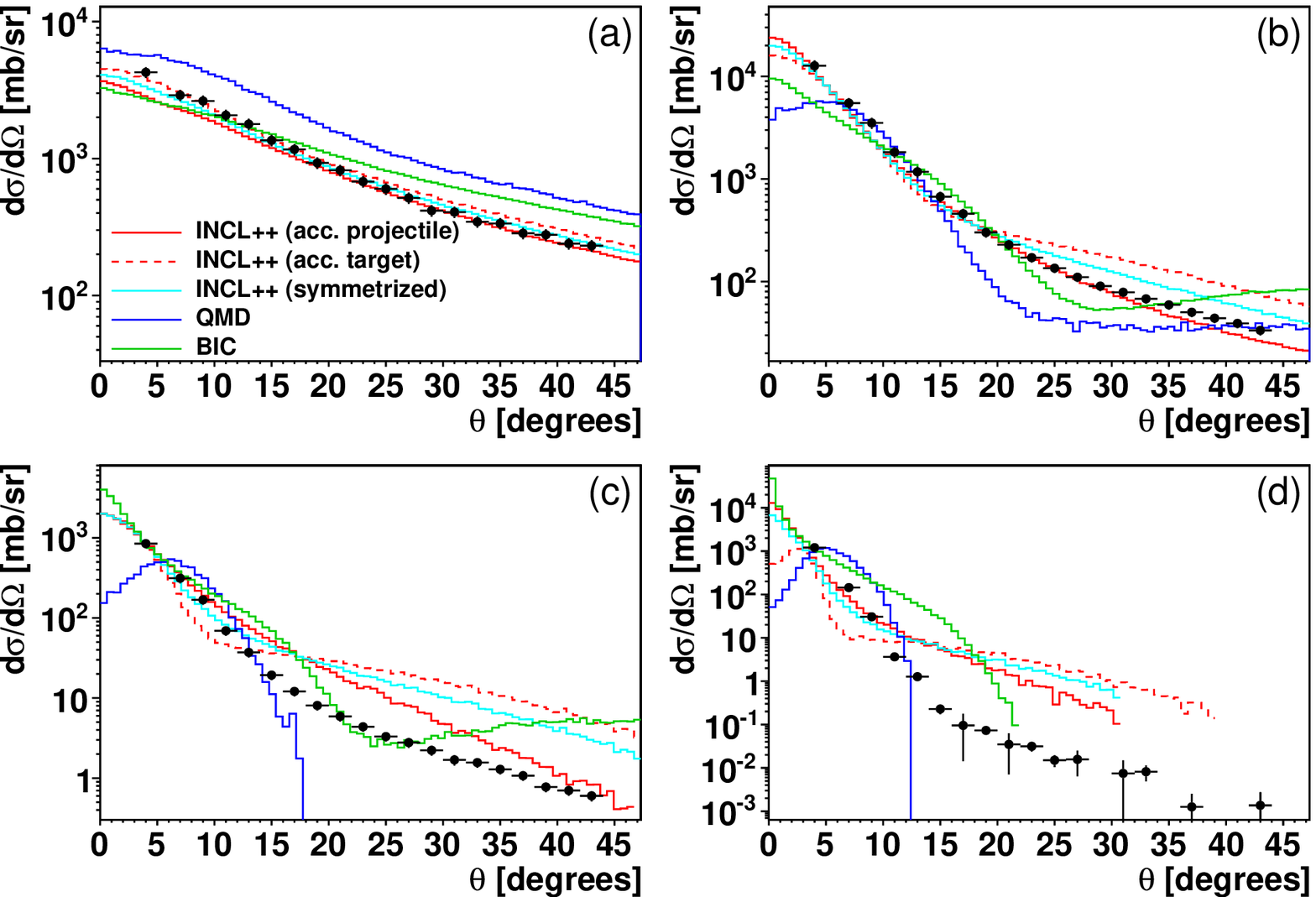}
  \caption{(Color online) Angle-differential cross section for the production of (a) protons,
    (b) $^4$He, (c) $^7$Li and (d) $^{11}$C from the 95-$A$MeV $^{12}$C+$^{12}$C
    reaction. Calculations with \inclxx\ (accurate-projectile mode, solid red
    lines; accurate-target mode, dashed red lines; randomly symmetrized, cyan
    lines), \qmd\ (blue lines) and \bic\ (green lines) are shown. Experimental
    data are taken from Ref.~\citenum{dudouet-C+C}.\label{fig:angdist_dudouet}}
\end{figure*}

Figures~\ref{fig:ddxs_C_C_n_inclxx} and \ref{fig:ddxs_a_Cu_n_inclxx} suggest
that \inclxx\ generally succeeds to capture the essential aspects of the
experimental data. This conclusion is corroborated by
Figures~\ref{fig:ddxs_C_C_n_models} and \ref{fig:ddxs_a_Cu_n_models}, which show
a comparison of the \inclxx\ result (in accurate-projectile mode) to
calculations performed by other models available in \geant: \qmd\ model (blue),
\bic\ \cite{folger-g4bic} (green) and Bertini+PreCompound
\cite{heikkinen-bertini_g4,kelsey-bertini_preparation,gudima-precompound}
(cyan, only applicable for the $^4$He-induced reaction). All models use the
same de-excitation (\geh), except Bertini, which has its own internal
de-excitation module.

One notices that the \bic\ predictions are generally in less good agreement with
the experimental data than \inclxx. The \qmd\ results are everywhere comparable
to or worse than the \inclxx\ calculation, except at the forward-most angles,
which were shown to be improvable in \inclxx\ by using the empirical Fermi
momentum. Note also that the CPU time for \qmd\ is about two orders of magnitude
larger than for \inclxx. All the other models fail to describe the
$^4$He-fragmentation peak, which (in view of the above) might suggest that they
employ unrealistic Fermi momenta for this projectile. In addition, the \bic\
model shows some unphysical structures at small angles for the $^{4}$He+Cu
system.

We now turn to the production of charged particles. We focus in particular on a
recent experiment by \citet{dudouet-C+C,dudouet-C_frag}, who measured
double-differential cross sections for the production of several charged
particles from reactions induced by a 95-$A$MeV $^{12}$C beam on targets ranging
from hydrogen to titanium. We are mostly interested in the carbon-target data
for the purpose of verifying the \inclxx\ nucleus-nucleus extension and
assessing the severity of the projectile-target asymmetry (see
Section~\ref{sec:proj-targ-asymm}). Calculations with some \geant\ models have
been presented in Ref.~\citenum{dudouet-models}, where however the authors used
\inclxx\ \code{v5.1.9}, which was shown above to be affected by serious
drawbacks for the $^{12}$C+$^{12}$C reaction (Fig.~\ref{fig:reaction_C12}). Our
results can be reproduced using \geant\ \code{v10.0} and should be considered as
references.

First, we observe that the incident energy (95~$A$MeV) is rather low. The
conditions for the applicability of the intranuclear-cascade hypothesis
(independent binary nucleon-nucleon collisions) are not very well fulfilled
here. Figure~\ref{fig:reaction_C12} indicates that the reaction cross section
predicted by \inclxx\ is in excess of the experimental value by about 30\% at
this energy. Note also that \inclxx's low-energy fusion sector is responsible
for 43\% of the reaction cross section, which is far from negligible. Given the
empirical nature of the fusion sector, we do not expect very accurate
predictions.

Since the 95-$A$MeV $^{12}$C+C reaction is essentially symmetric, we shall use
this example to illustrate random symmetrization, as described in
Section~\ref{sec:proj-targ-asymm}.

Figure~\ref{fig:angdist_dudouet} shows angular-differential cross sections for
the production of protons, $^4$He, $^7$Li and $^{11}$C. For each angle, the
calculated ejectile energy distributions were integrated above the detection
thresholds reported by \citet[Table~IV in][]{dudouet-C_frag}.

It is striking that none of the considered models can accurately reproduce all
the experimental data (see however calculations with \laqgsm\
\cite{mashnik-fragments}). The proton angular distributions predicted by
\inclxx\ (either in accurate-projectile or in accurate-target mode) are quite
close to the experimental data; the accurate-projectile and accurate-target
predictions are again very similar, which confirms the remark made about
Fig.~\ref{fig:ddxs_C_C_n_inclxx}.

The agreement progressively degrades as the mass of the ejectile increases,
especially for the calculation in accurate-target mode. \citet{dudouet-C_frag}
showed that the experimental angular distributions can be represented as a sum
of a Gaussian and an exponential contribution and claimed \cite{dudouet-models}
that no model can reproduce this trend. Figure~\ref{fig:angdist_dudouet} shows
that this is incorrect: although the exponential tail of the angular
distribution might be quantitatively incorrect (especially for $^{11}$C),
\inclxx\ in accurate-projectile mode is clearly the only model that can capture
the trend of the experimental data. In spite of the crudeness of the model
ingredients, the agreement with the experimental data is remarkable, except for
the case of $^{11}$C.

Since the accurate-projectile results are generally closer to the experimental
data than the accurate-target calculations, there is not much to be gained here
by applying random symmetrization. The results of randomly symmetrized
calculations, which are shown in Fig.~\ref{fig:angdist_dudouet} and which are
simply averages of the accurate-projectile and accurate-target results, are
\emph{a fortiori} in good agreement with the experimental data for protons, but
they are less good than the results in accurate-projectile mode for all the
other ejectiles.

As far as the other models are concerned, \qmd\ seems to systematically
underpredict the fragment yields at small angles. In general, the shape of the
angular distribution is very different from the experimental result. Even for
protons one can observe a sizable overestimation of the yield. The \bic\ results
manage to capture at least some qualitative features of the experimental data,
but its predictions are in general less accurate than those of \inclxx.

\begin{figure*}
  \includegraphics[width=\linewidth]{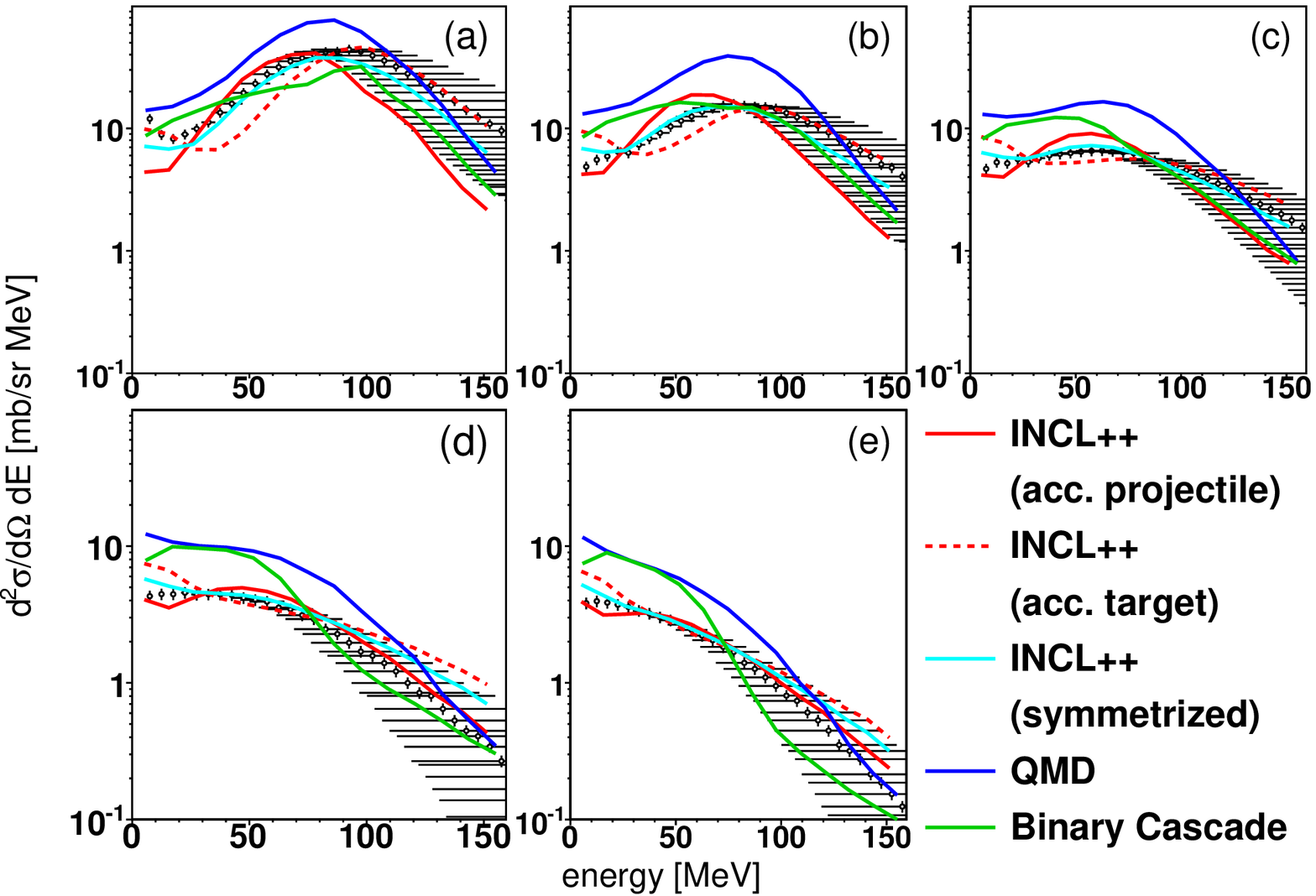}
  \caption{(Color online) Double-differential cross sections for the production
    of protons at (a) $4^\circ$, (b) $13^\circ$, (c) $21^\circ$, (d) $29^\circ$
    and (e) $43^\circ$, from the 95-$A$MeV $^{12}$C+$^{12}$C
    reaction. Calculations with \inclxx\ (accurate-projectile mode, solid red
    lines; accurate-target mode, dashed red lines), \qmd\ (blue lines) and \bic\
    (green lines) are shown. Experimental data are taken from
    Ref.~\citenum{dudouet-C+C}.\label{fig:ddxs_C12_C_95_p}}
\end{figure*}

\begin{figure*}
  \includegraphics[width=\linewidth]{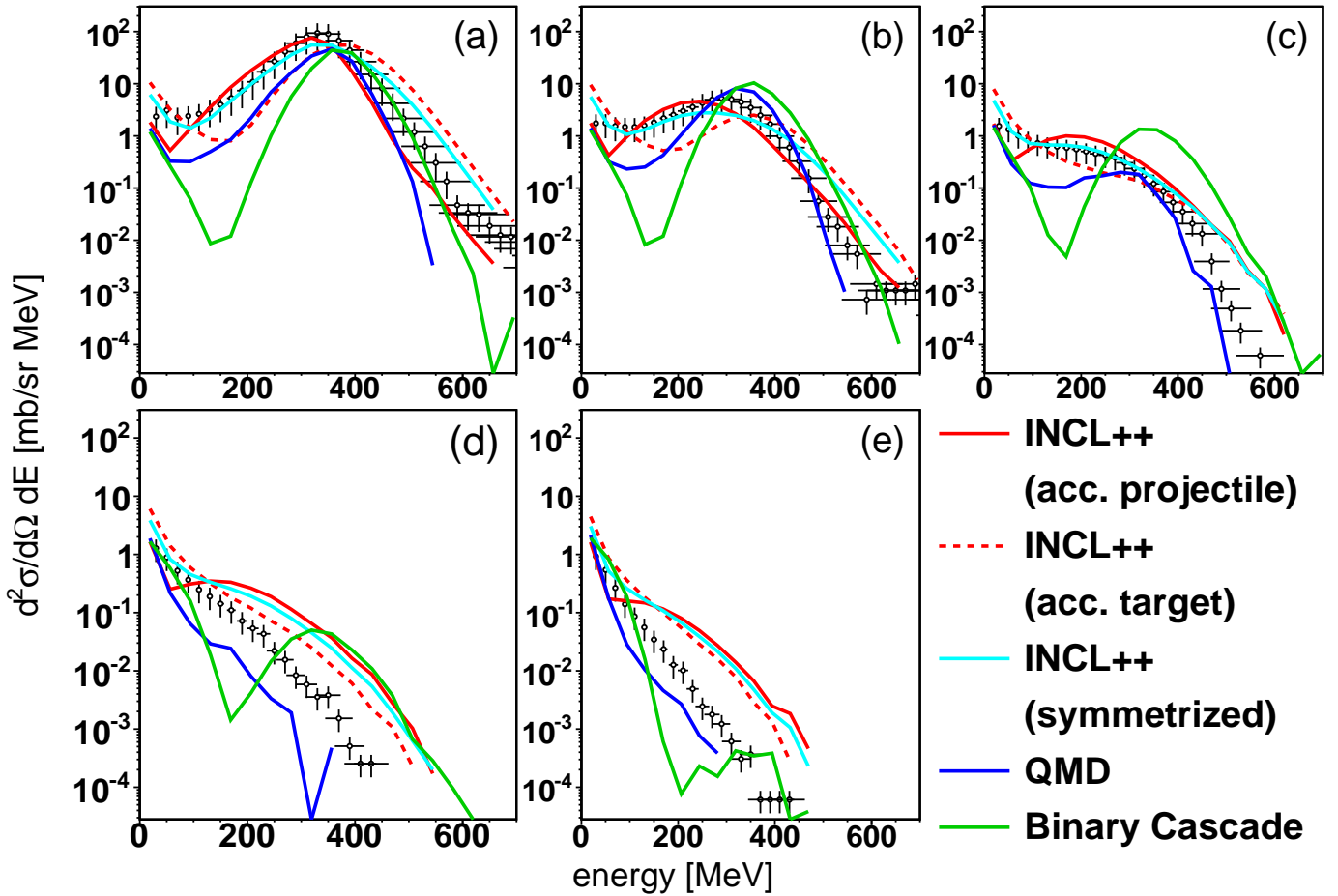}
  \caption{(Color online) Same as Fig.~\ref{fig:ddxs_C12_C_95_p}, for $^4$He
    ejectiles.\label{fig:ddxs_C12_C_95_a}}
\end{figure*}

\begin{figure*}
  \includegraphics[width=\linewidth]{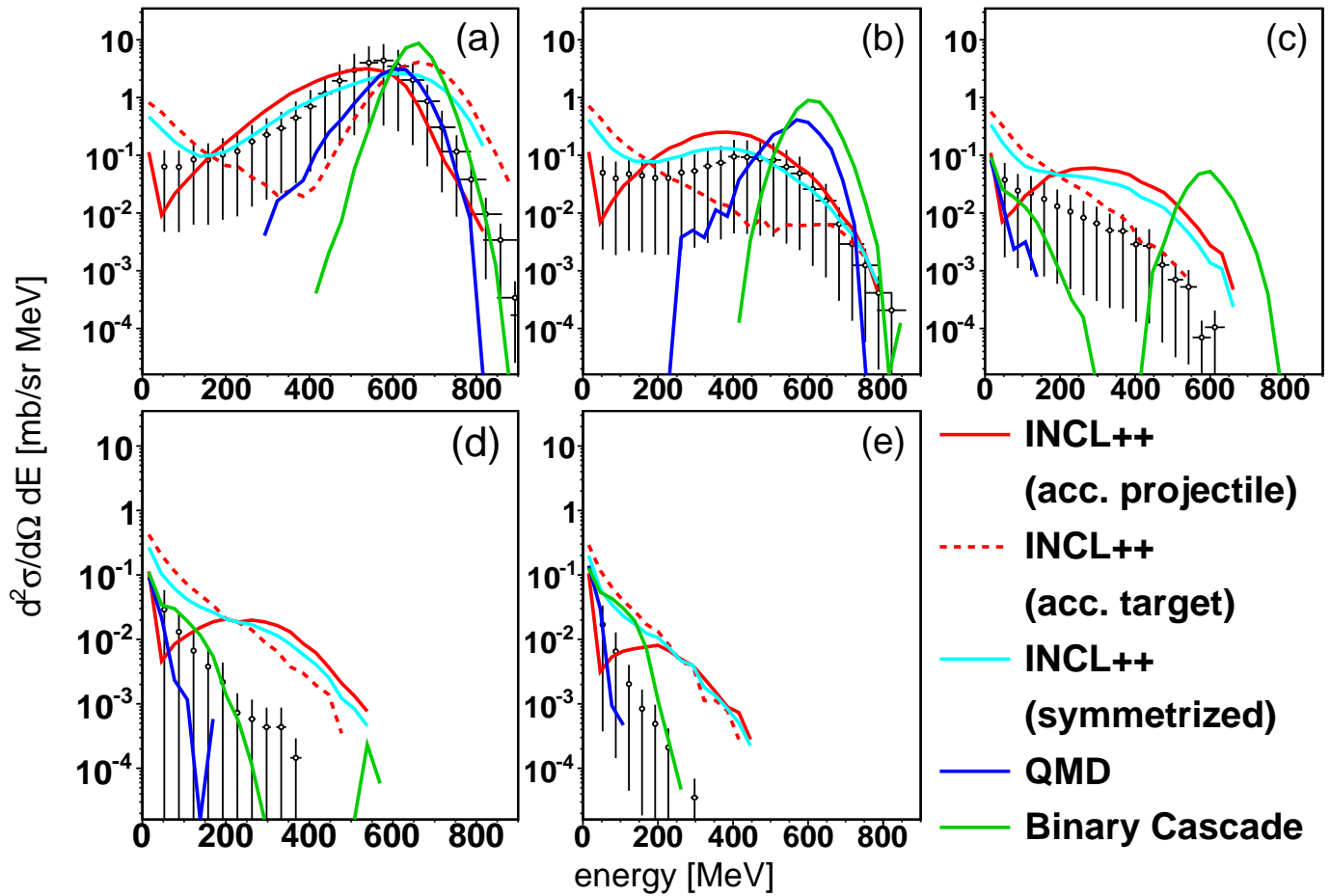}
  \caption{(Color online) Same as Fig.~\ref{fig:ddxs_C12_C_95_p}, for $^7$Li
    ejectiles.\label{fig:ddxs_C12_C_95_Li7}}
\end{figure*}

\begin{figure*}
  \includegraphics[width=\linewidth]{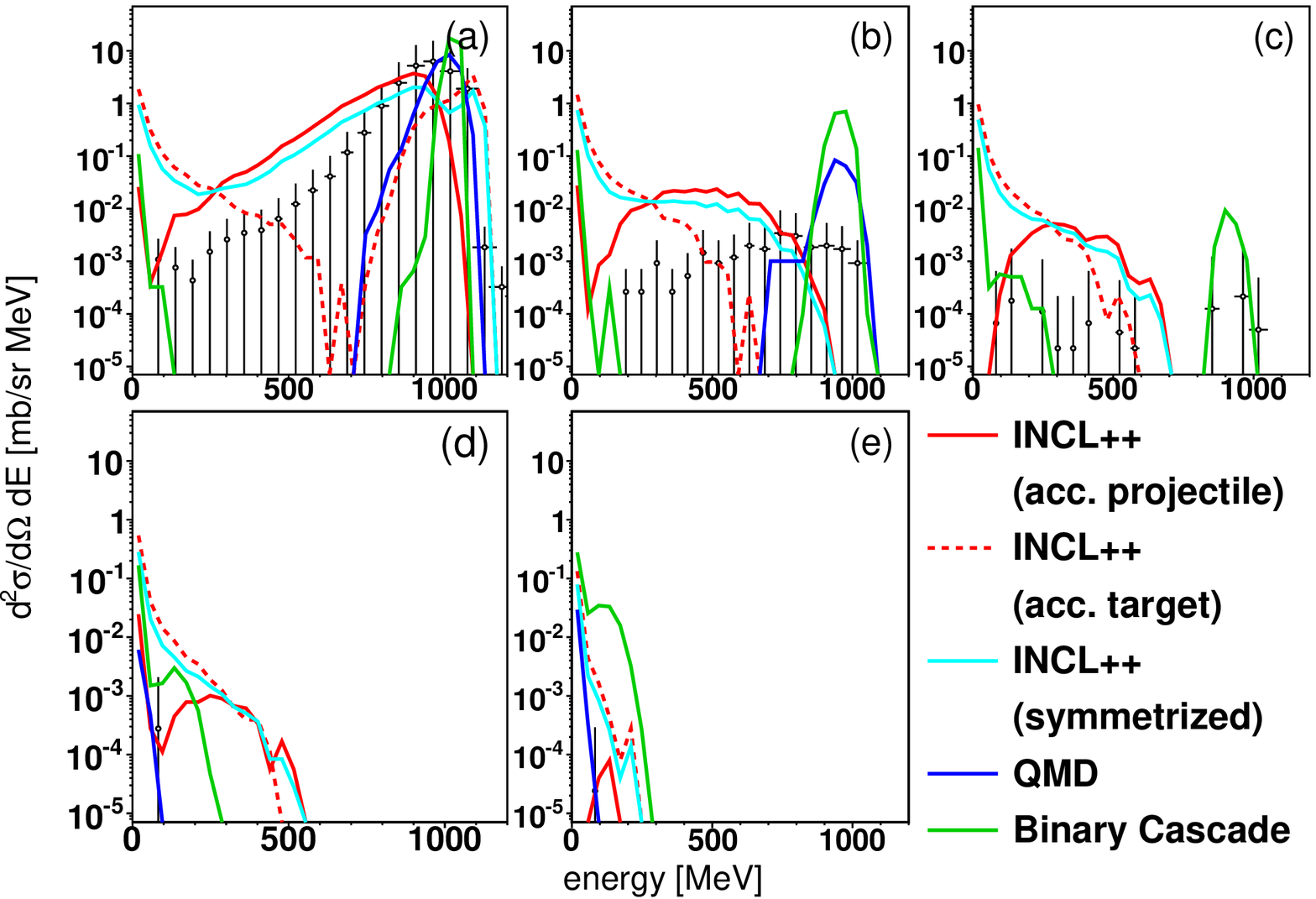}
  \caption{(Color online) Same as Fig.~\ref{fig:ddxs_C12_C_95_p}, for $^{11}$C
    ejectiles.\label{fig:ddxs_C12_C_95_C11}}
\end{figure*}

Double-differential spectra for the same ejectiles are shown in
Figs.~\ref{fig:ddxs_C12_C_95_p}--\ref{fig:ddxs_C12_C_95_C11}. Here we notice
larger discrepancies than in Fig.~\ref{fig:angdist_dudouet}, even for the
\inclxx\ calculation in accurate-projectile mode. For example, no model can
reproduce the slope of the high-energy tail of the proton spectra at all
angles. Experimental fragment spectra show a mid-rapidity component that is not
reproduced by any of the models, although \inclxx\ is much closer to the data
than the others. The randomly-symmetrized calculations are especially good on
the spectra for $^4$He nuclei, Fig.~\ref{fig:ddxs_C12_C_95_a}. At large angles,
the \inclxx\ spectra show a broad bump that is not seen in the data and that is
the continuation of the projectile-like fragmentation peak at $4^\circ$. In
other words, the projectile-like fragments seem to pick up too much transverse
momentum from the collision, which results in a too broad angular
distribution. This obviously indicates that the model fails to properly describe
some aspects of projectile fragmentation.

\subsection{Rapidity spectra and projectile-target asymmetry}

\begin{figure}
  \includegraphics[width=\linewidth]{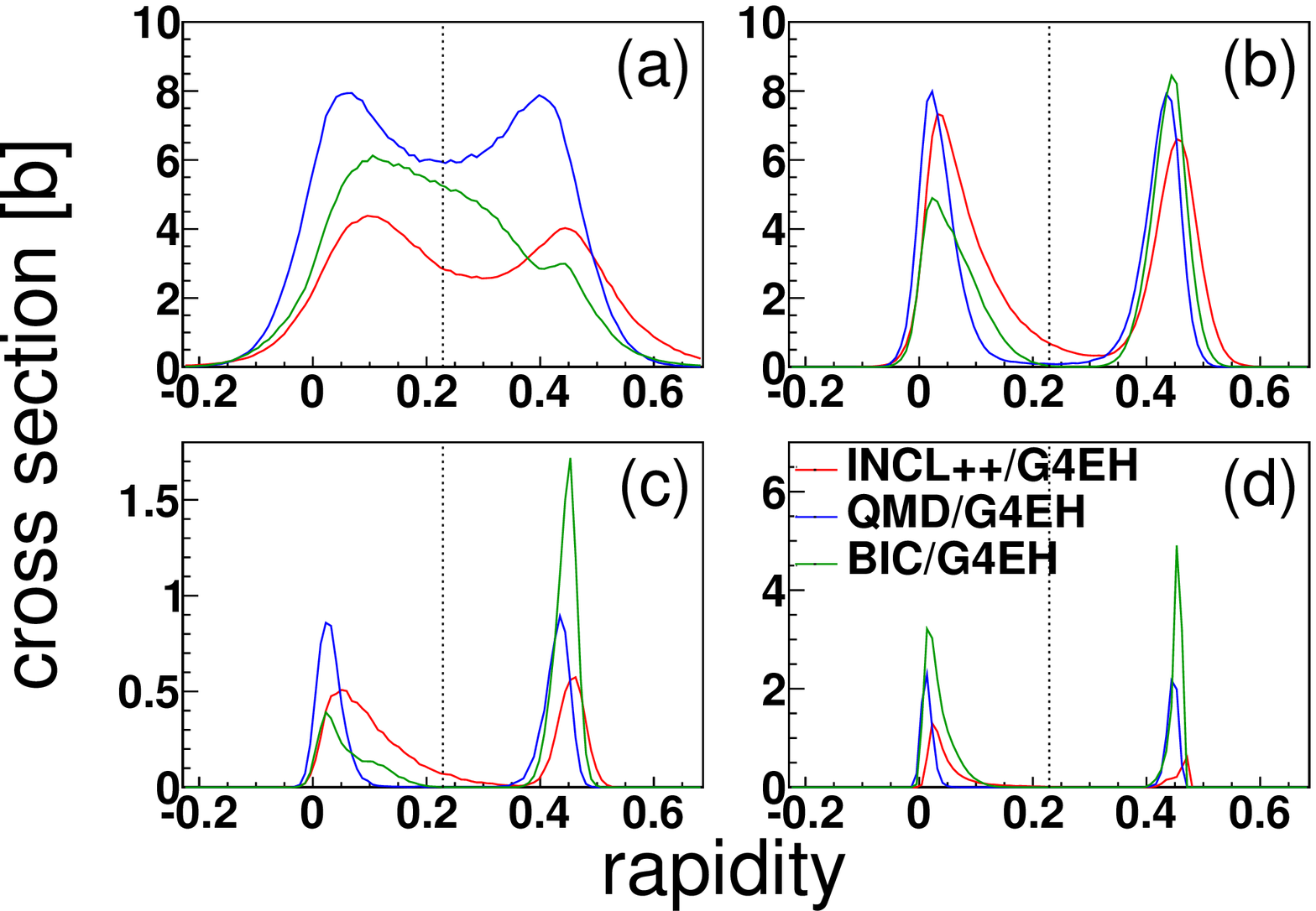}
  \caption{(Color online) Rapidity spectra of (a) protons, (b) $^4$He, (c)
    $^7$Li and (d) $^{11}$C fragments produced in 100-$A$MeV $^{12}$C+$^{12}$C,
    as calculated by \inclxx/\geh\ in accurate-target mode. The vertical dotted
    lines denote the central rapidity.\label{fig:rapidity_100MeV}}
\end{figure}

\begin{figure}
  \includegraphics[width=\linewidth]{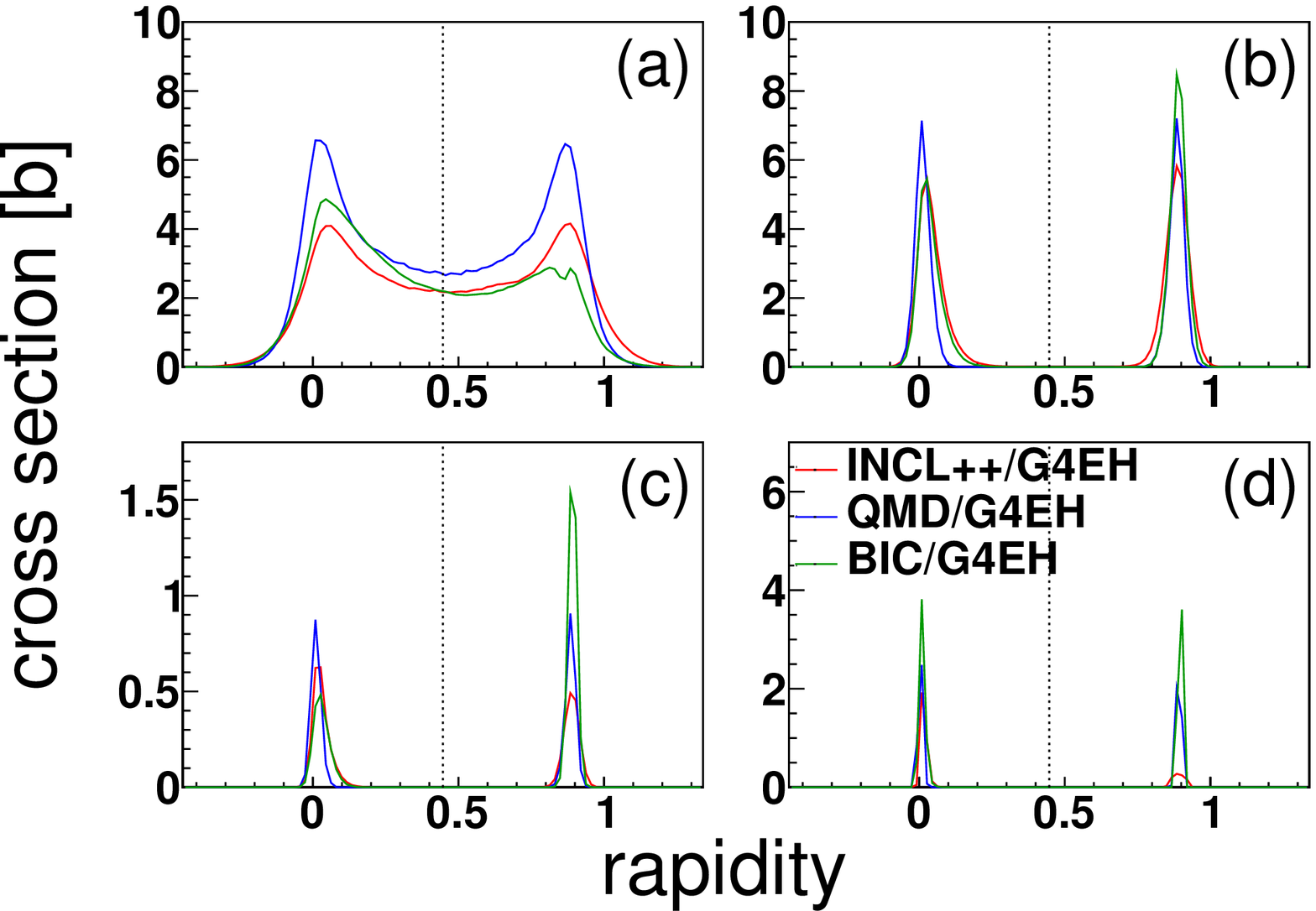}
  \caption{(Color online) Same as Fig.~\ref{fig:rapidity_100MeV}, for 400-$A$MeV
    $^{12}$C+$^{12}$C.\label{fig:rapidity_400MeV}}
\end{figure}

In addition to and independently of the comparison with the experimental data,
we present in Figs.~\ref{fig:rapidity_100MeV} and \ref{fig:rapidity_400MeV} the
rapidity spectra of particles emitted in $^{12}$C+$^{12}$C respectively at 100
and 400~$A$MeV laboratory energy, which we can use to assess the severity of the
projectile-target asymmetry in \inclxx. The results of calculations performed
with \bic\ and \qmd\ are also shown. A perfectly projectile-target symmetric
model produces distributions that are symmetric around the dotted mid-rapidity
line, which is located at half of the nominal rapidity of the projectile.

Visual inspection of Figs.~\ref{fig:rapidity_100MeV}--\ref{fig:rapidity_400MeV}
reveals violations of the projectile-target symmetry in \inclxx\ and
\bic. Still, it is clear that \inclxx\ is approximately symmetric for protons
(as discussed above) and progressively degrades as the ejectile mass
increases. The results of \qmd\ are fully symmetric, which is a consequence of
the fact that the model treats all nucleons on the same footing.

\subsection{Fragmentation cross sections}\label{sec:fragm-cross-sect}

We finally turn to the analysis of fragmentation cross section. In keep with our
approach to the validation of nucleon-induced reactions, we focus on
measurements of isotopic cross sections in inverse kinematics. The advantage of
such data sets is that they provide a comprehensive picture of the reaction
mechanism. The accurate fragmentation data on hydrogen targets taken using the
Fragment Separator at GSI (Darmstadt, Germany)
\citep[e.g.][]{napolitani-fe,benlliure-gold,enqvist-lead} have often proved
invaluable for the study of the nucleon-nucleus reaction mechanism and for the
optimization of de-excitation models.

Unfortunately, the coverage for reactions on light nuclei is not as extensive as
for hydrogen. Beryllium is often exploited as a production target in the search
for exotic neutron-rich
\cite[e.g.][]{stolz-sn,*benlliure-south,*benlliure-xe,*alvarez_pol-u238} or
neutron-poor \cite[e.g.][]{blank-proton_rich,*kurcewicz-u238} nuclei, but there
exist only few experiments where essentially all projectile-like fragments were
covered.
\modified{We chose to limit our comparison to such extensive experimental
  datasets.}

\begin{figure}
  \includegraphics[width=\linewidth]{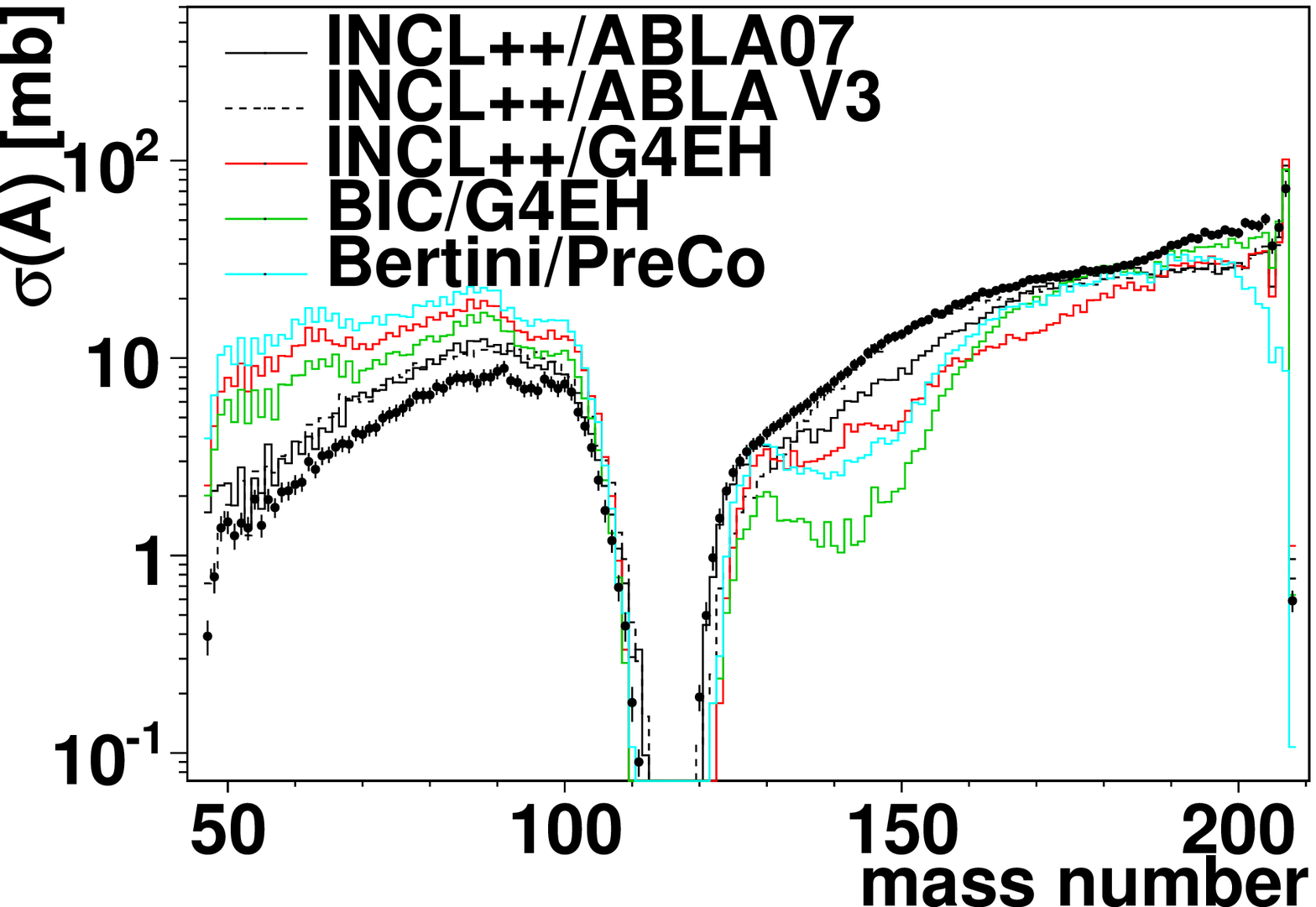}
  \caption{(Color online) Fragmentation cross sections for the 1-$A$GeV
    $^{208}$Pb+$^2$H reaction, as a function of the fragment mass number. Model
    calculations are compared to the data taken from
    Refs.~\citenum{enqvist-lead_d} and \citenum{kelic-bi}. In the plot legend,
    \emph{G4EH} stands for \geh.\label{fig:isot_Pb_d}}
\end{figure}

The data for 1~$A$GeV $^{208}$Pb on deuterium \cite{enqvist-lead_d}, although
only marginally relevant for the assessment of \inclxx's nucleus-nucleus
extension, are perhaps the most complete. Figure~\ref{fig:isot_Pb_d} shows the
mass distributions of the fragments. Note that the model predictions are
obtained by summing up the isotopic cross sections only over the isotopes that
were detected in the experiment; this is the reason of the dip around $A=115$.

One immediately observes that the model predictions are very sensitive to the
choice of the de-excitation model. The distribution of spallation residues
($A>115$) is accurately described only by \inclxx/\abla\ and \inclxx/\ablaold\
(except very close to the projectile mass 208). Models coupled with \geh\
systematically underestimate the yields for deep spallation residues
($115<A\lesssim160$). All the models overestimate the cross sections for the
fission products ($A<115$) by a factor of 2--4. This was already the case with
\code{INCL4.2} \cite[Fig.~23 in][]{boudard-incl}. The overestimation of
\inclxx/\abla's and \inclxx/\ablaold's predictions should probably be related to
the underestimation around $A=195$; it has been shown \cite{cugnon-fission} that
fissioning nuclei belong exactly to this mass range.

\begin{figure*}
  \includegraphics[width=\linewidth]{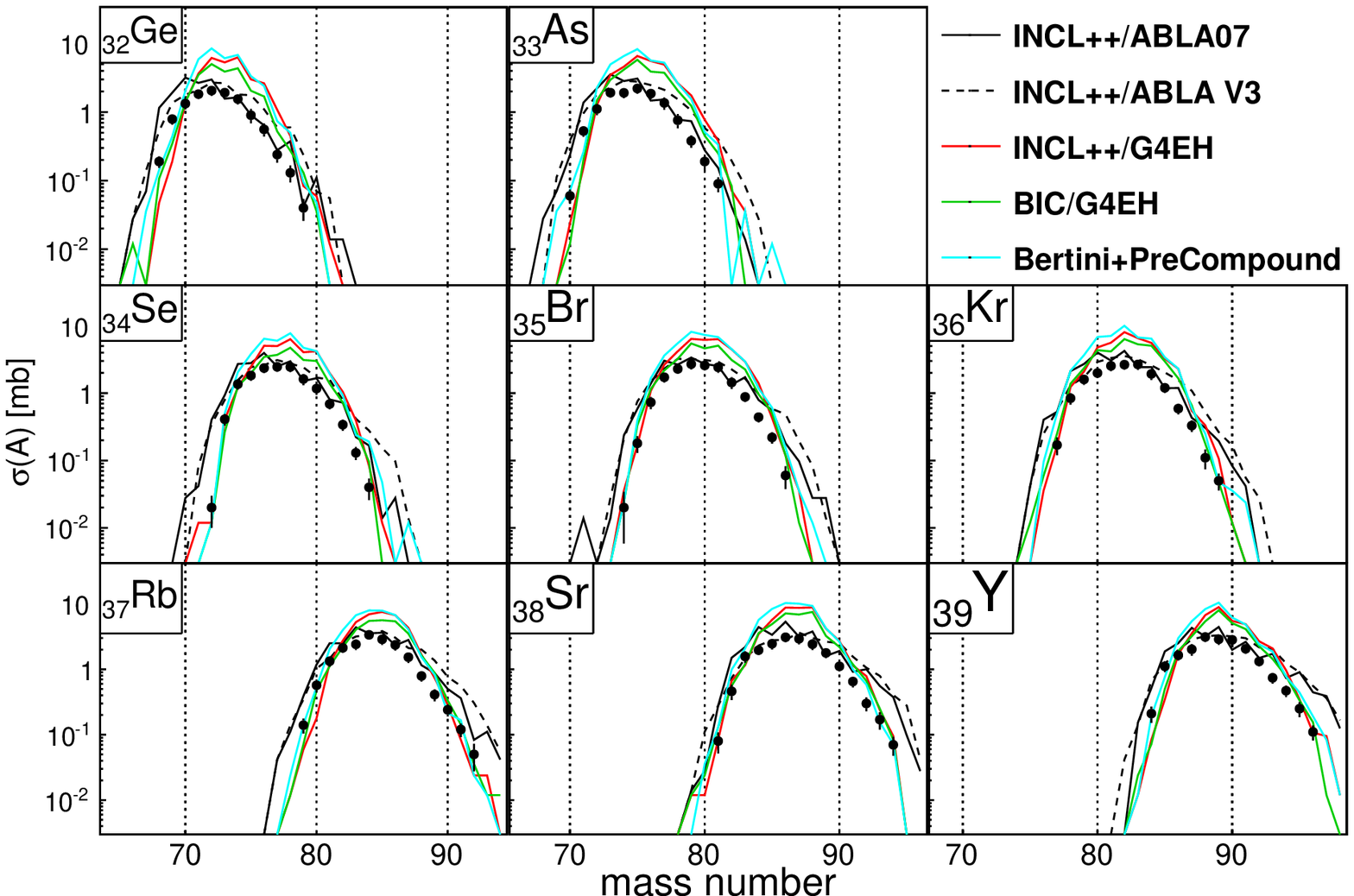}
  \caption{(Color online) Isotopic distributions from the fission region
    ($32\leq Z\leq39$) for the 1-$A$GeV $^{208}$Pb+$^2$H reaction. Model
    calculations are compared to the data taken from
    Ref.~\citenum{enqvist-lead_d}. In the plot legend, \emph{G4EH} stands for
    \geh.\label{fig:pb_d_isot_fiss}}
\end{figure*}

\begin{figure*}
  \includegraphics[width=\linewidth]{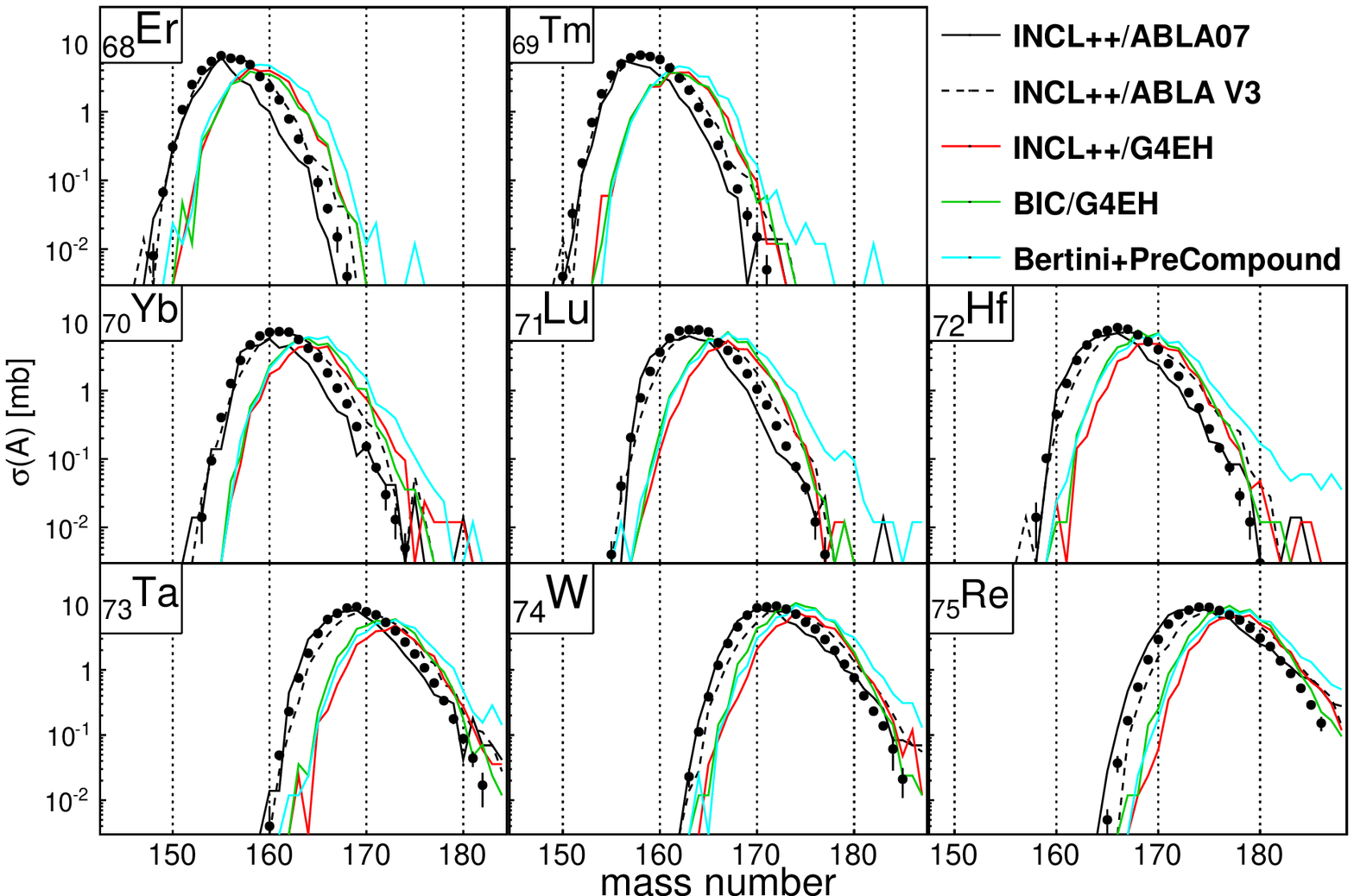}
  \caption{(Color online) Same as Fig.~\ref{fig:pb_d_isot_fiss}, for isotopic
    distributions from the spallation region ($68\leq
    Z\leq75$).\label{fig:pb_d_isot_spal}}
\end{figure*}

Figure~\ref{fig:pb_d_isot_fiss} shows a few isotopic distributions from the
fission region.
\modified{The distributions predicted by \geh\ and Bertini's fission module
  systematically overpredict the peak height; if one rescaled the distributions
  to match the experimental peak height, the tails would be underestimated,
  i.e. the distributions are too narrow. Moreover, the peak position is slightly
  shifted to the neutron-rich side. On the other hand, the \inclxx/\abla\ and
  \inclxx/\ablaold\ predictions have more or less the correct shape.}
This suggests that it should be possible to reproduce the data in
Fig.~\ref{fig:isot_Pb_d} by acting on the competition between fission and
evaporation in \abla\ or \ablaold. Compared to the $^{208}$Pb+$^1$H data in
Fig.~\ref{fig:isot_Pb}, the $^{208}$Pb+$^2$H reaction explores higher excitation
energies and should be more sensitive to dissipative effects in the fission
dynamics \cite{jurado-fissionabla3}, for instance.

Figure~\ref{fig:pb_d_isot_spal} shows some isotopic distributions in the region
of the spallation residues. Again, the predictions by \inclxx/\abla\ and
\inclxx/\ablaold\ are rather close to the experimental data, while the other
models systematically overestimate the $N/Z$ ratio of the residues.

\begin{figure}
  \includegraphics[width=\linewidth]{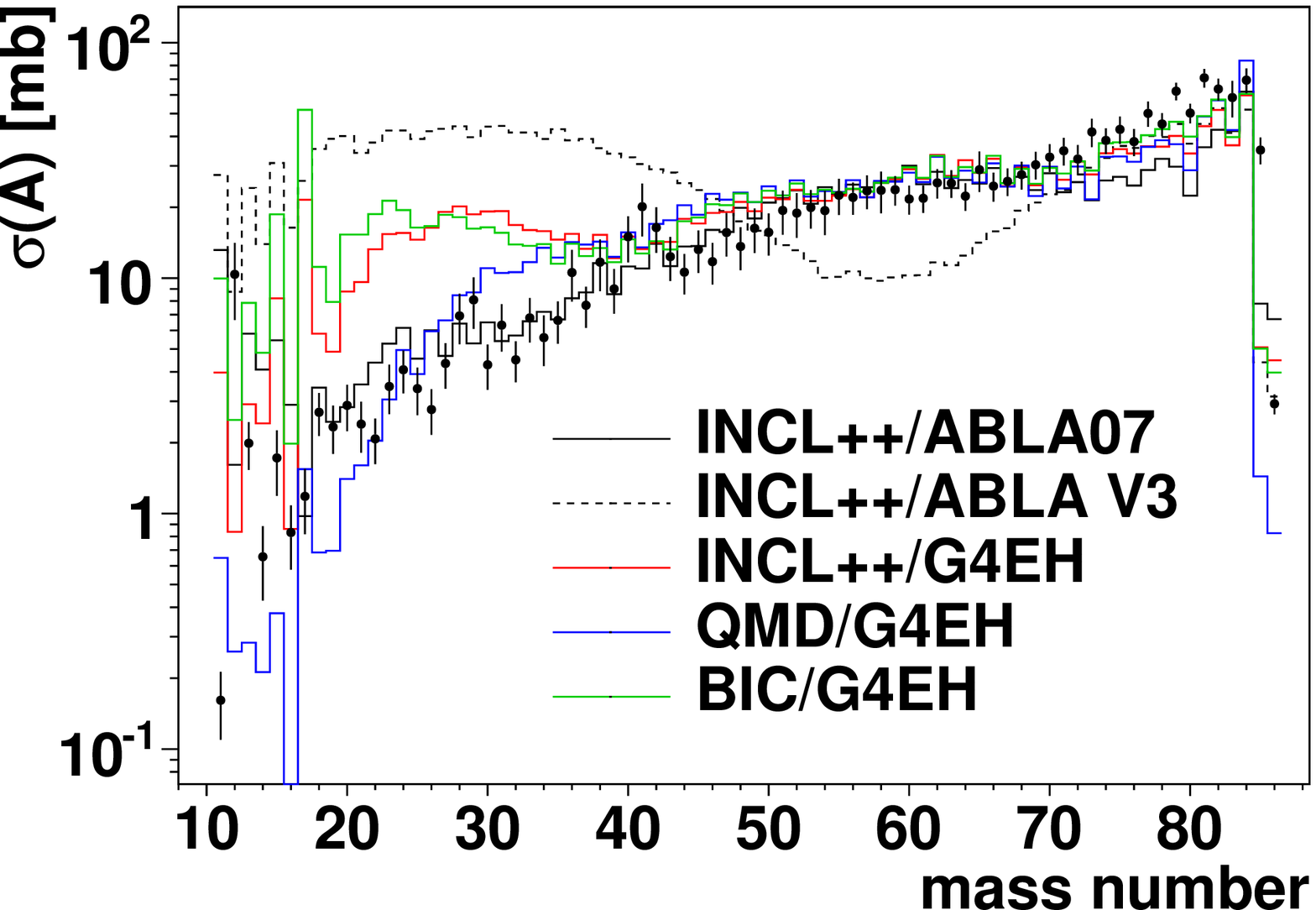}
  \caption{(Color online) Fragmentation cross sections for the 500-$A$MeV $^{86}$Kr+$^9$Be
    reaction, as a function of the fragment mass number. Model calculations are
    compared to the data taken from Refs.~\citenum{weber-kr+be}. In the plot
    legend, \emph{G4EH} stands for \geh.\label{fig:Kr_Be_mass}}
\end{figure}

\begin{figure*}
  \includegraphics[width=\linewidth]{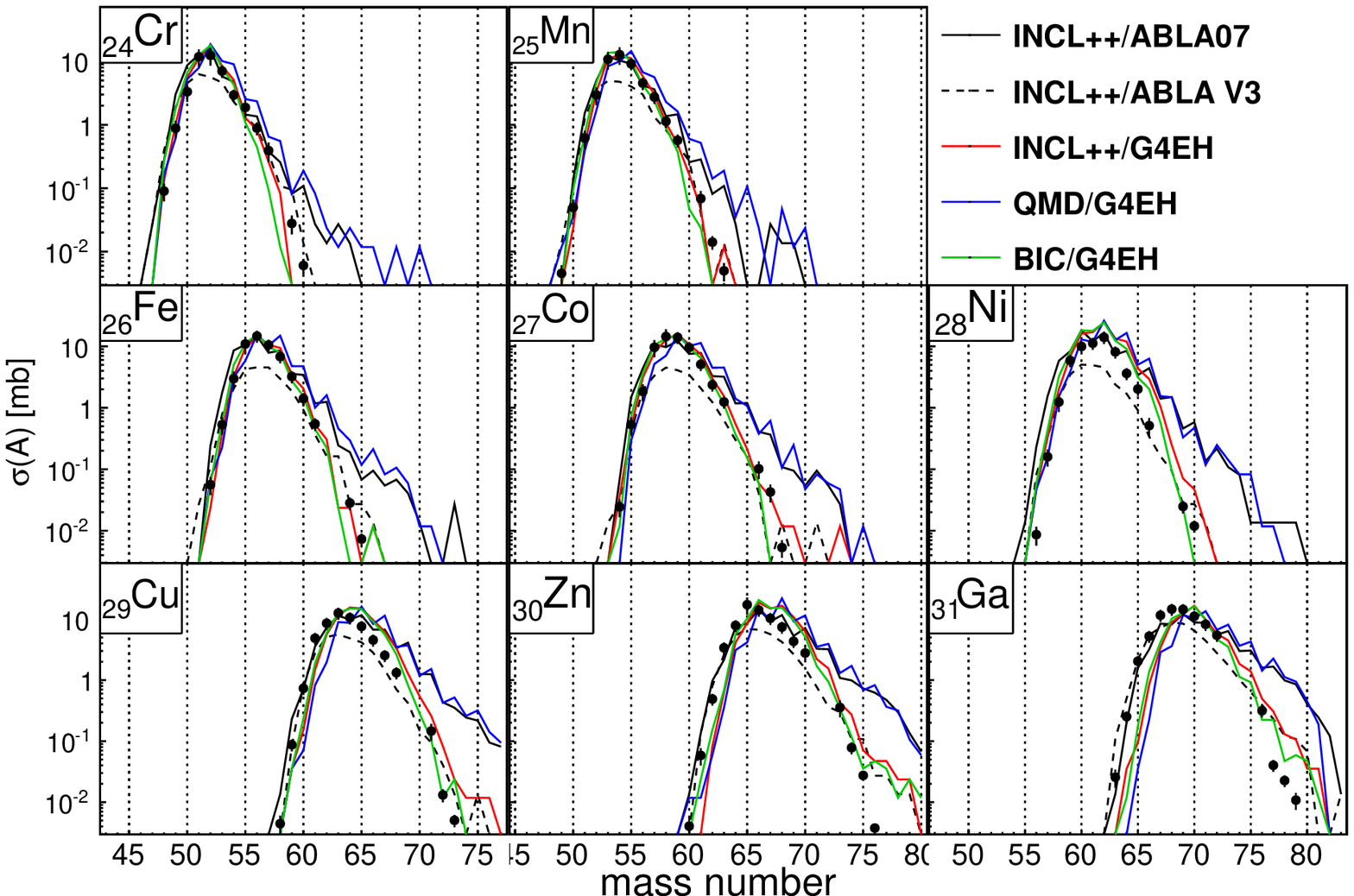}
  \caption{(Color online) Isotopic distributions ($24\leq Z\leq31$) for the
    500-$A$MeV $^{86}$Kr+$^9$Be reaction. Model calculations are compared to the
    data taken from Ref.~\citenum{weber-kr+be}. In the plot legend, \emph{G4EH}
    stands for \geh.\label{fig:Kr_Be_isot}}
\end{figure*}

\modified{As far as reactions on light nuclei are concerned, isotopic
  fragmentation cross sections have been measured by \citet{weber-kr+be} for
  500~$A$MeV $^{86}$Kr+$^9$Be.}
The mass distribution and some isotopic distributions are shown in
Figs.~\ref{fig:Kr_Be_mass} and \ref{fig:Kr_Be_isot}. Again, note that only the
\emph{measured} isotopes contribute to the model predictions for the mass
distribution.

The mass distribution of fragments is again mostly sensitive to the choice of
the de-excitation model. The \inclxx/\abla\ can reproduce most of the
experimental data fairly well, but it underestimates the production of fragments
close to the projectile $^{86}$Kr. \qmd\ performs slightly better close to the
projectile but slightly worse at intermediate mass ($A\simeq35$). The
\inclxx/\geh\ and \bic/\geh\ couplings well reproduce the data for $A>40$, but
overestimate the cross sections for lighter fragments. The \inclxx/\ablaold\
coupling, finally, largely overestimates the cross section for the lightest
fragments.

The large difference between \ablaold\ and \abla\ can be explained by the fact
that evaporation channels in \ablaold\ are limited to proton, neutron and
alpha. \abla, on the other hand, can simulate the emission of any fragment up to
half of the mass of the excited nucleus. Also, \geh\ can evaporate fragments up
to $^{28}$Mg and can be considered to be intermediate between \ablaold\ and
\abla.  Thus, the predicted cross sections in the $A<40$ region seem to
correlate well with the models' maximum ejectile mass. The \qmd/\geh\ coupling
respects this trend to a degree for fragment masses above $\sim25$.

The isotopic distributions in Fig.~\ref{fig:Kr_Be_isot} illustrate that
\inclxx/\abla\ is affected by a defect. The yields for neutron-rich isotopes of
$Z>25$ nuclei are systematically overestimated. This defect might be correlated
with the underestimation of the cross sections for the heaviest fragments. Given
that \abla\ is probably the most sophisticated of the de-excitation models
considered, one might be tempted to conclude that defects in the predicted
isotopic yields are actually due to intranuclear cascade; however, \inclxx/\geh\
does not exhibit the same defect, but \qmd/\geh\ does. The emerging picture is
unclear and no conclusion can be drawn. We have anyway verified that the
overestimation of the neutron-rich isotopes is not due to the neglect of Pauli
blocking on the first collision in the projectile. This is reasonable in the
light of the \qmd/\geh\ results, which are surprisingly similar to those of
\inclxx/\abla\ on the neutron-rich sides of the isotopic distributions, but must
be generated by a completely different dynamics.

\begin{figure}
  \includegraphics[width=\linewidth]{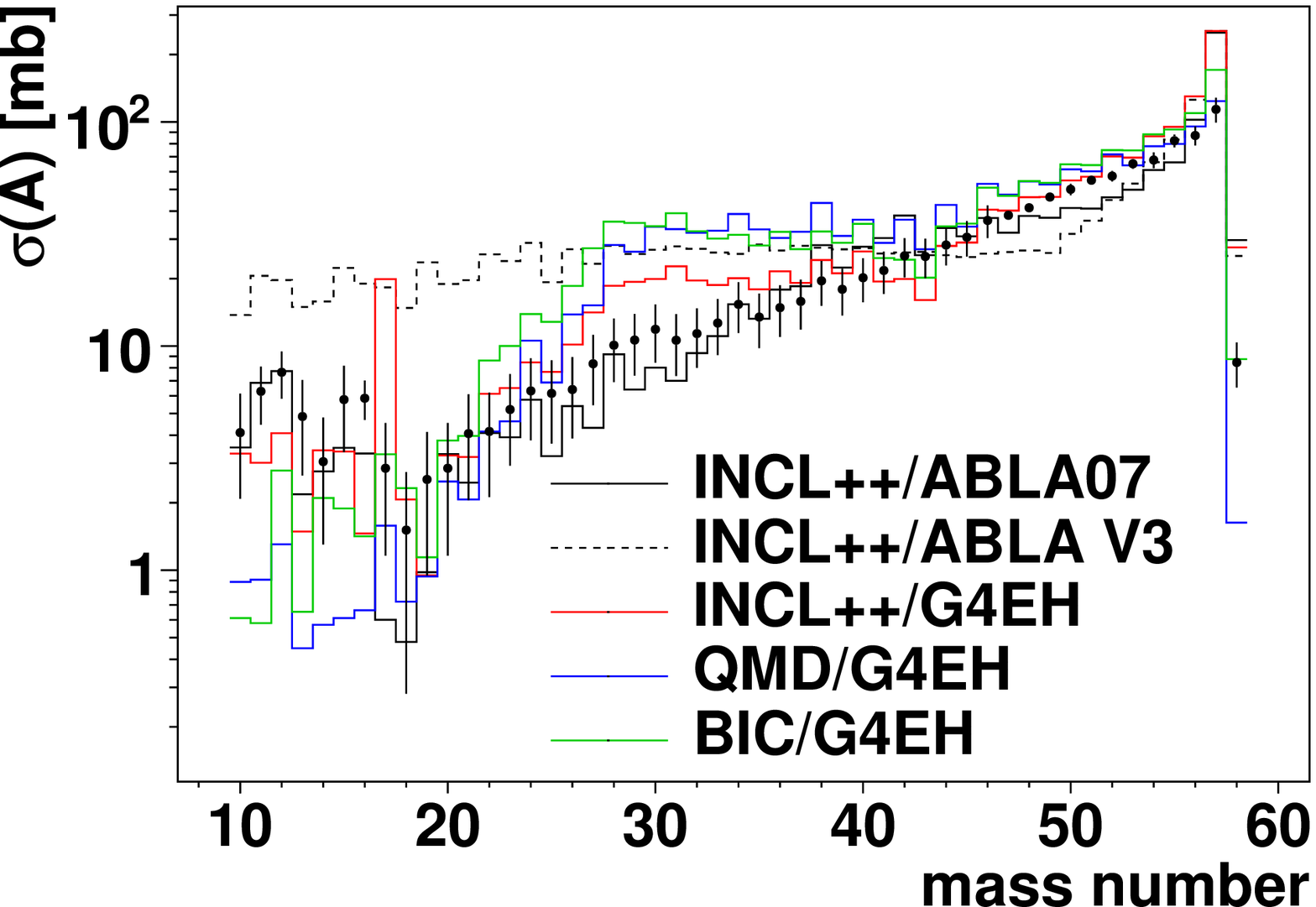}
  \caption{(Color online) Fragmentation cross sections for the 140-$A$MeV $^{58}$Ni+$^9$Be
    reaction, as a function of the fragment mass number. Model calculations are
    compared to the data taken from
    Refs.~\citenum{mocko-Ca40_Ca48_Ni58_Ni64,mocko-phd}. In the plot legend,
    \emph{G4EH} stands for \geh.\label{fig:Ni_Be_mass}}
\end{figure}

\begin{figure*}
  \includegraphics[width=\linewidth]{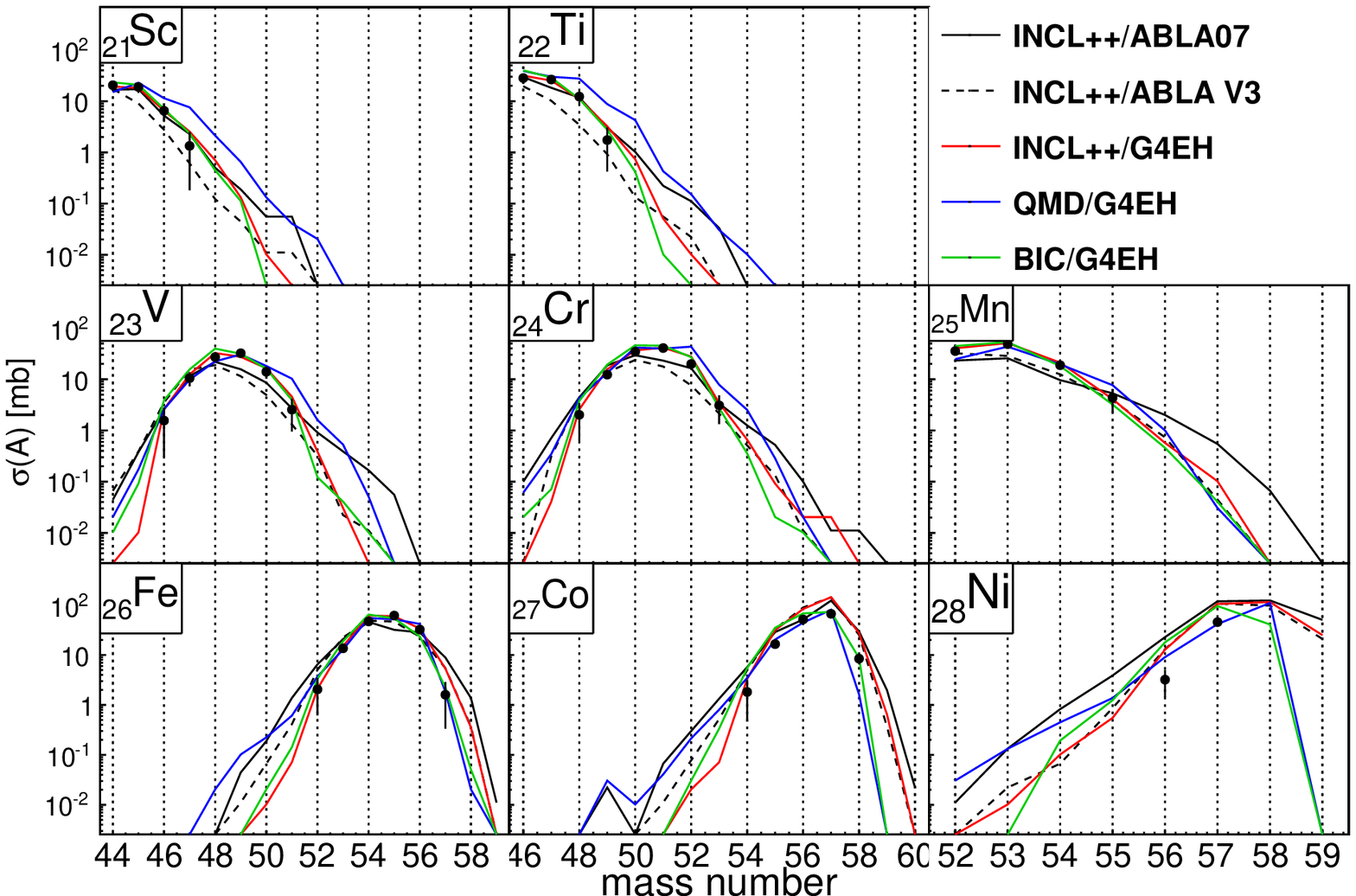}
  \caption{(Color online) Isotopic distributions ($21\leq Z\leq28$) for the 140-$A$MeV
    $^{58}$Ni+$^9$Be reaction. Model calculations are compared to the data taken
    from Refs.~\citenum{mocko-Ca40_Ca48_Ni58_Ni64,mocko-phd}. In the plot
    legend, \emph{G4EH} stands for \geh.\label{fig:Ni_Be_isot}}
\end{figure*}

Similar conclusions can be drawn from the results at lower beam energy. We show
in Figs.~\ref{fig:Ni_Be_mass} and \ref{fig:Ni_Be_isot} the comparison between
the model predictions and the experimental data for 140-$A$MeV $^{58}$Ni+$^9$Be
\cite{mocko-Ca40_Ca48_Ni58_Ni64,mocko-phd}. Note that at this energy only about
10\% of the reaction cross section is generated by \inclxx's low-energy fusion
sector.

Again, most of the mass distribution is best predicted by \inclxx/\abla, with
the exception of nuclei close to the projectile $^{58}$Ni. The \inclxx/\geh\
result is similar but slightly less good (the $A=17$ cross section is largely
overestimated, but all the yield comes from the single isotope $^{17}$O). The
\bic/\geh\ and \qmd/\geh\ couplings are yet less good, and \inclxx/\ablaold\ is
overall the worst. This is easy to understand if one remembers that \ablaold\
cannot evaporate intermediate-mass fragments, which occur most abundantly in
light systems (such as $^{58}$Ni and $^{86}$Kr).

The \inclxx-based calculations systematically overestimate the cross sections
for very small mass losses ($\Delta A=1$ or $2$). It would be tempting to
interpret this in terms of lacking Pauli blocking on the first collision in the
Fermi sea of the cascade projectile. We have indeed verified that these cross
sections are decreased by roughly 10--20\% if Pauli blocking in the projectile
is introduced (not shown in Figs.~\ref{fig:Ni_Be_mass} and
\ref{fig:Ni_Be_isot}). This is however insufficient to cure the overestimation,
which in the worst case (the yield for $A=57$) is close to a factor of $2.5$.

The isotopic distributions in Fig.~\ref{fig:Ni_Be_isot} are qualitatively
similar to those of Fig.~\ref{fig:Kr_Be_isot}, but one has to bear in mind that
the experimental coverage is less extensive here. It is difficult to verify if
\inclxx/\abla\ and \qmd/\geh\ overestimate the yields for neutron-rich residues,
as they do in 500-$A$MeV $^{86}$Kr+$^9$Be.

\begin{figure}
  \includegraphics[width=\linewidth]{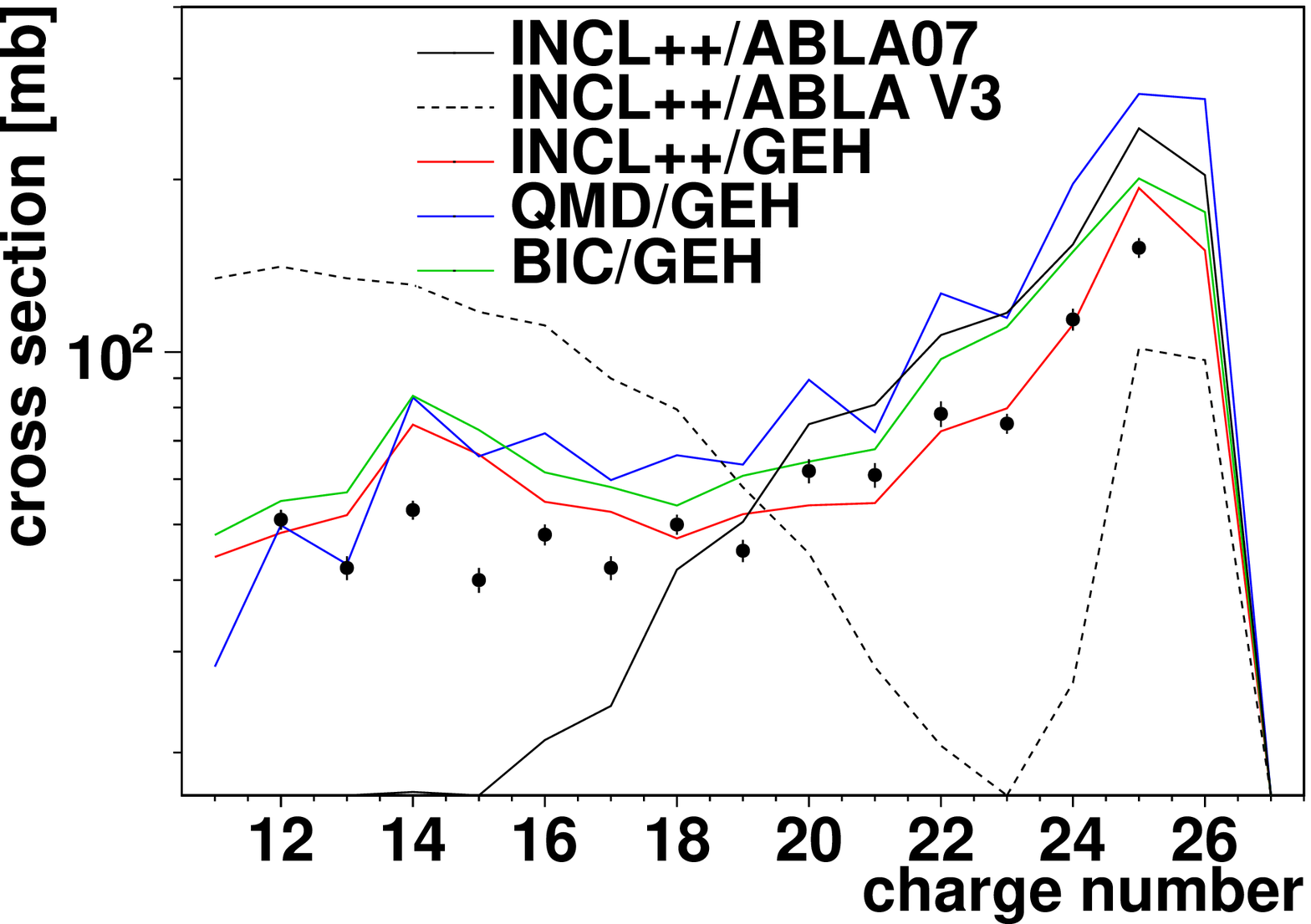}
  \caption{(Color online) Partial charge-changing cross sections for the 1.05-$A$GeV
    $^{56}$Fe+$^{12}$C reaction. Model calculations are compared to the data
    taken from Refs.~\citenum{zeitlin-fe1000}.\label{fig:Fe_C_cc}}
\end{figure}

\begin{figure}
  \includegraphics[width=\linewidth]{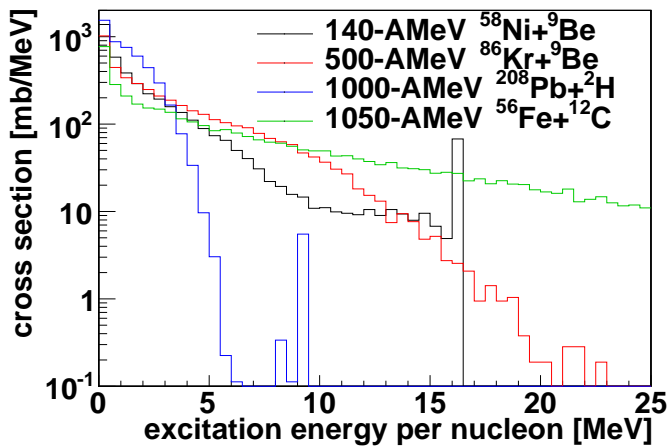}
  \caption{(Color online) Distributions of the excitation energy of the cascade remnants, as
    calculated by \inclxx, for the reactions shown in this
    section.\label{fig:AA_EStarRem}}
\end{figure}

\begin{table}
  \caption{Average characteristics of the projectile-like cascade remnant for
    the reactions studied in Section~\ref{sec:fragm-cross-sect}.\label{tab:AA_remnants}}
  \begin{ruledtabular}
    \begin{tabular}{c|c|c|c|c}
      \multirow{2}{*}{reaction} &\multirow{2}{*}{$A$} & \multirow{2}{*}{$Z$} & excitation energy & spin\\
      & & & per nucleon (MeV) &($\hbar$) \\
      \hline
      140-$A$MeV $^{58}$Ni+$^9$Be & 56.6 & 27.2 & 2.3 & 33.6\\
      500-$A$MeV $^{86}$Kr+$^9$Be & 76.6 & 32.0 & 3.0 & 37.1\\
      1000-$A$MeV $^{208}$Pb+$^2$H & 199.6 & 78.7 & 1.2 & 22.9\\ 
      1050-$A$MeV $^{56}$Fe+$^{12}$C & 44.9 & 20.8 & 6.8 & 60.0\\
    \end{tabular}
  \end{ruledtabular}
\end{table}

We conclude this section by discussing the model predictions for partial
charge-changing cross sections for a 1.05~$A$-GeV $^{56}$Fe projectile colliding
with a $^{12}$C target \cite{zeitlin-fe1000}. Of all the reactions so far
considered, $^{56}$Fe+$^{12}$C is the one that leads to the highest excitation
energies per nucleon, due to the high kinetic energy and the relatively small
size of the projectile nucleus. This is illustrated by
Fig.~\ref{fig:AA_EStarRem}, which compares the distributions of the excitation
energy of the projectile-like cascade remnants, as calculated by \inclxx, for
all the reactions studied in this section. The average excitation energies are
reported in Table~\ref{tab:AA_remnants}. At sufficiently large excitation
energy, multifragmentation is expected to become the dominant de-excitation
mechanism. Among the considered de-excitation models, \abla\ is the only one to
feature a semi-empirical treatment of multifragmentation. The \geh\ model does
include a multifragmentation module, but it is deactivated by default.

The model calculations are compared with the experimental data in
Fig.~\ref{fig:Fe_C_cc}. One remarks that the \inclxx/\ablaold\ prediction is
poor. We have already observed above that \ablaold\ is not suitable for systems
for which there is a large probability of evaporating intermediate-mass
fragments. The 1.05~$A$GeV $^{56}$Fe+$^{12}$C reaction surely falls within this
category. The \inclxx/\geh\ and \bic/\geh\ predictions are quite similar and in
good agreement with the data, while the \qmd/\geh\ cross sections are slightly
too large. Finally, \inclxx/\abla\ is close to the experimental data for
$Z\gtrsim19$, but severely underpredicts the data for the smallest charges.

It is perhaps surprising to observe that the cross sections for large charge
losses are best reproduced using de-excitation models that \emph{neglect}
multifragmentation. \abla\ is the only model that somehow tries to handle this
mechanism, but the comparison with the data seems to indicate that its
semi-empirical treatment is inadequate for the very large excitation energies
that can be reached in this reaction. On the other hand, it is known that
sequential binary decay can generate fragment partitions that are similar to
those generated by multifragmentation \cite{mancusi-multifragmentation}. More
discriminating observables would be needed to illustrate the difference between
the two de-excitation modes.

\begin{figure}
  \includegraphics[width=\linewidth]{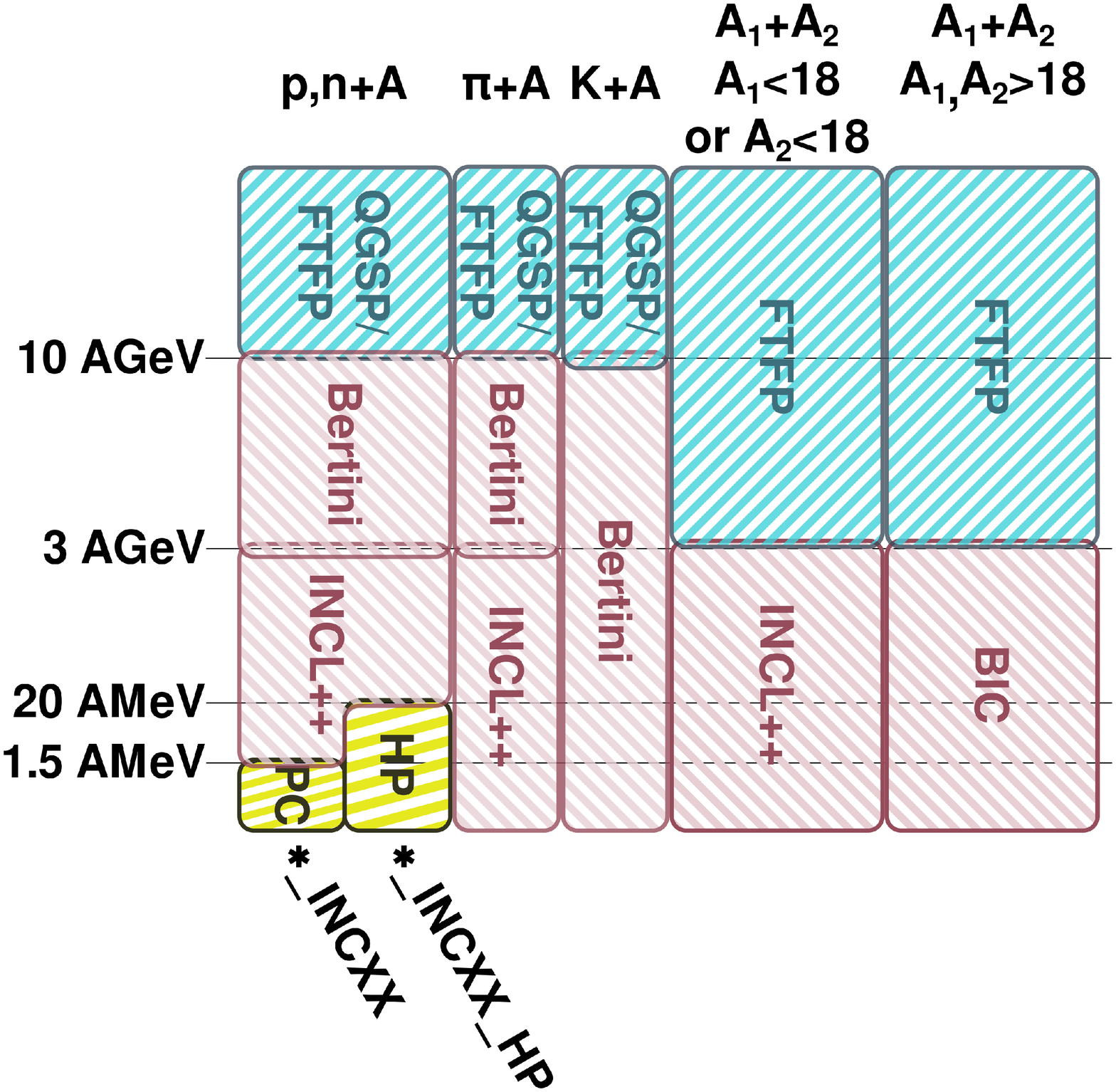}
  \caption{(Color online) Map of models used by the \inclxx-based physics lists
    in \geant\ \code{v10.0}. Physics lists whose name ends in \code{\_HP} use
    the \code{NeutronHP} model for neutron transport at low energies
    (represented as ``HP'' on the map); those starting with \code{QGSP\_}
    (\code{FTFP\_}) use the Quark-Gluon-String model (the Fritiof model) at high
    energy. ``PC'' stands for PreCompound.\label{fig:hadmodmap}}
\end{figure}

\begin{figure}
  \includegraphics[width=\linewidth]{{{hadmodmap_10.1beta_4cols}}}
  \caption{(Color online) Same as Fig.~\ref{fig:hadmodmap} for the \inclxx-based physics lists
    in \geant\ \code{v10.1$\beta$}.\label{fig:hadmodmap_10.1beta}}
\end{figure}

\begin{figure*}
  \includegraphics[width=\linewidth]{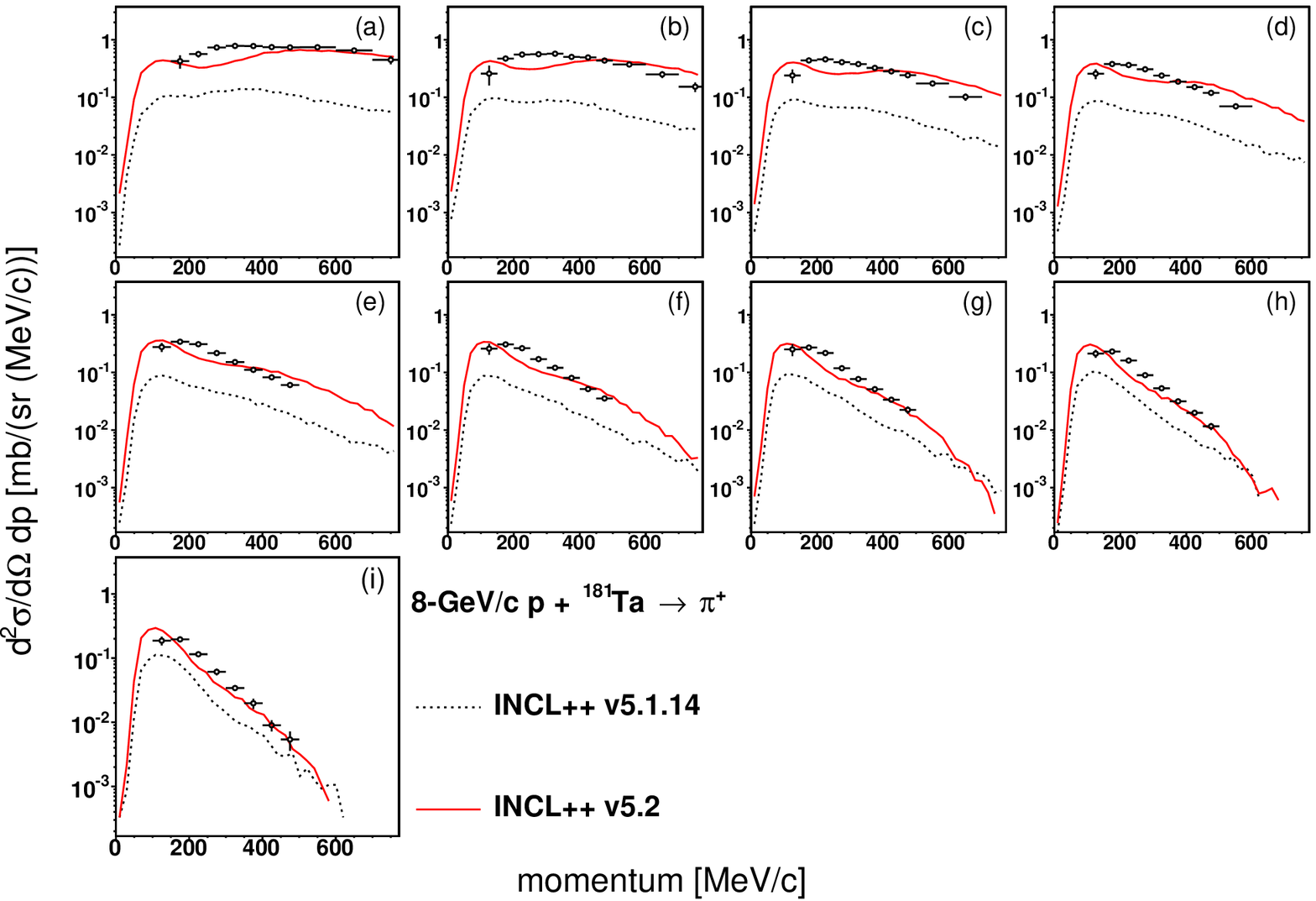}
  \includegraphics[width=\linewidth]{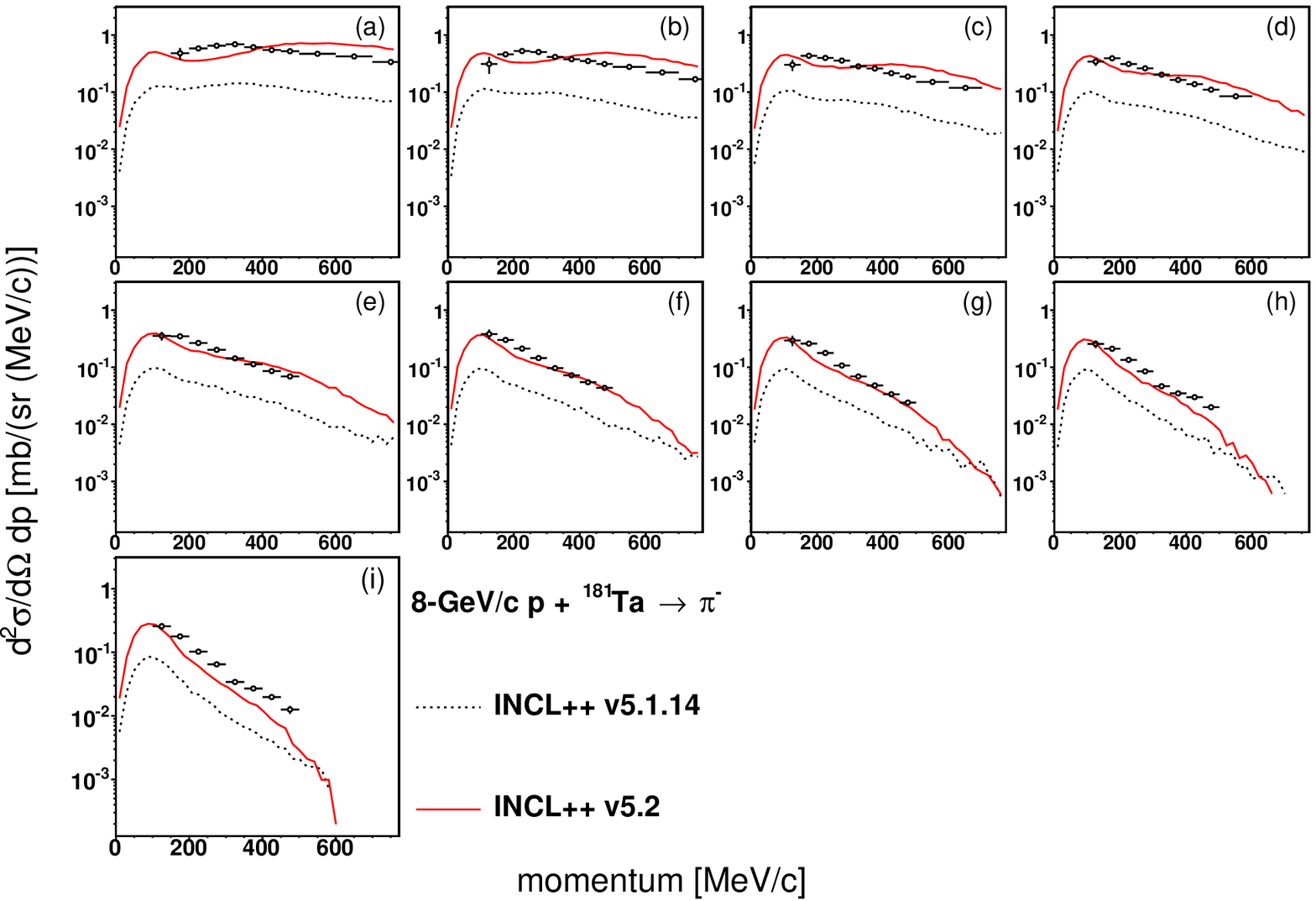}
  \caption{(Color online) Double-differential cross sections for the production
    of $\pi^+$ (top) and $\pi^-$ (bottom) at (a) $25^\circ$, (b) $37^\circ$, (c)
    $48^\circ$, (d) $60^\circ$, (e) $71^\circ$, (f) $83^\circ$, (g) $94^\circ$,
    (h) $105^\circ$ and (i) $117^\circ$ from a 8-GeV${}/c$
    \proton+$^{181}$Ta reaction. The \inclxx\ calculations are compared to the
    data from Ref.~\citenum{catanesi-harp}.\label{fig:harp_pipm}}
\end{figure*}

\subsection{Summary}

For the benefit of the reader, we shall now try to condense the vast amount of
information about nucleus-nucleus reactions that we have presented in this
section into a few general observations. Our summary might contain some degree
of subjectivity, but it should be read as an attempt to guide users of reaction
models towards the best choice for their needs. Our conclusions are also based
on the results of other comparisons with experimental data that have been
omitted for the sake of conciseness.

As far as neutron production is concerned, the choice of the de-excitation model
generally represents a second-order effect, except for the case of very light
projectiles (say alphas and lighter). The most accurate reproduction of the
experimental data is probably guaranteed by \qmd. Still, \inclxx\ is very close
to \qmd\ in terms of quality (and sometimes better in specific kinematical
regions; Figs.~\ref{fig:ddxs_C_C_n_models}--\ref{fig:ddxs_a_Cu_n_models}), while
being much faster. For very light projectiles, the choice of the Fermi momentum
can play an important role in the projectile-fragmentation kinematical region
(Fig.~\ref{fig:ddxs_a_Cu_n_inclxx}).

The scenario for proton production is less detailed than the neutron case
because of the limited experimental coverage. Nevertheless, for reasons that are
not clear to us, \qmd\ seems to perform less well than for neutrons, which is
rather surprising. The \inclxx\ predictions are of rather good quality
(Figs.~\ref{fig:angdist_dudouet}--\ref{fig:ddxs_C12_C_95_p}).

Light charged particles are best reproduced by \inclxx\
(Figs.~\ref{fig:angdist_dudouet} and \ref{fig:ddxs_C12_C_95_a}). This is
somewhat expected, since \inclxx\ is the only dynamical model considered to
include a dedicated mechanism for LCP production. \qmd's nucleon-nucleon
interaction is able to coalesce escaping nucleons, but it is known that the
resulting LCP spectra are not in good agreement with the experimental data
\cite{watanabe-qmd_clusters}. Note that the choice of the de-excitation model
\modified{is very} important for this observable, insofar as \ablaold\ can only
evaporate neutrons, protons and alphas.

The production of residual nuclei is in general sensitive to the choice of the
de-excitation model, except for small mass losses with respect to the
pre-fragment. Generally speaking, \ablaold\ and \abla\ have a long historical
record of applications to the de-excitation of heavy nuclei (say $A\gtrsim150$)
\cite[e.g.][]{boudard-incl}; as a consequence, the treatment of fission is quite
sophisticated in both versions of the model. It should be stressed, however,
that fission models typically contain a great deal of free parameters, which are
sometimes adjusted in relation to a specific dynamical model. Because of this,
the quality of the predictions of the very same fission model can vary wildly if
different dynamical models are used in the entrance channel. Models of the
\ablabare\ family have often been used in conjunction with the \incl\ cascades
and generally perform rather well in reactions induced by LCPs
(Fig.~\ref{fig:isot_Pb_d}) and especially in nucleon-nucleus reactions
(Fig.~\ref{fig:isot_Pb}). The fission parameters of the \geh\ model have been
adjusted in conjunction with \bic\ \cite{quesada-g4excitationhandler}; thus, the
fission cross sections in Fig.~\ref{fig:isot_Pb} are correctly reproduced by
\bic/\geh\ (not shown in the figure, see
Ref.~\citenum{quesada-g4excitationhandler}), but they are overestimated by
\inclxx/\geh.
\modified{We are considering the possibility to restore the previous parameter
  values when coupling \geh\ with \inclxx; this should already yield better
  results in view of the fact that \geh\ is rather similar to Furihata's \gem\
  model \cite{furihata-gem} and the latter performs reasonably well with
  \inclf. Another option would be to perform an \incl-specific adjustment of the
  fission parameters. Finally, for nucleus-nucleus reactions,}
the possibility of large mass losses during the dynamical reaction stage makes
the validation of fission models considerably more difficult; for these cases,
the \ablaold/\abla\ fission models should globally offer acceptable
performances.

The fragmentation of light nuclei ($A\lesssim150$) and the production of deep
spallation residues from heavy nuclei generally require de-excitation mechanisms
other than the conventional neutron-, proton- and alpha-evaporation
channels. This is especially true in high-energy nucleon-nucleus reactions and
even more so in nucleus-nucleus reactions, where the large pre-fragment
excitation energies can favor the emission of small nuclei and/or induce
multifragmentation. The \ablaold\ model is severely limited in this respect,
insofar as it does not include evaporation of any particle with $A>4$. Indeed,
our comparison shows that \inclxx/\ablaold\ is unsuitable for the description of
the fragmentation of light systems
(Figs.~\ref{fig:Kr_Be_mass}--\ref{fig:Fe_C_cc}), even for small mass losses.

The other de-excitation models considered here (\abla\ and \geh) do not suffer
from this limitation. The best agreement is generally observed for
\inclxx/\abla, except for systems where multifragmentation plays a major
role. The \bic/\geh, \inclxx/\geh\ and \qmd/\geh\ couplings produce predictions
of similar fair quality and are all reasonable choices within the \geant\
framework. The CPU time is roughly of the same order of magnitude for \bic\ and
\inclxx, but it is typically much larger for \qmd.

\section{Conclusions}\label{sec:conclusions}

We have presented for the first time the new \cxx\ incarnation of the Li\`ege
Intranuclear Cascade model, a solid, modern code that is intended to be used as
the base for any future development. The \inclxx\ code is feature-wise and
physics-wise equivalent to its \fortran\ counterpart as far as nucleon- and
pion-induced reactions are concerned. Small differences exist for reactions
induced by light charged particles. The new code can be used for thick-target
calculations through the \geant\ toolkit for particle transport.

The new \inclxx\ code can also accommodate reactions induced by light ions (up
to $A=18$). We have described the crucial elements of the extension and we have
discussed the limitations of our
approach\modified{, which is admittedly more phenomenological than the core of
  the model.}
A broad comparison with heterogeneous observables has shown that, in spite of
the conceptual difficulties, the extended \inclxx\ model yields predictions in
fair agreement with the considered experimental data. In comparison to other
models for nucleus-nucleus reactions available in \geant, \inclxx\ stands out as
one of the most viable options; it is however crucial (and we have issued
recommendations in this sense) to make a suitable choice for the coupling with
the statistical de-excitation model. We conclude that our extended model is
successful at capturing the physics that is essential for the description of
inclusive observables from reactions induced by light nuclei.

Future work on \inclxx\ will proceed along several directions. First, we shall
try to improve on the limits of the present nucleus-nucleus collision model,
starting with the inclusion of Pauli blocking in the Fermi sea of the
projectile.
\modified{Second, we will work on providing an all-round well-performing model
  for \geant\ users, which should ideally combine the advantages of \geh\ and
  \abla. Third,}
we will perform an extensive verification of the newly extended model in the
$3$--$15$~$A$GeV incident-energy range. Fourth, we plan to introduce the
strangeness degree of freedom. This will provide the means to develop
predictions for the production of kaons and hyperons and to simulate
kaon-induced reactions. We ultimately aim at making predictions for the
production of hypernuclei, although this also requires a strangeness-aware
de-excitation model.

\section*{Acknowledgments}

This work has been partially supported by the EU ENSAR FP7 project (grant
agreement 262010). We thank Dr.~Stepan Mashnik for kindly directing us to the
manuscript of Ref.~\citenum{mocko-phd}.

\appendix*

\section{\inclxx\ in \geant}
\label{sec:inclxx-geant}

A stand-alone version of the \inclxx\ code is available on request via
\href{http://irfu.cea.fr/Sphn/Spallation/incl.html}{the official \incl\ web
  site}.
This code can simulate any thin-target reaction and produces output in ROOT
format.  \modified{Couplings to de-excitation models are also provided.}

However, if one needs to simulate reactions in a thick absorber, the stand-alone
code is not sufficient and one needs to turn to full-fledged particle-transport
simulations. The \geant\ toolkit for particle transport has been including some
version of the Li\`ege Intranuclear-Cascade model since \code{v9.1} (released in
December 2007). The \inclxx\ code was first introduced in \code{v9.5} (December
2011).

In recent versions of \geant, it is possible to use the \inclxx\ model by
selecting one of the following dedicated physics lists:
\begin{itemize}
\item \code{QGSP\_INCLXX} (available since \code{v9.5})
\item \code{QGSP\_INCLXX\_HP}  (since \code{v10.0})
\item \code{FTFP\_INCLXX}  (since \code{v10.0})
\item \code{FTFP\_INCLXX\_HP}  (since \code{v10.0})
\end{itemize}
The \code{*\_HP} variants use the \code{NeutronHP} model below 20 MeV to
simulate neutron elastic and inelastic scattering using evaluated data
libraries. The \code{QGSP\_*} and \code{FTFP\_*} variants respectively use the
Quark-Gluon String model (QGS) and the Fritiof model (FTF) at high energy. For
low-energy nucleon-induced reactions, the Precoumpound model is used below 1~MeV
and \inclxx\ fades in between 1 and 2~MeV; the Binary-Cascade model is used for
reactions between heavy nuclei. Bertini is used for reactions induced by kaons,
which cannot be treated by \inclxx\ at the moment.  A map of models (accurate as
of \geant\ \code{v10.0}) is shown in Fig.~\ref{fig:hadmodmap}. For further
details about all the \geant\ models, the reader is referred to the \geant\
Physics Reference Manual \cite{g4-phys-ref-man}.

\subsection{Recommendations for the choice of the de-excitation model in \geant}

We have shown (Sections~\ref{sec:comp-with-inclf} and
\ref{sec:comparison-exp-data}) that fragmentation cross sections are very
sensitive to the choice of the de-excitation model. Since \geant\ \code{v10.0},
it is possible to choose to couple \inclxx\ to \geh\ (default) or to \ablaold;
therefore, we provide some guidelines for the users hereafter.

We can summarize some of the results presented in Secs.~\ref{sec:part-prod} and
\ref{sec:fragm-cross-sect}. \ablaold\ describes rather well most of the
observables connected with the de-excitation of heavy nuclei (say
$A\gtrsim150$); this conclusion relies partly on the results of the present work
(Figs.~\ref{fig:isot_Pb_d}--\ref{fig:pb_d_isot_spal}) and mostly on a large body
of validation for nucleon-induced reactions \cite[e.g.][]{boudard-incl}. There
are however a few observables that are not accounted for by \ablaold, even on
heavy systems, such as those connected with evaporation of deuterons, tritons,
$^3$He or fragments with $A>5$.

For light systems, we show below that \geh\ often provides a better description
of de-excitation than \ablaold. In addition, \geh\ provides de-excitation
mechanisms for the evaporation of any fragment up to $^{28}$Mg. On the other
hand, \geh's fission sector performs less well than \ablaold's fission module
when coupled with \inclxx. Again, this is shown in the present paper
(Fig.~\ref{fig:isot_Pb_d}) but is also confirmed by an extensive private
intercomparison.

We therefore recommend that users employ \geh\ (the default choice) when they
expect emphasis to be put on the de-excitation of light nuclei ($A\lesssim150$)
and/or on specific observables that are most probably incorrectly described by
\ablaold\ (such as tritium production). We recommend use of \ablaold\ when
emphasis is to be put on fission. The \geant\ Application Developer Guide
\cite[section~5.2.2.4]{g4-app-dev-guide} describes the steps necessary to couple
\inclxx\ to \ablaold\ within \geant.

\modified{The present situation is clearly unsatisfactory insofar as none of the
  available de-excitation models yields good overall performance. The
  development of such a model is one of our development goals.}

\subsection{Newer \inclxx\ version in \geant\ \code{v10.1$\beta$}}
\label{sec:inclxx-geant-v10.1b}

At the time of writing, a newer version of \inclxx\ (\code{v5.2}) has been
distributed with the latest public release of \geant\ (\code{v10.1$\beta$}). The
essential difference with the model described by the present paper
(\code{v5.1.14}) consists in the extension towards higher incident energies
\cite{pedoux-phd,pedoux-pions}. This work had initially been carried out in the
framework of an old \fortran\ version of \incl\ and has recently been merged in
the \fortran\ development version and converted to \cxx\ for inclusion in
\inclxx. The essential ingredient for the extension is the inclusion of new
inelastic channels in the elementary nucleon-nucleon and pion-nucleon
collisions. We do not introduce additional baryonic resonances (besides the
narrow $\Delta(1232)$) because they are largely overlapping and very
short-lived, compared to the time between subsequent cascade
collisions. Instead, the inelastic collisions are assumed to proceed directly to
multiple-pion production. Final-state pion multiplicities up to 4 are
considered, which pushes the high-energy limit of \inclxx\ \code{v5.2} up to
$\sim$15~GeV in nucleon- and pion-induced reactions. Note that the high-energy
extension does not substantially modify the results of the code below
$\sim1$~GeV.  Further details are available in Refs.~\citenum{pedoux-phd} and
\citenum{pedoux-pions}.

Figure~\ref{fig:hadmodmap_10.1beta} above shows a map of models for the
\inclxx-based physics lists in \geant\ \code{v10.1$\beta$}. \inclxx\ is used up
to 15~GeV for pion- and nucleon-induced reactions, and it is gradually replaced
by the relevant high-energy string model between 15 and 20~GeV.

An extensive verification of the new \inclxx\ version has been performed and
will be the object of a future publication. As a sample of the quality of the
new predictions, we show in Fig.~\ref{fig:harp_pipm} the calculations results
for double-differential cross sections for pion production from 8-GeV${}/c$
\proton+$^{181}$Ta.

\clearpage


\input{inclxx-hi.bbl}

\end{document}

%% file: inclxx-hi.bbl
%